\documentclass[acmsmall,xcolor=table]{acmart}

\usepackage{etoolbox}

\usepackage[inline]{enumitem}
\usepackage{mdframed}
\usepackage{xcolor}
\usepackage{caption}
\usepackage{subcaption}
\usepackage{listings}
\usepackage{xspace}
\usepackage{color}
\usepackage{multirow}
\usepackage{comment}
\usepackage{pifont}
\definecolor{deepblue}{rgb}{0,0,0.5}
\definecolor{deepred}{rgb}{0.6,0,0}
\definecolor{deepgreen}{rgb}{0,0.5,0}
\definecolor{darkgreen}{RGB}{43,163,39}
\definecolor{bluesquare}{rgb}{126,166,224}
\def\tightcol{\hskip 6pt}

\usepackage[utf8]{inputenc}
\usepackage[T1]{fontenc}
\usepackage{zlmtt}
\usepackage{fancyvrb}

\usepackage{siunitx}
\usepackage{longtable}
\usepackage{lipsum}

\usepackage{ltxtable}
\usepackage{filecontents}

\usepackage{listings}
\usepackage{parcolumns}

\newcommand{\partic}[1]{\textcolor{black!70!white}{\textsf{\smaller (P#1)}}}

\newcommand{\inlineQuote}[2]{{\smaller \fontfamily{cmss}\selectfont ``#1'' \partic{#2}}}

\renewenvironment{quote}
               {\list{}{\rightmargin\leftmargin}%
                \fontfamily{cmss}\item\relax\smaller\ignorespaces}
               {\unskip\unskip\endlist}

\usepackage{ragged2e}
\usepackage{booktabs, tabularx}
\newcolumntype{L}{>{\setlength{\baselineskip}{0.5\baselineskip}\RaggedRight\hangafter=1\hangindent=1em\arraybackslash}X}

\usepackage{array}
\newcolumntype{P}[1]{>{\setlength{\baselineskip}{0.7\baselineskip}\raggedright\everypar{\hangindent=1em\hangafter=1}\arraybackslash}p{#1}}
\newcolumntype{R}{>{\everypar\expandafter{\the\everypar\dohang\everypar{\dohang}\dohang\raggedright{\hangindent=4em}\arraybackslash}\arraybackslash}p{6cm}}
\newcolumntype{Q}{>{\hangindent=1em\hangafter=1}X}

\usepackage{graphicx,caption}

\lstdefinestyle{pythoncode}{
  language=Python,
  morekeywords={self,join,append,split,write,chain,sub,lower,strip,decode,choice,listdir,savefig,expanduser,translate,read_csv,drop},             
  keywordstyle=\bfseries\color{deepblue},
  emphstyle=\color{deepred},    
  showstringspaces=false,
  breaklines=true,
  escapeinside=||,
  columns=fullflexible,
  basicstyle=\fontfamily{cmtt}\small,
}

\newcommand{\inlineCode}{\lstinline[mathescape,style=pythoncode]}

\newcommand{\refmark}{\ding{51}\xspace}
\newcommand{\cmark}{$\clubsuit$\xspace}
\newcommand{\xmark}{$\spadesuit$\xspace}

\usepackage{dcolumn}
\usepackage{xspace}

\newcommand{\ie}{{\emph{i.e.}},\xspace}

\newcommand{\eg}{{\emph{e.g.}},\xspace}
\newcommand{\etc}{\emph{etc.}\xspace}

\newcommand\numparticipants{31\xspace}
\newcommand\numapplicants{64\xspace}
\newcommand\telemetrydata{170\xspace}
\newcommand\numresults{7\xspace}
\newcommand\numoraclequeries{50\xspace}
\newcommand\numqueries{397\xspace} 

\usepackage{soul}
\usepackage{bm}

\newcounter{RQCounter}
\newcounter{HCounter}
\newcounter{RSCounter}
\newcommand{\RQ}[2]{%
\refstepcounter{RQCounter} \label{#1}
\textbf{RQ}$_{\bm{\arabic{RQCounter}}}$.~\emph{#2}
}

\newcommand{\RQref}[1]{%
\textbf{RQ}$_{\textbf{\ref{#1}}}$\xspace
}

\usepackage{framed}

\newcommand{\mysec}[1]{\vspace{0.1cm} \noindent \textbf{#1.}}

\usepackage{soul}
\definecolor{almond}{rgb}{0.94, 0.87, 0.8}
\definecolor{antiquewhite}{rgb}{0.98, 0.92, 0.84}
\definecolor{electriclavender}{rgb}{0.96, 0.73, 1.0}
\definecolor{rdd}{rgb}{0.85, 0.8, 0.95} 

\usepackage{quoting}
\quotingsetup{leftmargin=10pt,rightmargin=10pt,vskip=3pt}

\AtBeginDocument{%
  \providecommand\BibTeX{{%
    \normalfont B\kern-0.5em{\scshape i\kern-0.25em b}\kern-0.8em\TeX}}}

\setcopyright{acmlicensed}
\acmJournal{TOSEM}
\acmYear{2021} \acmVolume{1} \acmNumber{1} \acmArticle{1} \acmMonth{1} \acmPrice{15.00}\acmDOI{10.1145/3487569}

\acmJournal{TOSEM}
\acmVolume{37}
\acmNumber{4}
\acmArticle{111}
\acmMonth{8}

\begin{document}

\title[In-IDE Code Generation from Natural Language: Promise and Challenges]{In-IDE Code Generation from Natural Language: \\ Promise and Challenges}

\author{Frank F. Xu}
\email{fangzhex@cs.cmu.edu}
\affiliation{%
  \institution{Carnegie Mellon University}
  \streetaddress{5000 Forbes Ave.}
  \city{Pittsburgh}
  \state{PA}
  \postcode{15213}
}

\author{Bogdan Vasilescu}
\email{vasilescu@cs.cmu.edu}
\affiliation{%
  \institution{Carnegie Mellon University}
  \streetaddress{5000 Forbes Ave.}
  \city{Pittsburgh}
  \state{PA}
  \postcode{15213}
}

\author{Graham Neubig}
\email{gneubig@cs.cmu.edu}
\affiliation{%
  \institution{Carnegie Mellon University}
  \streetaddress{5000 Forbes Ave.}
  \city{Pittsburgh}
  \state{PA}
  \postcode{15213}
}


\begin{abstract}
A great part of software development involves conceptualizing or communicating the underlying procedures and logic that needs to be expressed in programs.
One major difficulty of programming is turning \emph{concept} into \emph{code}, especially when dealing with the APIs of unfamiliar libraries.
Recently, there has been a proliferation of machine learning methods for code generation and retrieval from \emph{natural language queries}, but these have primarily been evaluated purely based on retrieval accuracy or overlap of generated code with developer-written code, and the actual effect of these methods on the developer workflow is surprisingly unattested.
In this paper, we perform the first comprehensive investigation of the promise and challenges of using such technology inside the PyCharm IDE, asking
``at the current state of technology does it improve developer productivity or accuracy, how does it affect the developer experience, and what are the remaining gaps and challenges?''
To facilitate the study, we first develop a plugin for the PyCharm IDE that implements a hybrid of code generation and code retrieval functionality, and orchestrate virtual environments to enable collection of many user events (e.g. web browsing, keystrokes, fine-grained code edits).
We ask developers with various backgrounds to complete 7 varieties of 14 Python programming tasks ranging from basic file manipulation to machine learning or data visualization, with or without the help of the plugin.
While qualitative surveys of developer experience are largely positive, quantitative results with regards to increased productivity, code quality, or program correctness are inconclusive.
Further analysis identifies several pain points that could improve the effectiveness of future machine learning based code generation/retrieval developer assistants, and demonstrates when developers prefer code generation over code retrieval and vice versa.
We release all data and software to pave the road for future empirical studies on this topic, as well as development of better code generation models.
\end{abstract}



\maketitle

\section{Introduction}
\label{sec:intro}


One of the major hurdles to programming is the time it takes to turn 
ideas into code~\cite{mohagheghi2007quality}.
All programmers, especially beginners but even experts, frequently reach points 
in a program where they understand conceptually what must be done next, but do 
not know how to create a concrete implementation of their idea, or would rather 
not have to type it in if they can avoid it.
The popularity of the Stack Overflow Q\&A website is a great example of this need. 
Indeed, developers ask questions about how to transform ideas into code all the 
time, \eg ``How do I check whether a file exists without exceptions?,''\footnote{\url{https://stackoverflow.com/q/82831}} 
``How can I merge two Python dictionaries in a single expression?,''\footnote{\url{https://stackoverflow.com/q/38987}} 
etc.
Moreover, this need is likely to continue in the future, as new APIs appear continuously
and existing APIs change in non-backwards compatible ways~\cite{murphy2018api}, 
requiring recurring learning effort~\cite{ko2004six, myers2016improving}.


Despite early skepticism towards the idea of ``natural language programming''%
~\cite{dijkstra1979foolishness}, researchers now widely agree on a range of scenarios
where it can be useful to be able to formulate instructions
using natural language and have the corresponding source code snippets automatically
produced.
For example, software developers can save keystrokes or avoid writing dull pieces 
of code~\cite{wei2015building, franks2015cacheca, raychev2014code, nam2019marble}; 
and non-programmers and practitioners in other fields, who require computation 
in their daily work,
can get help with creating data manipulation scripts~\cite{gulwani2011automating, le2014flashextract}. 

Given a natural language query carrying the intent of a desired step in a 
program, there are two main classes of methods to obtain code implementing 
this intent, corresponding to two major research thrusts in this area. 
On the one hand, \emph{code retrieval} techniques aim to search for and 
retrieve an existing code fragment in a code base; given the abundance of 
code snippets online, on platforms such as Stack Overflow, it is plausible that a 
lot of the code that one might write, especially for lower level functionality 
and API usage primitives, already exists somewhere, therefore the main 
challenge is search.
On the other hand, \emph{code generation} techniques aim to synthesize 
code fragments given natural language descriptions of intent. 
This is typically a harder challenge than retrieval and therefore more 
ambitious, but it may be particularly useful in practice if those exact 
target code fragments do not exist anywhere yet and can be generated instead.

The early attempts at general-purpose code generation from natural language 
date back to the early to mid 2000s, and resulted in
groundbreaking but relatively constrained grammatical and template-based 
systems, \eg converting English into Java~\cite{price2000naturaljava} 
and Python~\cite{vadas2005programming}.
Recent years have seen an increase in the scope and diversity of such
programming assistance tools, as researchers have devised 
code generation techniques that promise to be more flexible and expressive 
using machine (deep) learning models trained on data from ``Big Code'' 
repositories like GitHub and Stack Overflow; 
see \citet{allamanis2018survey} for an excellent survey of such techniques.
Code retrieval systems have also improved dramatically in recent years,
thanks to the increasing availability of source code online and more 
sophisticated information retrieval and machine learning techniques;
perhaps the most popular current code retrieval system is Microsoft's 
Bing Developer Assistant~\cite{wei2015building}, which is an adaptation
of the Bing search engine for code.

While both types of methods (generation and retrieval) for producing appropriate 
code given natural language intents have received significant interest in machine 
learning circles, there is a surprising paucity of research using human-centered approaches~\cite{myers2016programmers} to evaluate the usefulness and impact of
these methods \emph{within the software development workflow}.
An important open question is to what extent the typically high accuracy scores 
obtained during automatic evaluations on benchmark datasets will translate to 
real-world usage scenarios, involving software developers completing actual 
programming tasks.
The former does not guarantee the latter.
For example, an empirical study on code migration by \citet{tran2019does} 
showed that the BLEU~\cite{papineni-etal-2002-bleu} accuracy score commonly used in 
natural language machine translation has only weak correlation with the semantic 
correctness of the translated source code~\cite{tran2019does}.

In this paper, we take one step towards addressing this gap.
We implemented two state-of-the-art systems for natural language to code 
(NL2Code) generation and retrieval as in-IDE developer assistants, and
carried out 
a controlled human study with \numparticipants participants 
assigned to complete a range of Python programming tasks \textit{with and without} 
the use of the two varieties of NL2Code assistance.
Our results reveal that while participants in general enjoyed interacting 
with our IDE plugin and the two code generation and retrieval systems,
surprisingly \textit{there were 
no statistically significant gains in any measurable outcome when using the plugin}. 
That is, tasks with code fragments automatically generated or retrieved 
using our plugin were, on average, neither completed faster nor more correctly than tasks where participants did
not use any NL2Code assistant.
This indicates that despite impressive improvements in the intrinsic performance 
of code generation and retrieval models, there is a clear need to further improve the accuracy of code generation, and we may need to consider other extrinsic factors (such as providing documentation for the generated code) before such models can make sizable impact on the developer workflow.

In summary, the \textbf{main contributions} of this paper are: 
(i)~A hybrid code generation and code retrieval plugin for the Python PyCharm IDE,
that takes as input natural language queries.
(ii)~A controlled user study with \numparticipants participants observed across
7 types of programming tasks (14 concrete subtasks). 
(iii)~An analysis of both quantitative and qualitative empirical data collected
from the user study, revealing how developers interact with the NL2Code assistant 
and the assistant's impact on developer productivity and code quality.
(iv)~A comparison of code snippets produced by the two models, generation versus 
retrieval. 
(v)~An anonymized dataset of events from our instrumented IDE and virtual 
environment, capturing multiple aspects of developers' activity during the 
programming tasks, including plugin queries and edits, web browsing activities, 
and code edits.

\section{Overview of Our Study}
\label{sec:overview}

The goal of our research is to elucidate to what extent and in what ways current 
natural language programming techniques for code generation and retrieval 
can be useful within the development workflow as NL2Code developer assistants.
Our main interest is evaluating the usefulness in practice of state-of-the-art 
NL2Code \textit{generation} systems, which have been receiving significant 
attention from researchers in recent years, but have so far only been evaluated 
on benchmark datasets using standard NLP metrics.
However, as discussed above, code generation and code retrieval are closely
related problems, with increasingly blurred lines between them; \eg
recent approaches to align natural language intents with their corresponding
code snippets in Stack Overflow  for retrieval purposes~\cite{yin2018mining}
use similar deep learning technology as some code generation techniques~\cite{yin2017syntactic}.
Therefore, it is important to also consider code retrieval systems when 
experimenting with and evaluating code generation systems.

Given this complementarity of the two tasks, we select as a representative 
example of state-of-the-art techniques for code generation the semantic parsing 
approach by \citet{yin2017syntactic}. 
In short, the approach is based on a tree-based neural network model that 
encodes natural language utterances and generates corresponding syntactically 
correct target code snippets; for example, the model can generate the Python 
code snippet ``\texttt{x.sort(reverse=True)}'' given the natural language 
input ``sort list \texttt{x} in reverse order''.
We chose the approach by \citet{yin2017syntactic} over similar approaches
such as those of \citet{Iyer2018MappingLT} and \citet{Agashe2019JuICeAL} as it is the most 
general purpose and most naturally comparable to code retrieval approaches; 
see Section~\ref{sec:relatedwork} for a discussion.
For code retrieval, the closest analogue is Microsoft's proprietary Bing 
Developer Assistant~\cite{wei2015building}, which takes English queries 
as input and returns existing matching code fragments from the Web, using 
the Bing search engine. 
However, given the proprietary nature of this system, we build a custom 
Stack Overflow code search engine inspired by it rather than use the system 
itself.

\begin{figure}[t]
    \centering
    \includegraphics[width=\textwidth, clip=true, trim=0 0 155 0]{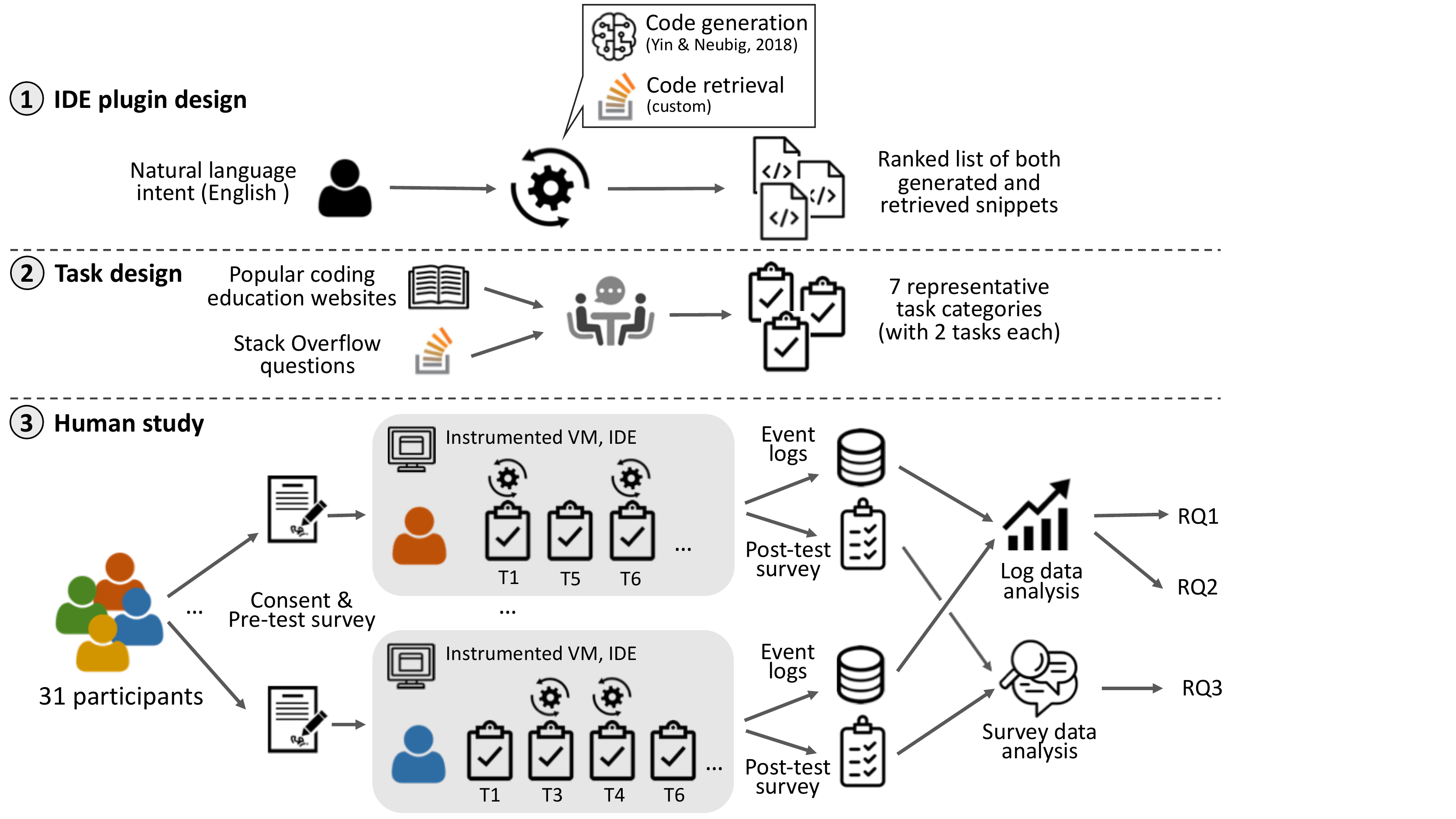}
    \caption{Overview of our study.}
    \label{fig:overview}
\end{figure}

We then designed and carried out the controlled human study summarized
in Figure~\ref{fig:overview}.
First, we implement the two code generation and retrieval techniques as a 
custom plugin for 
the PyCharm\footnote{\url{https://www.jetbrains.com/pycharm/}} IDE, which takes 
as input natural language text intents and displays as output the corresponding 
code snippets generated and retrieved by the respective underlying models.
Second, we compile 14 representative Python programming tasks across 7 task categories 
with varying difficulty, ranging from basic Python to data science topics.
Third, we recruit \numparticipants participants with diverse experience in 
programming in Python and with the different task application domains.
Then, using an instrumented virtual environment and our IDE plugin,
we collect quantitative and qualitative data about task performance and 
subjective tool use from each participant, as well as over \telemetrydata
person hours of telemetry data from the instrumented environment.

Finally, we analyze these data to answer three research questions, as follows. 

\smallskip
\RQ{task-performance}{How does using a NL2Code developer assistant 
    affect task completion time and program correctness?}
This research question investigates quantitatively differences in outcome variables between tasks completed in the treatment and control conditions. To this end, we use the log data from our instrumented virtual environment to compute task completion times, and rubric-based manual scoring of the solutions submitted by study participants to evaluate program correctness. Then, we use multivariate mixed-effects regression modeling to analyze the data. We expect that using the plugin developers can complete tasks faster, without compromising solution quality.

\smallskip
\RQ{types-of-code}{How do users query the NL2Code assistant, and how does that associate with their choice of generated vs retrieved code?}
This research question investigates quantitatively three dimensions of the inputs and outputs of the NL2Code plugin. 
Again using log data from our instrumented virtual environment, we first model how the natural language input queries differ when study participants favor the code snippets returned by the code generation model over those returned by the code retrieval model. Second, we evaluate the quality of the natural language queries input by study participants in terms of their ability to be answerable by an oracle (human expert), which is also important for the success of NL2Code systems in practice, in addition to the quality of the underlying code generation or retrieval systems.
Third, we study how the length and the frequency of different types of tokens changes after study participants edit the candidate code snippets returned by the NL2Code plugin, which could indicate ways in which even the chosen code snippets are still insufficient to address the users' needs.

\smallskip
\RQ{opinions}{How do users perceive the usefulness of the in-IDE 
    NL2Code developer assistant?}
Finally, this research question investigates qualitatively the experience of the study participants interacting with the NL2Code plugin and underlying code generation and retrieval models.


\smallskip
In the remainder of this paper, Sections~\ref{sec:plugin}--%
\ref{sec:study-design} describe our study setup in detail;
then Sections~\ref{sec:results:task-performance}--\ref{sec:perceptions} 
present our answers to the research questions;
Section~\ref{sec:discussionimplications} discusses implications; and Section~\ref{sec:relatedwork} discusses related work.

Following best practices for empirical software engineering research~\cite{wohlin2012experimentation,shull2007guide},
we make our study replicable, publishing our plugin prototype, 
instrumented virtual environment, data extraction 
and analysis scripts, and the obtained anonymized raw data;
see the online appendices at \url{https://github.com/neulab/tranX-plugin} and \url{https://github.com/neulab/tranX-study}.

\section{NL2Code IDE Plugin Design}
\label{sec:plugin}

We designed and built a joint NL2Code generation and retrieval plugin for 
PyCharm, a popular Python IDE.
Our plugin is open source and available online.\footnote{At  \url{https://github.com/neulab/tranX-plugin}}
As mentioned above, the plugin takes as input an English query describing the 
user's intent, and gives as output a ranked list of the most relevant code 
snippets produced by each of the two underlying code generation and retrieval 
systems.
Using IDE plugins to query Web resources such as Stack Overflow is expected to
be less disruptive of developers' productivity than using an external Web browser, 
since it reduces context switching~\cite{bacchelli2012harnessing,ponzanelli2013seahawk}.
Moreover, there exist already a number of IDE plugins for Web / Stack Overflow
search and code retrieval~\cite{ponzanelli2013seahawk,campbell2017nlp2code,rahman2014towards,wei2015building}, therefore the human-computer interaction modality should
feel at least somewhat natural to study participants.

\mysec{The Underlying Code Generation System}
For code generation, we use the model by \citet{xu2020incorporating} 
(available online\footnote{\url{https://github.com/neulab/external-knowledge-codegen}}),
which is an improved version of the tree-based semantic parsing model by 
\citet{yin2018tranx}, further pre-trained on official API documentation in
addition to the original training on Stack Overflow questions and answers.%
\footnote{
We deployed the model on an internal research server and exposed a HTTP 
API that the plugin can access; queries are fast enough for the plugin to
be usable in real time.
}

This model reports state-of-the-art accuracy on the CoNaLa benchmark 
dataset~\cite{yin2018mining}, a benchmark dataset of intent/code pairs mined
from Stack Overflow and standardly used to evaluate code generation models. 
Accuracy is computed using the BLEU score~\cite{papineni-etal-2002-bleu}, 
a standard metric used in the NLP community, that measures the token-level 
overlap between the generated code and a reference implementation.
As discussed above, the BLEU score (and similar automated metrics) are typically 
not sufficiently sensitive to small lexical differences in token sequence that
can greatly alter the semantics of the code~\cite{tran2019does}, hence our current 
human-centered study.
Still, qualitatively, it appears that the model can generate reasonable code 
fragments given short text inputs, as shown in Table~\ref{tab:case}. 
Note how the model can generate syntactically correct code snippets by
construction; demonstrates ability to identify and incorporate a wide variety 
of API calls; and also has the ability to copy important information like 
string literals and variable names from the input natural language intent, in 
contrast to the code retrieval results.
When displaying multiple generation results in the plugin described below, these results are ordered by the conditional probability of the generated code given the input command.


\begin{table}[t]
\begin{center}
\begin{tabular}{l@{\tightcol}p{12cm}}
\toprule
\multicolumn{2}{l}{Open a file ``\texttt{f.txt}'' in write mode.} \\
\refmark & \inlineCode+f = open('f.txt', 'w')+  \\
\cmark & \inlineCode+f = open('f.txt', 'w')+  \\
\xmark & \inlineCode-with open("users.txt", "a") as f: f.write(username + "\n") -  \\
\midrule
\multicolumn{2}{l}{Remove first column of dataframe \textit{df}.} \\
\refmark & \inlineCode+df = df.drop(df.columns[[0]], axis=1)+  \\
\cmark & \inlineCode+df.drop(df.columns[[0]])+  \\
\xmark & \inlineCode-del df['column_name']-  \\
\midrule
\multicolumn{2}{p{12cm}}{Lower a string \textit{text} and remove non-alphanumeric characters aside from space.} \\
\refmark & \inlineCode+re.sub(r'[^\sa-zA-Z0-9]', '', text).lower().strip()+ \\
\cmark & \inlineCode+re.sub(r'[^\sa-zA-Z0-9]', '', text)+ \\
\xmark & \inlineCode+re.sub(r'[^\sa-zA-Z0-9]', '', text).lower().strip()+ \\
\bottomrule
\end{tabular}
\end{center}
\caption{Examples, where
\refmark is the ground-truth code snippet, \cmark is the output from the state-of-the-art code generation model, and \xmark is the first candidate retrieved from Stack Overflow using Bing Search.
}
\label{tab:case}
\end{table}

\mysec{The Underlying Code Retrieval System}
For code retrieval, similarly to a number of recent works on the subject \cite{ponzanelli2013seahawk,wei2015building,campbell2017nlp2code}, we implement a 
wrapper around a general-purpose search engine, specifically the Bing\footnote{\url{https://www.bing.com/}} search engine.%
\footnote{We chose Bing rather than other alternatives such as Google due to the availability of an easily accessible search API.}
The wrapper queries this search engine for relevant questions on Stack Overflow,\footnote{\url{https://stackoverflow.com/}} the dominant programming 
Q\&A community, and retrieves code from the retrieved pages.
A dedicated search engine already incorporates advanced indexing and ranking mechanisms in its algorithms, driven by user interaction data, therefore it is preferable to using the internal Stack Overflow search engine directly~\cite{wei2015building}.

Specifically, we add the ``Python'' prefix to all user queries to confine the search to the Python programming language domain, and add ``site:stackoverflow.com'' to confine the results to the Stack Overflow platform.
We do not structurally alter the queries otherwise, \eg we do not remove variables referenced therein, if any, although we do strip away grave accents that are part of the code generation model's syntax.\footnote{
To mitigate concerns that user queries using the specified syntax (command form sentences and including variable names) may adversely affect the retrieval results, after the full study was complete we modified 59 user-issued queries that were indeed complete sentences with full variable names, converting them into short phrases without variable names and re-ran the retrieval.
We then compared the results and manually annotated the number of times the search engine returned a result that we judged was sufficient to understand how to perform the programming task specified by the user's intent.
As a result, the user-written full intent resulted in a sufficient answer 34/59 times, and the simplified intent without variable names returned a sufficient answer 36/59 times, so it appears that including variable names has a marginal to no effect on whether the search engine was able to provide a good top-1 result. We also measured the exact-match overlap between the top-1 results, and found it to be 22/59, and overlap between the top-7 result lists was 182/(59*7).}
For the query example mentioned above, the actual query string for Bing search would become ``Python reverse a list x site:stackoverflow.com''.
For each Stack Overflow question page retrieved, we then extract the code snippets from the top 3 answers into a ranked list, sorted descending by upvotes.
The code snippet extraction procedure follows~\citet{yin2018mining} for identifying the code part of the answer, based on Stack Overflow-specific syntax highlighting and heuristics.
When displaying multiple retrieval results, these results are ordered by the order they appeared in Bing search engine results and the ordering of answers inside SO posts is done by upvotes.

Table~\ref{tab:case} shows a few example outputs.
Note how the retrieval results sometimes contain spurious code, not part of the 
natural language intent (first example), and otherwise seem to complement the 
generation results. 
Indeed, in the second example the generation result is arguably closer to 
the desired answer than the retrieval result, with the opposite situation in 
the third example.


\begin{figure}[t]
     \centering
     \begin{subfigure}[b]{0.48\textwidth}
         \centering
         \includegraphics[width=0.9\textwidth]{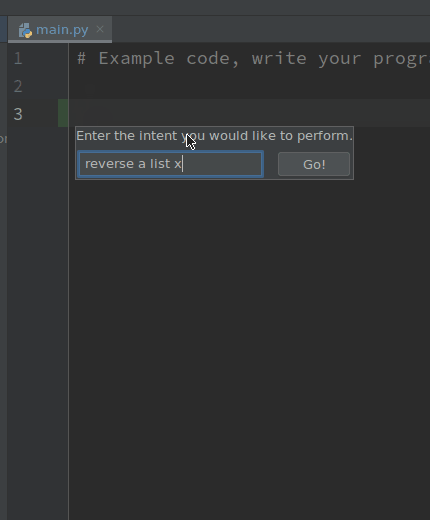}
         \caption{Query input interface}
         \label{fig:plugin_input}
     \end{subfigure}
     \quad
     \begin{subfigure}[b]{0.48\textwidth}
         \centering
         \includegraphics[width=0.9\textwidth]{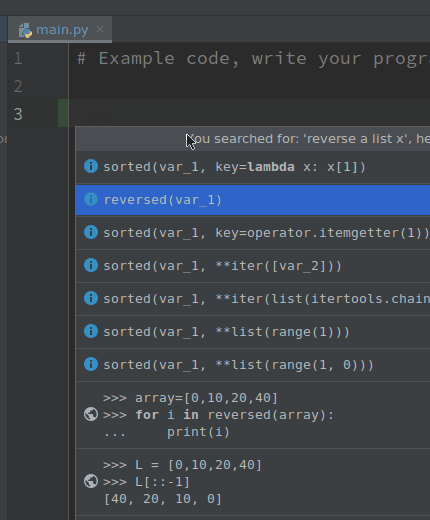}
         \caption{Code snippet candidates}
         \label{fig:plugin_candidates}
     \end{subfigure}
        \caption{Screenshots of the in-IDE plugin taking a natural language query as input and listing code snippet candidates from both code generation and code retrieval.}
        \label{fig:plugin_query}
\end{figure}

\begin{figure}[t]
     \centering
     \begin{subfigure}[b]{0.48\textwidth}
         \centering
         \includegraphics[width=\textwidth]{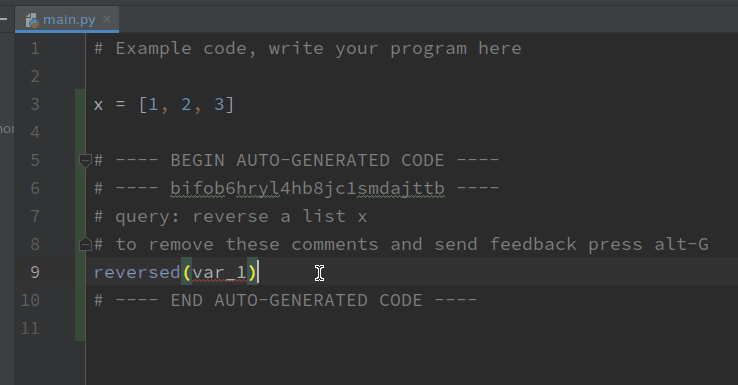}
         \caption{Generated code with errors in the context}
         \label{fig:plugin_error}
     \end{subfigure}
     \quad
     \begin{subfigure}[b]{0.48\textwidth}
         \centering
         \includegraphics[width=\textwidth]{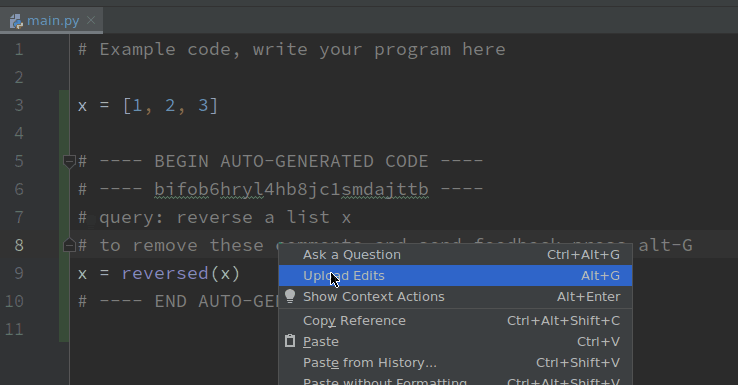}
         \caption{The user fixes the error and uploads}
         \label{fig:plugin_upload}
     \end{subfigure}
        \caption{Screenshots of fixing the small errors in generated code and upload the correct snippet.}
        \label{fig:plugin_fix}
\end{figure}

\mysec{Interacting With the Plugin}
Figure~\ref{fig:plugin_query} illustrates the plugin's user interface.
The user first activates the query interface by pressing a keyboard shortcut 
when the cursor is in the IDE's editor.
A popup appears at the current cursor position (Figure~\ref{fig:plugin_input}), 
and the user can enter a command in natural language that they would like to 
be realized in code (\eg ``reverse a list \`{}x\`{}''\footnote{Note the
special syntax used to mark explicit variables; see Appendix~\ref{app:plugin-syntax} 
for full syntax details.}).
The plugin then sends the request to the underlying code generation and code 
retrieval systems, and displays a ranked list of results, with the top \numresults  
code generation results at the top, followed by the top \numresults code
retrieval results (Figure~\ref{fig:plugin_candidates});
14 results are displayed in total.%
\footnote{
We note that the main motivation for this ordering is that the generation results tend to be significantly more concise than the retrieval results (Figure~\ref{fig:edit_comparison}).
If we put the retrieval results first it is likely that the users would rarely scroll past the retrieval results and view the generation results due to issues of screen real-estate.
It is important to consider that alternative orderings may result in different experimental results, although examining alternate orderings was not feasible within the scope of the current study.}

The number 7 was chosen subjectively, trying to maximize the amount and diversity 
of resulting code snippets while minimizing the necessary screen space to display
them and, therefore, the amount of scrolling expected from study participants
looking to inspect all the plugin-returned results.
After completing the current study, we found that the most relevant code snippets are typically within the top
3 results, and thus a smaller number of candidates may be sufficient.
While the number and ordering of candidates has the potential to have a significant impact on the efficiency and efficacy of the developer assistant, a formal evaluation of this impact is beyond the scope of this work.


If a code snippet is selected, the code snippet is then inserted in the current cursor's position in the code editor. The user's selection is also recorded by our instrumentation in the back end. 
Understandably, some returned code snippets may not be directly suitable for the context inside the editor, so the user is welcome (and encouraged by the instructions we give as part of our human study) to edit the auto-inserted code snippets to fit their specific intent.
After the edit is done, the user is asked to upload their edits to our server, along with the context of the code, using a dedicated key combination or the IDE's context menu.
The process is illustrated in Figure~\ref{fig:plugin_fix}.
The edit data enable us to analyze how many and what kind of edits the users need to make to transform the auto-generated code to code that is useful in their context.%
\footnote{
The edit data may also be helpful as training data for improving code generation and retrieval models. We release our data publicly to encourage this direction in future work.}

\section{Human Study Design}
\label{sec:study-design}

Given our NL2Code joint code generation and retrieval IDE plugin above,
we designed and carried out a human study with \numparticipants participants
assigned to complete a range of Python programming tasks in both control 
(no plugin) and treatment (plugin) conditions.

\subsection{Task Design}
\label{sec:tasks}

To emulate real world Python development activities, but also fit within 
the scope of a user study, we compiled a set of 14 reasonably sized Python 
programming tasks, organized into 7 categories (2 tasks per category) that
span a diversity of levels of difficulty and application domains. 

We started by identifying representative task categories that many users
would encounter in practice.
To that end, we analyzed two sources.
First, we manually reviewed all the Python programming courses listed
on three popular coding education websites,  Udacity,\footnote{\url{https://www.udacity.com/courses/all}} Codecademy,\footnote{\url{https://www.codecademy.com/catalog}} and Coursera,\footnote{\url{https://www.coursera.org/}}
to identify modules commonly taught across all websites that indicate 
common usage scenarios of the Python language.
Second, we cross checked if the previously identified use cases are 
well represented among frequently upvoted questions with the [python] tag 
on Stack Overflow, which would further indicate real programmer needs.
By searching the category name, we found that each of our identified categories covers more than 300 questions with more than 10 upvotes on Stack Overflow.
We iteratively discussed the emerging themes among the research team,
refining or grouping as needed, until we arrived at a diverse but relatively 
small set of use cases, covering a wide range of skills a Python developer 
may need in practice.

In total, we identified 7 categories of use cases, summarized in Table~\ref{tab:tasks}.
For each of the 7 categories, we then designed 2 tasks covering use 
cases in the most highly upvoted questions on Stack Overflow.
To this end, we searched Stack Overflow for the ``python'' keyword together
with another keyword indicative of the task category (\eg ``python matplotlib,'' 
``python pandas''), selected only questions that were asking how to do something 
(\ie excluding questions that ask about features of the language, or about 
how to install packages), and drafted and iteratively refined after discussion
among the research team tasks that would cover 3-5 of the most frequently 
upvoted questions.

We illustrate this process with the following example task for the 
``Data visualization'' category:\footnote{Corresponding to the search \url{https://stackoverflow.com/search?tab=votes\&q=python\%20matplotlib}.}

{\footnotesize
\begin{mdframed}[backgroundcolor=blue!10] 
By running \texttt{python3 main.py}, draw a scatter plot of the data in \texttt{shampoo.csv} and save it to \texttt{shampoo.png}. 
The plot size should be 10 inches wide and 6 inches high. 
The \texttt{Date} column is the x axis (some dates are missing from the data and in the plot the x axis should be completed with all missing dates without sales data).
The date string shown on the plot should be in the format \texttt{(YYYY-MM-DD)}. 
The \texttt{Sales} column is the y axis. 
The graph should have the title ``Shampoo Sales Trend''. 
The font size of the title, axis labels, and x \& y tick values should be 20pt, 16pt, and 12pt respectively. 
The scatter points should be colored \texttt{purple}.
\end{mdframed}
}

This task covers some of the top questions regarding data visualization with \texttt{matplotlib} found on Stack Overflow through the approach described above:
\begin{enumerate}
    \item How do you change the size of figures drawn with matplotlib?\footnote{\url{https://stackoverflow.com/questions/332289/how-do-you-change-the-size-of-figures-drawn-with-matplotlib}}
    \item How to put the legend out of the plot?\footnote{\url{https://stackoverflow.com/questions/4700614/how-to-put-the-legend-out-of-the-plot}}
    \item Save plot to image file instead of displaying it using Matplotlib?\footnote{\url{https://stackoverflow.com/questions/9622163/save-plot-to-image-file-instead-of-displaying-it-using-matplotlib}}
    \item How do I set the figure title and axes labels font size in Matplotlib?\footnote{\url{https://stackoverflow.com/questions/12444716/how-do-i-set-the-figure-title-and-axes-labels-font-size-in-matplotlib}}
\end{enumerate}

For each task designed, we also provide the user with required input data or directory structure for their program to work on, as well as example outputs (console print-outs, output files \& directories, etc.) so that they could verify their programs during the user study.

Table~\ref{tab:tasks} summarizes the 14 tasks.
The full task descriptions and input/output examples can be found online, as part of our replication package at \url{https://github.com/neulab/tranx-study}.
The tasks have varying difficulties, and on average each task would take about 15-40 minutes to complete.

{\small \begin{table}[t]
\centering
\caption{Overview of our 14 Python programming tasks.}\label{tab:tasks}
\begin{tabular}{ccl}
\toprule
\textit{Category} & \multicolumn{2}{c}{\textit{Tasks}} \\ 
\cmidrule(lr){1-1} \cmidrule(lr){2-3}
\multirow{2}[0]{*}{Basic Python} & T1-1 & Randomly generate and sort  numbers and characters with dictionary \\ 
& T1-2 & Date \& time format parsing and calculation with timezone \\
\multirow{2}[0]{*}{File} & T2-1 & Read, manipulate and output CSV files \\ 
& T2-2 & Text processing about encoding, newline styles, and whitespaces \\
\multirow{2}[0]{*}{OS} & T3-1 & File and directory copying, name editing \\ 
& T3-2 & File system information aggregation \\
\multirow{2}[0]{*}{Web Scraping} & T4-1 & Parse URLs and specific text chunks from web page \\ 
& T4-2 & Extract table data and images from Wikipedia page \\
\multirow{2}[0]{*}{Web Server \& Client} & T5-1 & Implement an HTTP server for querying and validating data \\ 
& T5-2 & Implement an HTTP client interacting with given blog post APIs \\
\multirow{2}[0]{*}{Data Analysis \& ML} & T6-1 & Data analysis on automobile data of performance metrics and prices \\ 
& T6-2 & Train and evaluate a multi-class logistic regression model given dataset \\
\multirow{2}[0]{*}{Data Visualization} & T7-1 & Produce a scatter plot given specification and dataset \\ 
& T7-2 & Draw a figure with 3 grouped bar chart subplots aggregated from dataset \\
\bottomrule
\end{tabular}
\end{table}}

\subsection{Participant Recruitment \& Task Assignments}
\label{sec:participants}

Aiming to recruit participants with diverse technical backgrounds but at
least some programming experience and familiarity with Python so as to 
be able to complete the tasks, we 
advertised our study in two ways: (1)~inside the university community 
through personal contacts, mailing lists, and Slack channels, hoping to
recruit researchers and students in computer science or related areas;
(2)~on the freelancer platform Upwork,\footnote{\url{https://www.upwork.com/}} 
hoping to attract participants with software engineering and data science
experience.
We promised each participant US\$5 per task as compensation; each participant
was expected to complete multiple tasks.

To screen eligible applicants, we administered a pre-test survey to collect 
their self-reported levels of experience with Python and with each of the 
7 specific task categories above; see Appendix~\ref{app:pre_survey_details} for 
the actual survey instrument.
We only considered as eligible those applicants who reported at least 
some experience programming in Python,
\ie a score of 3 or higher given the answer range [1: very inexperienced] to 
[5: very experienced]; 
\numapplicants applicants satisfied these criteria.

We then created personalized task assignments for each eligible applicant
based on their self reported levels of experience with the 7 specific task 
categories (see Appendix~\ref{app:particpants_experience} for the distributions 
of participants' self reported experience across tasks), using the following 
protocol:
\begin{enumerate}
    \item To keep the study relatively short, we only assign participants 
    to a total of 4 task categories (8 specific tasks, 2 per category) out
    of the 7 possible.
    \item Since almost everyone eligible for the study reported being at 
    least somewhat experienced with the first 2 task categories (Basic 
    Python and File), we assigned everyone to these 2 categories (4 specific 
    tasks total). Moreover, we assigned these 2 categories first and second,
    respectively.
    \item For the remaining 5 task categories, sorted in increasing complexity
    order,\footnote{The task identifiers in Table~\ref{tab:tasks} reflect
    this order.} we rank them based on a 
    participant's self reported experience with that task genre, and then 
    assign the participant to the top 2 task categories with most experience
    (another 4 specific tasks total). 
\end{enumerate}
Note that this filtering by experience is conducive to allowing participants to finish the tasks in a reasonable amount of time, and reflective of a situation where a developer is working in their domain of expertise.
However, at the same time it also means that different conclusions might be reached if novice programmers or programmers without domain expertise used the plugin instead.

Next, we randomly assigned the first task in a category to either the
treatment condition, \ie the NL2Code plugin is enabled in the virtual
environment IDE and the participants are instructed to use it,%
\footnote{Despite these instructions, some participants did not use the 
plugin even when it was available and when instructed. We discovered 
this while analyzing the data collected from the study and filtered out 
8 participants that did not use the plugin at all. They do not count
towards the final sample of \numparticipants participants we analyze data from, even
though they completed tasks.}
or the control condition, \ie 
the NL2Code plugin is disabled.
The second task in the same category is then automatically assigned to
the other condition,
\eg if the plugin is on for task1-1, it should be off for task1-2.
Therefore, each participant was asked to complete 4 tasks out of 8 total
using the NL2Code plugin, and 4 without.

Finally, we invited all eligible applicants to read the detailed study 
instructions, access the virtual environment, and start working 
on their assigned tasks.
Only \numparticipants out of the \numapplicants eligible applicants
after the pre-test survey actually completed their assigned tasks.%
\footnote{Note that 4 of the \numparticipants participants did not 
complete all 8 of their assigned tasks. We include their data from
the tasks they completed and do not consider the tasks they did not 
finish.}
Their backgrounds were relatively diverse; 
of the \numparticipants participants, 
12 (39\%) were software engineers and 11 (35\%) were computer 
science students, with the rest being researchers (2, 6\%), and other occupations (6, 19\%).
Our results below are based on the data from these \numparticipants
participants.

\subsection{Controlled Environment}
\label{sec:controlled-environment}

Participants worked on their assigned tasks inside a custom 
instrumented online virtual environment, accessible remotely.
Our virtual machine 
is preconfigured with the PyCharm Community Edition 
IDE\footnote{\url{https://www.jetbrains.com/pycharm/download/}} and the
Firefox Web browser; and it has our NL2Code plugin either enabled or
disabled inside the IDE, depending on the condition.
See Appendix~\ref{app:environment_design} for complete technical details.

In addition, the environment logs all of the user's interactions with 
the plugin in the PyCharm IDE, including queries, candidate selections, 
and edits; all of the user's fine-grained IDE editor activities;
the user's Web search/browsing activities inside Firefox;
all other keystrokes inside the VM;
and the source code for each one of the user's completed tasks.

To get a sense of how the source code evolves, whenever the user does 
not make modifications to the code for at least 1.5 seconds, the plugin 
also automatically uploads the current snapshot of the code to our server.
The intuition behind this heuristic is that after a user makes some 
type of meaningful edit, such as adding or modifying an argument, variable, 
or function, they usually pause for a short time before the next edit.
This edit activity granularity can be more meaningful than keystroke/character 
level, and it is finer grained than intent level or commit level edits.

Given that it is identifiable, we record participants' contact information 
(only for compensation purposes) separately from their activity logs.
This Human Subjects research protocol underwent review and was approved 
by the Carnegie Mellon University Institutional Review Board.

\subsection{Data Collection}
\label{sec:data}

To answer our research questions (Section~\ref{sec:overview}), we collect
the following sets of data.

\begin{figure}[t]
     \centering
    \includegraphics[width=\textwidth]{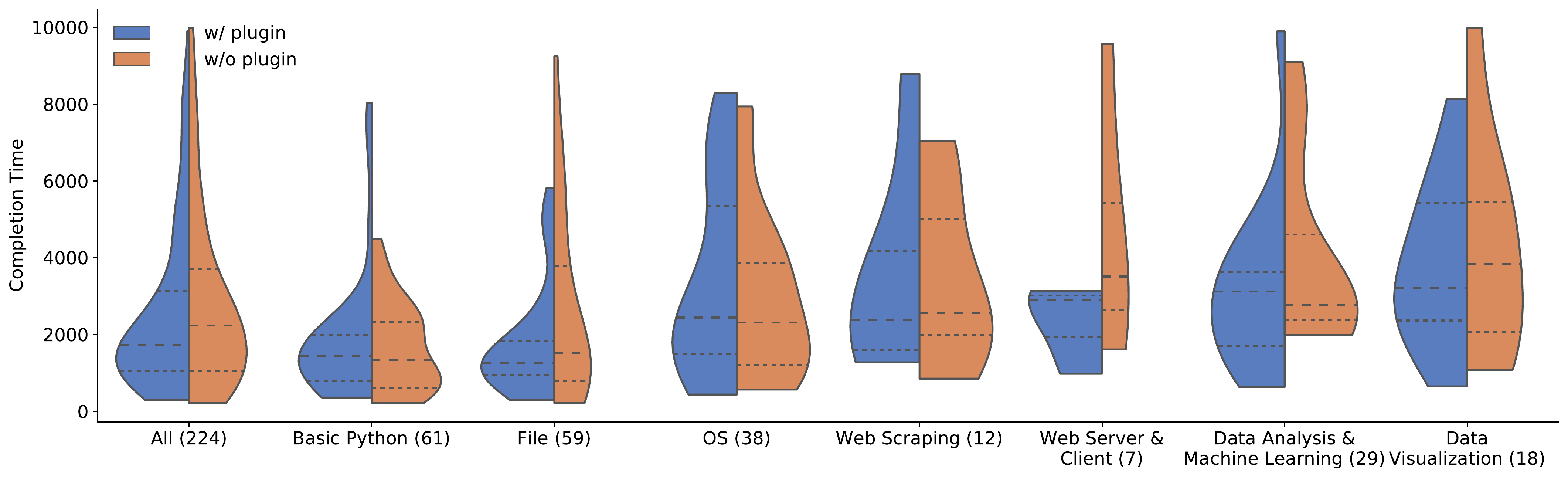}
    \caption{Distributions of task completion times (in seconds) across
    tasks and conditions (w/ and w/o using the plugin).
    The horizontal dotted lines represent 25\% and 75\% quartiles, and the dashed lines represent medians.
    }
    \label{fig:completion_time_category}
\end{figure}
    
\mysec{Task Performance Data (\RQref{task-performance})}
The first research question compares measurable properties of the tasks
completed with and without the help of or NL2Code IDE plugin and its 
underlying code generation and code retrieval engines.
One would expect that if such systems are useful in practice, developers
would be able to complete programming tasks faster without compromising
on output quality.
To investigate this, we measure two variables related to how well study 
participants completed their tasks and the quality of the code they produced:
\begin{itemize}[listparindent=\parindent]
    \item \textit{Task Completion Time.}
    Since all activity inside the controlled virtual environment is logged,
    including all keystrokes and mouse movements,
    we calculate the time interval between when a participant started working
    on a task (first keystroke inside the IDE) and when they uploaded their
    final submission to our server.
    
    Recall that participants worked asynchronously and they may have decided
    to take breaks; we designed our virtual environment to account for this,
    with explicit pause/resume functionality.
    To account for possible breaks and obtain more accurate estimates of time
    spent on task, we further subtract the time intervals when participants used 
    our explicit pause/resume functionality, as well as all intervals of idle 
    time in which participants had no mouse or keyboard activity for 2 minutes or 
    more (they may have taken a break without recording it explicitly).
    
    Figure~\ref{fig:completion_time_category} shows the distributions
    of task completion times across the two conditions (with and without the
    plugin).
    

    \smallskip 
    \item \textit{Task Correctness.}
    Following the common practice in computer science 
    education~\cite{Dawood2013RubricBA,Catet2017ApplicationOT,Grover2018WhatWC}, 
    we design a rubric for each task concurrently with designing the task, and 
    later score each submission according to that rubric.
    We weigh all tasks equally, assigning a maximum score of 10 points to each.
    For each task, the rubric covers both basic aspects (\eg runs without 
    errors/exceptions; produces the same output as the example output provided 
    in the task description) as well as implementation details regarding 
    functional correctness (\eg considers edge cases, implements all required 
    functionality in the task description).

    For example, for the data visualization task described in 
    Section~\ref{sec:tasks}, we created the following rubric, with the number 
    in parentheses representing the point value of an item, for a total of
    10 points:
    \begin{enumerate*}[label=(\roman*)]
        \item Runs without errors (2);
        \item Correct image output format (png) (2);
        \item Read in the raw data file in correct data structure (1);
        \item Correct plot size (1);
        \item Correctly handle missing data points (1);
        \item Date (x axis) label in correct format (1);
        \item Title set correctly (1);
        \item Font size and color set according to specification (1).
    \end{enumerate*}

    To reduce subjectivity, we graded each submission blindly (\ie not knowing
    whether it came from the control or treatment condition) and we automated
    rubric items when possible, \eg using input-output test cases 
    for the deterministic tasks and checking if the abstract syntax tree contains
    nodes corresponding to required types (data structures) such as dictionaries.
    See our online appendix\footnote{\url{https://github.com/neulab/tranx-study/blob/master/rubrics.md}} 
    for the complete rubrics and test cases for all tasks.
    
    Figure~\ref{fig:score_category} shows the distributions of scores
    across tasks, between the two conditions.
    
\begin{figure}[t]
     \centering
    \includegraphics[width=\textwidth]{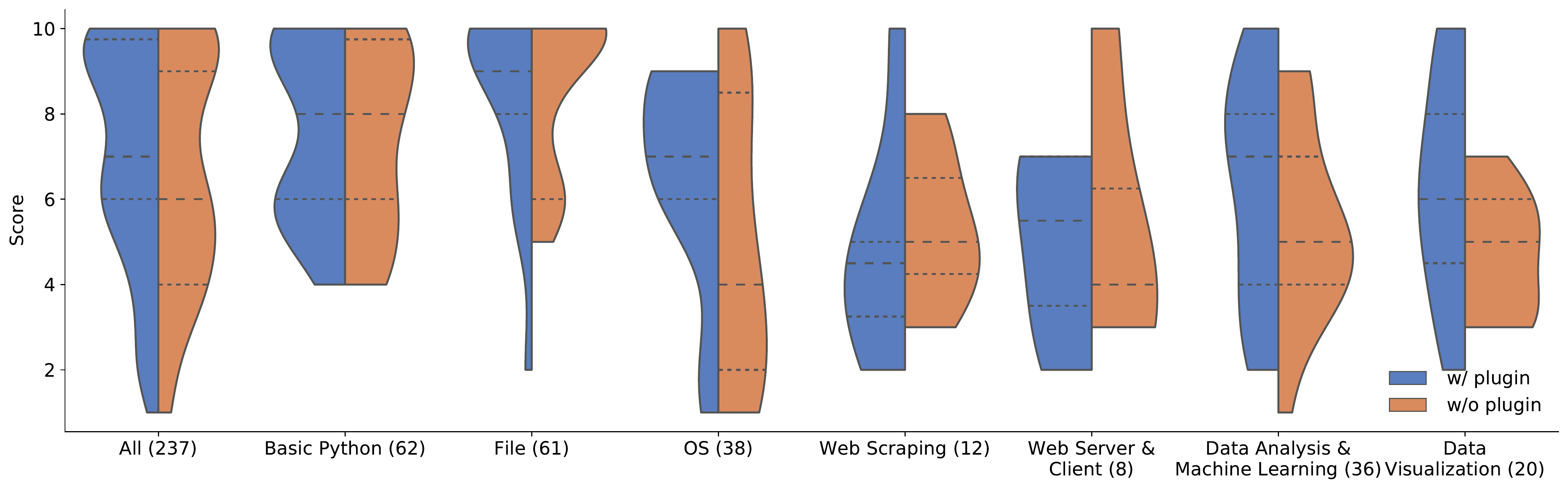}
        \caption{Distributions of task correctness scores (0--10 scale) 
        across tasks and conditions.
         The horizontal dotted lines represent 25\% and 75\% quartiles, and the dashed lines represent medians.}
        \label{fig:score_category}
\end{figure}

\end{itemize}

\mysec{Plugin Queries, Snippets and User Edits (\RQref{types-of-code})}
We record user queries using the plugin, both the generated and retrieved code snippet candidates returned for the query, and the user selection from the candidates to insert into their source code.
We use the data to analyze the NL queries and whether users preferred to use generated vs. retrieved code.
In addition, we also record the user edits after inserting the code snippet from the plugin, along with the code context for the analysis on post edits required after using the plugin.

\mysec{Participant Perceptions of Tool Use (\RQref{opinions})}
We ran short post-test surveys after every task and a final post-test
survey at the end of the study as a whole (see Appendix~\ref{app:post_survey_details}
for instruments) to collect data on the participants' subjective impressions
of using the NL2code plugin and interacting with the code generation and
code retrieval systems.
We asked Likert-style and open-ended questions about aspects of using the plugin the participants enjoyed, and aspects they wish to see improved.

Next we describe how we analyzed these data and we answer each of our research questions.

\section{\RQref{task-performance}: NL2Code Plugin Effects on Task Completion Time and Program Correctness}
\label{sec:results:task-performance}

We start by describing our shared data analysis methodology, applied similarly to 
both variables corresponding to \RQref{task-performance}, then 
present our results for each variable.

\mysec{Methodology}
Recall, we assign each participant a total of 8 tasks, 2 per task category,
based on their experience levels with those categories; 
in each category, we randomly assign one 
of the 2 tasks to the NL2Code plugin (treatment) condition and the other
task to the no plugin (control) condition.
We then compute the three sets of outcome variables above.

The key idea behind our analysis is to compare the distributions of outcome
variables between tasks completed in the treatment and control conditions.
However, this comparison is not straightforward.
First, our study design imposes a hierarchical structure during data 
collection, therefore the individual observations are not independent---by 
construction, the same participant will have completed multiple tasks over
the course of the study. 
Moreover, tasks vary in difficulty, again by construction,
therefore it is expected that their corresponding response variables, \eg
task completion times, can be correlated with the tasks themselves; 
\eg on average, more complex tasks will take longer to complete.
Finally, the participants vary in their self reported levels of Python 
and individual task category experience; 
we should separate experience-related effects from effects of using the plugin,
if any.

Therefore, we use mixed-effects~\cite{gelman2006data} as opposed to 
the more common fixed-effects regression models to analyze our data. 
Fixed-effects models assume that 
residuals are independently and identically distributed, which is an
invalid assumption in our case given the hierarchical nature of our data:
\eg responses for the different measurement occasions (tasks) within a 
given individual are likely correlated; a highly experienced Python programmer
completing one task quickly is more likely to complete other tasks quickly 
as well.
Mixed-effects models address this issue by having a residual
term at each level, \eg the observation level and the study participant 
level, in which case the individual participant-level residual is the 
so-called random effect.
This partitions the unexplained residual variance into two components: 
higher-level variance between higher-level entities (study participants) 
and lower-level variance within these entities, between measurement 
occasions (tasks).

We consider two model specifications for each response variable.
Our default model includes random effects for the individual
and task, per the rationale above, a fixed effect for task category
experience (\eg participants with more machine learning experience should
complete the machine learning task faster, on average),
and a dummy variable to indicate the condition (plugin vs no plugin).
For example, for the task completion time response, we estimate the model:%
\footnote{We are using the R syntax to specify random effects.}
{\small
\begin{align}
    \texttt{completion\_time} = \texttt{experience} + \texttt{uses\_plugin} + (1 \vert \texttt{user}) + (1 \vert \texttt{task})
\end{align}
\vspace{-12pt}
}

As specified, our default model may suffer from heterogeneity bias~\cite{bell2019fixed}.
Task category experience, a higher-level (\ie individual-level as opposed
to observation-level) predictor, varies both within and across study participants:
within participants, {\small \texttt{experience}} can vary across the
4 task categories---a user may be more experienced with basic Python than 
with data science; and across participants, {\small \texttt{experience}}
with any given task category is likely to vary as well---some participants 
report higher experience with data science-related tasks than others.
This means that {\small \texttt{experience}} (a fixed effect) and 
{\small \texttt{user}} (a random effect) may be ``correlated.''
In turn, this may result in biased estimates, because both the within- and 
between-effect are captured in one estimate.

There are two sources of variation that can be used to explain changes 
in the outcome: (1)~overall, more experienced programmers may be more efficient
at completing tasks (group-level pattern); and (2)~when becoming more
experienced, programmers may also become more efficient at completing
tasks (individual-level pattern).
Therefore, to address potential heterogeneity bias, we split our fixed effect 
({\small \texttt{experience}}) into two variables, each representing a 
different source of variation: a participant's average experience across 
all task categories ({\small \texttt{experience\_btw}}), and the 
deviation for each task from the participants’s overall mean experience
({\small \texttt{experience\_wi}}).
This process is known as de-meaning or person-mean centering~\cite{gelman2006data}.
This way, mixed-effects models can model both within- and between-subject 
effects~\cite{bell2019fixed}, as recommended for a long time by \citet{mundlak1978pooling}.
Taking the same task completion time response variable as an example
(other variables are modeled analogously), our refined model becomes:
{\small
\begin{align}
    \texttt{completion\_time} = \texttt{experience\_btw} + \texttt{experience\_wi} + \texttt{uses\_plugin} + (1 \vert \texttt{user}) + (1 \vert \texttt{task})
\end{align}
\vspace{-12pt}
}

In both cases, the estimated coefficient for {\small \texttt{uses\_plugin}}
indicates the effect of using the plugin, \emph{while holding fixed the
effects of experience and other random user and task effects}.



For estimation we used the functions {\small \texttt{lmer}} and {\small \texttt{lmer.test}} in R.
We follow the traditional level for statistical significance when interpreting coefficient estimates, \ie $p < 0.05$.
As indicators of goodness of fit, we report a marginal ($R^2_m$) and a conditional
($R^2_c$) coefficient of determination for generalized mixed-effects 
models~\cite{nakagawa2013general, johnson2014extension},
as implemented in the {\small \texttt{MuMIn}} package in R: $R^2_m$ describes the 
proportion of variance explained by the fixed effects alone; $R^2_c$ describes 
the proportion of variance explained by the fixed and random effects together.

\mysec{Threats to Validity}
Besides potential threats to statistical conclusion validity arising from the very nature of the data we are regressing over, discussed above and mitigated through our choice of mixed-effects regression models and their specific designs, we note the standard threats to statistical conclusion validity affecting linear regression models in general.
To mitigate these, we take standard precautions.
First, we removed as outliers the top 1\% most extreme values.
Second, we checked for collinearity among the predictors we use the variance inflation 
factor (VIF)~\cite{cohen2003applied}; all were below 3, \ie multicollinearity
is not an issue~\cite{kock2012lateral}.
Finally, we acknowledge that additional time may be spent as the users are asked to upload their edits, increasing the amount of time necessary in the plugin setting.
However the time spent for uploading is minimal as the plugin automatically helps the user to remove the auto-generated comments with only a press of a keyboard shortcut.


{\footnotesize 
\begin{table}[t] \centering 
  \caption{LMER task performance models (default specification).} 
  \label{tbl:task-performance} 
\begin{tabular}{@{\extracolsep{5pt}}lD{.}{.}{-2} D{.}{.}{-2} } 
\\[-1.8ex]\toprule
\\[-1.8ex] 
 & \multicolumn{2}{c}{\textit{Dependent variable:}} \\ 
\cline{2-3} 
\\[-1.8ex] & \multicolumn{1}{c}{Completion time} & \multicolumn{1}{c}{Correctness score} \\ 
\\[-1.8ex] & \multicolumn{1}{c}{(1)} & \multicolumn{1}{c}{(2)}\\ 
\midrule \\[-1.8ex] 
 Experience & -195.62 & 0.07 \\ 
  & (183.11) & (0.24)  \\ 
  Uses plugin & 15.76 & 0.44 \\ 
  & (196.11) & (0.30) \\ 
  Constant & 3,984.51^{***} & 5.88^{***} \\ 
  & (838.07) & (1.03) \\ 
 \midrule \\[-1.8ex] 
Observations & 224 & 237 \\ 
Num users & 31 & 31 \\ 
Num tasks & 14 & 14 \\ 
sd(user) & 1489.25 & 0.82  \\ 
sd(task) & 1104.7 & 1.14 \\ 
R2m & 0.004 & 0.008  \\ 
R2c & 0.642 & 0.289 \\ 
Akaike Inf. Crit. & \multicolumn{1}{c}{3,987.14} & \multicolumn{1}{c}{1,106.66} \\ 
Bayesian Inf. Crit. & \multicolumn{1}{c}{4,007.61} & \multicolumn{1}{c}{1,127.46} \\ 
\bottomrule \\[-1.8ex] 
\textit{Note:}  & \multicolumn{2}{r}{$^{*}$p$<$0.1; $^{**}$p$<$0.05; $^{***}$p$<$0.01} \\ 
\end{tabular} 
\end{table}  }

\mysec{Results}
Table~\ref{tbl:task-performance} summarizes our default specification 
mixed-effects regressions for both response variables;
the models with our second specification (de-meaned task experience)
are equivalent, see Appendix~\ref{app:demeaned-models}.
All models include controls for the amount of users' experience with
the respective task categories as well as other random user and task effects.
In all cases, the models fit the data reasonably well (ranging from $R^2_c = 29\%$ for task correctness scores, to $R^2_c = 64\%$ for task completion time), 
with most of the variance explained attributable to the two random effects 
(task and user)---there is significant user-to-user and task-to-task 
variability in all response variables.

Analyzing the models we make the following observations.
First, looking at the completion time model~(1), there is no statistically 
significant difference between the two conditions.
Stated differently, we do not find sufficient evidence to conclude that users 
in the plugin condition complete their tasks with different speed on 
average than users in the control group, contrary to our expectation.

Second, and this time in line with our expectation, there is no statistically 
significant difference between the two conditions in task correctness scores
(model~(2)).
That is, the code written by users in the plugin condition appears statistically 
indistinguishably as correct from the code written by users 
in the control group.

We investigate more differences between the code written by study participants
in each of the two conditions in more detail in the next section.

\section{\RQref{types-of-code}: Comparison of Generated vs Retrieved Code}
\label{sec:rq2}


In this section we focus on \emph{how} study participants are interacting with the
code generation and retrieval systems.
Specifically, we dive deeper into both the inputs to and the outputs of the
plugin, \ie we analyze the quality of the queries issued by study participants 
and of the code snippets produced in return, contrasting code generation to retrieval throughout.
We analyze these data along three dimensions, detailed next.

\subsection{For What Queries do Users Tend to Favor Generation vs Retrieval Answers}
\label{sec:codedifferences:queries}

First, we investigate whether there are any discernible characteristics of the 
natural language queries (and therefore tasks) that associate with study participants 
tending to favor the code snippets returned by the code generation model over those 
returned by the code retrieval model.

\mysec{Methodology}
Using our instrumented environment, we collect all \textit{successful} queries issued 
by the study participants, \ie those for which a code snippet from among the listed 
candidates was selected, and we record which of the two sources (generation or retrieval)
the snippet came from.
See Table~\ref{tbl:queries} in Appendix~\ref{app:plugin-queries} for the complete set 
of queries from our \numparticipants participants, organized per task.
We then build a binary logistic regression model with snippet source as outcome
variable and bag-of-words features of the natural language input queries as predictors.

If this model is able to predict the source of the code snippet better than by chance, then we can conclude that there is some correlation between the type of input query and the users' preference for generated versus retrieved code snippets.
Moreover, the word feature weights in the logistic regression model could shed some light on what features are the most representative of queries that were effectively answered using generation or retrieval.
For our analysis, we manually review the top 20 (approx. 7\%) contributing query features for each value of the outcome variable (``generation'' vs ``retrieval'') and discuss patterns we observe qualitatively, after thematic analysis.

Specifically, for each query, we tokenize it, filter out English stop words, and compute a bag-of-words and bag-of-bigrams vector representation, with each element of the vector corresponding to the number of times a particular word or bigram (two-word sequence) occurred in the query.
The number of distinct words in all queries is 302, and the number of distinct bigrams in all queries is 491, and thus the dimensionality of the query vector is 793.\footnote{We also experimented with other features, \eg query length, query format compliance, etc., but did not notice a significant difference in prediction accuracy.} 
We then estimate the model: 
\begin{align}
\label{logmod} 
Pr(\text{chosen snippet is ``generated''})
& =\frac{\exp(\textbf{X}\beta)}{1+\exp(\textbf{X}\beta)}, 
\end{align}
where $\textbf{X}$ here represents a $k$-dimensional bag-of-word vector 
representation of a given query, and $\beta$ are the weights to be estimated.
To this end, we randomly split all the collected query and candidate selection pairs into training (70\% of the data) and held-out test (30\%) sets.
We then train the model using 5-fold cross-validation until it converges, and subsequently test it on the held-out set.
We use 0.5 as a cutoff probability for our binary labels.
In addition, we also build a trivial baseline model that always predicts ``retrieval.''

The baseline model is 55.6\% accurate (among the successful queries in our sample there are slightly more code snippets retrieved rather than generated).
Our main logistic regression model is 65.9\% accurate, \ie the model was able to learn some patterns of differences between those queries that result in code generation results being chosen over code retrieval ones and vice versa.

\begin{table}[t]
\centering
\caption{Most important 20 features and their weights from the logistic regression modeling whether successful plugin queries result in generated or retrieved code snippets.
}
\label{tab:query-class-features}
\small
\begin{tabular}{llllllll}
\toprule
\multicolumn{4}{c}{\emph{Generation}} & \multicolumn{4}{c}{\emph{Retrieval}} \\ 
\cmidrule(lr){1-4} \cmidrule(lr){5-8}
\multicolumn{1}{l}{Weight} & \multicolumn{1}{l}{Feature} & \multicolumn{1}{l}{Weight} & \multicolumn{1}{l}{Feature} & \multicolumn{1}{l}{Weight} & \multicolumn{1}{l}{Feature}  & \multicolumn{1}{l}{Weight} & \multicolumn{1}{l}{Feature} \\ 
\cmidrule(lr){1-2} \cmidrule(lr){3-4} \cmidrule(lr){5-6} \cmidrule(lr){7-8}
0.828  & open       & 0.352  & current & 0.471  & letters    & 0.294  & extract  \\
0.742  & time       & 0.345  & delete row & 0.442  & copy       & 0.289  & set    \\
0.676  & sort       & 0.345  & random number   & 0.438  & matplotlib & 0.289  & plt set    \\
0.590  & read csv     & 0.339  & trim  & 0.437  & datetime     & 0.282  & read file  \\
0.556  & list       & 0.330  & text file    & 0.410  & python   & 0.282  & cross validation   \\
0.507  & number       & 0.326  & keys & 0.365  & column csv        & 0.274  & scikit  \\
0.402  & search     & 0.310  & round  & 0.361  & bar        & 0.274  & dataframe csv   \\
0.399  & open file  & 0.293  & numbers  & 0.344  & copy files      & 0.274  & sklearn \\
0.385  & dictionary & 0.291  & row dataframe   & 0.334  & delete column & 0.272  & digit   \\
0.353  & read       & 0.290  & load csv    & 0.302  & write file    & 0.270  & folders \\
\bottomrule
\end{tabular}
\end{table}

\mysec{Threats to Validity}
One potentially confounding factor is that the plugin always displays code generation results first, before code retrieval. Ordering effects have been reported in other domains~\cite{richardson2007predicting} and could also play a role here. Specifically, users who inspect query results linearly, top-down, would see the code generation results first and might select them more frequently than if the results were displayed in a different order. That is, we might infer that users prefer code generation to retrieval only because they see code generation results first, thus overestimating the users' preference for code generation versus retrieval. 

Even though testing ordering effects experimentally was not practical with our study design, we could test a proxy with our log data---to what extent the code generation results overlap with the code retrieval ones. High overlap could indicate that code retrieval results might have been chosen instead of code generation ones, if presented earlier in the candidates list. Whenever study participants chose a snippet returned by the code generation model, we compared (as strings) the chosen snippet to all candidates returned by the code retrieval engine. 
Only 6 out of 173 such unique queries (\textasciitilde 3.5\%) also contained the exact chosen code generation snippet among the code retrieval results, therefore we conclude that this scenario is unlikely.%
\footnote{Note that this only considers exact substring matches. There may be additional instances of functionally equivalent code that is nonetheless not an exact match.}

Another potentially confounding factor is that an icon indicative of generation or retrieval is displayed next to each result in the plugin UI. This means that users know which model produced which candidate snippet and might choose a snippet because of that reason rather than because of the snippet's inherent usefulness. More research is needed to test these effects. We hypothesize that biases may occur in both directions. On the one hand, holding other variables like ordering fixed, users might prefer code generation results because of novelty effects. On the other hand, users might prefer code retrieval results because of general skepticism towards automatically generated code, as has been reported, \eg about automatically generated unit tests~\cite{fraser2015does,roy2020deeptc}.

Regarding the analysis, we use an interpretable classifier (logistic regression) and follow standard practice for training and testing (cross-validation, held-out test set, \etc), therefore we do not expect extraordinary threats to validity related to this part of our methodology.
However, we do note the typical threats to trustworthiness in qualitative research related to our thematic analysis of top ranking classifier features~\cite{nowell2017thematic}.
To mitigate these, we created a clear audit trail, describing and motivating methodological choices, and publishing the relevant data (queries, top ranking features after classification, \etc).
Still, we note potential threats to transferability that may arise if different features or different classifiers are used for training, or a different number/fraction of top ranking features is analyzed qualitatively for themes.


\mysec{Results}
In Table~\ref{tab:query-class-features}, we show the top features that contributed to predicting each one of the two categories, and their corresponding weights.
Inspecting the table we make two observations.

First, we observe that for code generation, the highest ranked features (most predictive tokens in the input queries) refer mostly to basic Python functionality, \eg ``open, read csv, text file'' (opening and reading a file), ``sort, list, number, dictionary, keys'' (related to basic data structures and operations in Python), ``random number'' (related to random number generation), ``trim'' (string operations), etc.
For example, some stereotypical queries containing these tokens that result in the code generation snippets being chosen are ``open a csv file \`{}data.csv\`{} and read the data'', ``get date and time in gmt'', ``list all text files in the data directory'', etc.

In contrast, we observe that many queries that are more likely to succeed through code retrieval contain terms related to more complex functionality, some usually requiring a series of steps to fulfill.
For example, ``datetime'' (regarding date and time operations), ``cross validation, sklearn, column csv'' (regarding machine learning and data analysis), ``matplotlib'' (data visualization), etc.\ are all among the top features for queries where users more often chose the code retrieval snippets.


In summary, it seems predictable (substantially more so than by random chance) whether natural language user queries to our NL2Code plugin are more likely to succeed through code generation vs code retrieval on average, given the contents (words) of the queries.

\subsection{How Well-Specified Are the Queries}
\label{sec:queries-oracle}

Search is a notoriously hard problem~\cite{husain2019codesearchnet,Liu2020OpportunitiesAC}, especially when users do not start knowing exactly what they are looking for, and therefore are not able to formulate clear, well-specified search queries.
In this subsection we investigate the quality of the input natural language queries, and attempt to delineate it from the quality of the underlying code generation and retrieval systems---either one or both may be responsible for failures to obtain desirable code snippets for a given task.

Anecdotally, we have observed that input queries to our NL2Code plugin are not always well-specified, even when the participants selected and inserted into their code one of the candidate snippets returned by the plugin for that query.
A recurring issue seems to be that study participants sometimes input only a few keywords as their query (\eg ``move file''), perhaps as they are used to interacting with general purpose search engines like Google, instead of more detailed queries as expected by our plugin.
For example, study participants sometimes omit (despite our detailed instructions) variable names part of the intent but defined elsewhere in the program (\eg ``save dataframe to csv'' omits the DataFrame variable name).
Similarly, they sometimes omit flags and arguments that need to be passed to a particular API method (\eg ``load json from a file'' omits the actual JSON filename).

\mysec{Methodology}
The key idea behind our investigation here is to replace the underlying code generation and retrieval systems with an oracle assumed to be perfect---a human expert Python programmer---and study how well the oracle could have produced the corresponding code snippet given a natural language input query.
If the oracle could successfully produce a code snippet implementing the intent, then we deem the query ``good enough'', or well-specified; otherwise, we deem the query under-specified.
The fraction of ``good enough'' queries to all queries can be considered as an upper bound on the success rate of a perfect code generation model.

Concretely, we randomly sampled \numoraclequeries queries
out of all successful queries issued during the user study (see Table~\ref{tbl:queries-sampled} in Appendix~\ref{app:oracle-queries} for the sample), and had the first author of this paper, an proficient programmer with 8 years of Python experience, attempt to generate code based on each of them.
The oracle programmer considered two scenarios: (1)~generating code given the input query as is, without additional context; (2)~if the former attempt failed, generating code given the input query together with the snapshot of the source file the study participant was working in at the time the query was issued, for additional context.

For each query, we record three binary variables: two indicating whether each of the oracle's attempts succeeded, without and with additional context, respectively,\footnote{The former implies the latter but not vice versa.} and the third indicating whether the code snippet actually chosen by the study participant for that query came from the code generation model or the code retrieval one; see Table~\ref{tbl:queries-sampled} in Appendix~\ref{app:oracle-queries}.%
\footnote{Note that on the surface, when looking at the data in Table~\ref{tbl:queries-sampled}, the values of the former two binary variables (the oracle's determination) may not always seem intuitive given the query. For example, the oracle determined the query ``pandas to csv'' to be \textit{not good enough}, even with context, while the query ``pandas output csv'', seemingly equivalent, was found to be \textit{good enough with context}. In both cases, the intent appears to be exporting a pandas dataframe (a popular data science Python library) as a csv file. However, in the first example the snapshot of the source file the study participant was working in at the time of the query did not yet include \textit{any} such dataframe objects; the user appears to have issued the query ahead of setting up the rest of the context. A context-aware code generation model would also not be able to extract any additional information in this case, similarly to the human oracle.}

We then measure the correlation, across the \numoraclequeries queries, between each of the two oracle success variables and the code snippet source variable, using the phi coefficient $\phi$~\cite{cramer1999mathematical}, a standard measure of association for two binary variables similar to the Pearson correlation coefficient in its interpretation.
This way we can assess how close the code generation model is from a human oracle (the \textit{good enough as is} scenario), and whether contextual information from the source code the developer is currently working on might be worth incorporating into code generation models in the future (the \textit{good enough with context} scenario);
note that the code generation model we used in this study~\cite{yin2018tranx,xu2020incorporating} does not consider such contextual information.



\mysec{Threats to Validity}
We follow standard practice for the statistical analysis in this section, therefore we do not anticipate notable threats to statistical conclusion validity.
Due to the limitations of our telemetry system, we did not record unsuccessful queries (i.e. queries that the user entered but no candidate is selected).
As a result, queries that favor neither generation nor retrieval cannot be compared.
However, we acknowledge three other notable threats to validity.
First, we used only one expert programmer as oracle, which may introduce a threat to construct validity given the level of subjectivity in determining which queries are ``good enough''. To mitigate this, we discussed among the research team, whenever applicable, queries for which the expert programmer was not highly confident in the determination.
Second, our random sample of \numoraclequeries queries manually reviewed by the expert programmer is only representative of the population of \numqueries queries with 95\% confidence and 13\% margin of error, which may introduce a threat to internal validity. However, the relatively small sample size was necessary for practical reasons, given the high level of manual effort involved in the review.
Finally, we note a potential threat to construct validity around the binary variable capturing the source (generation or retrieval) of the candidate code snippets selected by the study participants. There is an implicit assumption here that study participants know what the right answer (code snippet) should be given a natural language query, and are able to recognize it among the candidates provided by the NL2Code plugin; therefore, we assume that the snippet source variable captures actual quality differences between code snippets produced by the generation and retrieval models, respectively.
However, this may not be the case. To test this, we reviewed all the candidate snippets returned by the plugin for the first 6 among the \numoraclequeries queries analyzed.
Across the $6 \cdot 2 \text{ models (generation/retrieval)} \cdot \numresults \text{ candidates per model} = 84 \text{ candidate snippets}$, we only discovered one case where the study participant could have arguably chosen a more relevant snippet. Therefore, we expect the incidence of violations of this assumption to be rare enough to not materially affect our results.
{\small \begin{table}[t]
\centering
\captionsetup{width=.65\linewidth}
\caption{Contingency tables for the two oracle comparison scenarios in Section~\ref{sec:queries-oracle}; see Table~\ref{tbl:queries} in Appendix~\ref{app:plugin-queries} for the actual queries.
}\label{tab:contingency}
\begin{tabular}{ccccc}
\toprule
\textit{Snippet} & \multicolumn{4}{c}{\textit{Query}} \\ 
\cmidrule(lr){1-1} \cmidrule(lr){2-5}
\multirow{2}[3]{*}{Generation} & \multicolumn{2}{c}{Good enough as is} & \multicolumn{2}{c}{Good enough w/ context} \\ 
\cmidrule(lr){2-3} \cmidrule(lr){4-5}
 & False & True & False & True \\
\midrule
False & 23 & 8 & 15 & 16 \\
True & 7 & 12 & 1 & 18 \\
\bottomrule
\end{tabular}
\end{table} }

\mysec{Results}
Table~\ref{tab:contingency} shows contingency tables for each of the two oracle comparison scenarios. Note that the ``good enough with context'' category includes all queries that are ``good enough as is'', by construction.
Inspecting the results in the table, we make the following observations.

First, the natural language queries analyzed are more often than not insufficiently well-specified for even the human expert to be able to write code implementing those intents; only 20 out of \numoraclequeries queries (40\%) are deemed ``good enough as is'' by the oracle.
Representative examples of failures from Table~\ref{tbl:queries-sampled} are the queries consisting of a few keywords (\eg ``csv writer'', ``defaultdict'') rather than queries containing sufficient details about the user's intent (\eg ``remove first column from csv file'').
Considering the source file the user was editing at query time helps, with 34 (68\%) of the queries now being deemed ``good enough with context'' by the oracle. 

Second, there is moderately high and statistically significant association between the success of the code generation model (\ie the study participant choosing one of those candidate code snippets) and the quality of queries in both scenarios: $\phi = 0.37$ ($p = 0.008$) for already well-specified queries and $\phi = 0.45$ ($p = 0.001$) for queries that become informative enough given additional context.
This suggests that input query quality can have a big impact on the performance of the code generation model, and that incorporating additional contextual information may help.

Analyzing the failure rate of the code generation model (generation = False), we observe that it is relatively high in general (31 out of \numoraclequeries queries, or 62\%). However, most of these cases are in response to under-specified queries (23 out of the 31 failures; 74\%), for which even the human oracle failed to generate the corresponding code.
Still, there are 8 (26\%) failure cases where the human expert could directly implement the natural language intent without additional context: ``date now'', ``for loop on range 100'', ``generate random letters'', ``get now one week from now'', ``get time and date'', ``open "data.csv" file'', ``how to remove an item from a list using the index'', and ``plt create 3 subplots''. All but the last one seem to refer to basic Python functionality. These queries are targets where further improved code generation techniques could improve the utility of the plugin.

Interestingly, we also observe a non-trivial number of under-specified queries (7 out of 30; 23\%) for which the code generation model succeeded despite the human oracle failing: ``call \texttt{\`}pick$\_$with$\_$replacement\texttt{\`}'', ``copy a file to dist'', ``pandas round value'', ``pandas to csv'', ``rename column pandas'', ``plt ax legend'', and ``scatter''.

\subsection{How Much Are the Code Snippets Edited After Plugin Use}

Choosing (and inserting into the IDE source file) one of the candidate code snippets returned by the NL2Code plugin indicates that the code snippet was generally useful.
However, while useful, the code snippet may still be far from an ideal solution to the user's query.
To get a sense of how appropriate the accepted code snippets are given the user intent, we compare the distributions of snippet lengths before (\ie as returned by the plugin) and after potential edits in the IDE.

\mysec{Methodology}
When inserting a code snippet a user selected from among the plugin-returned candidates, we also insert special code comments in the source file around the snippet, to mark the start and end of the code fragment corresponding to that particular intent (as shown in \autoref{fig:plugin_fix}). 
Study participants are instructed to use a certain key combination when they are done editing that code fragment to remove the delimiters and submit the edited version of the code fragment back to our server.
Our analysis in this section compares the length of code snippets and types of tokens present between these two versions.

\begin{figure}[t]
    \centering
    \includegraphics[width=0.5\textwidth]{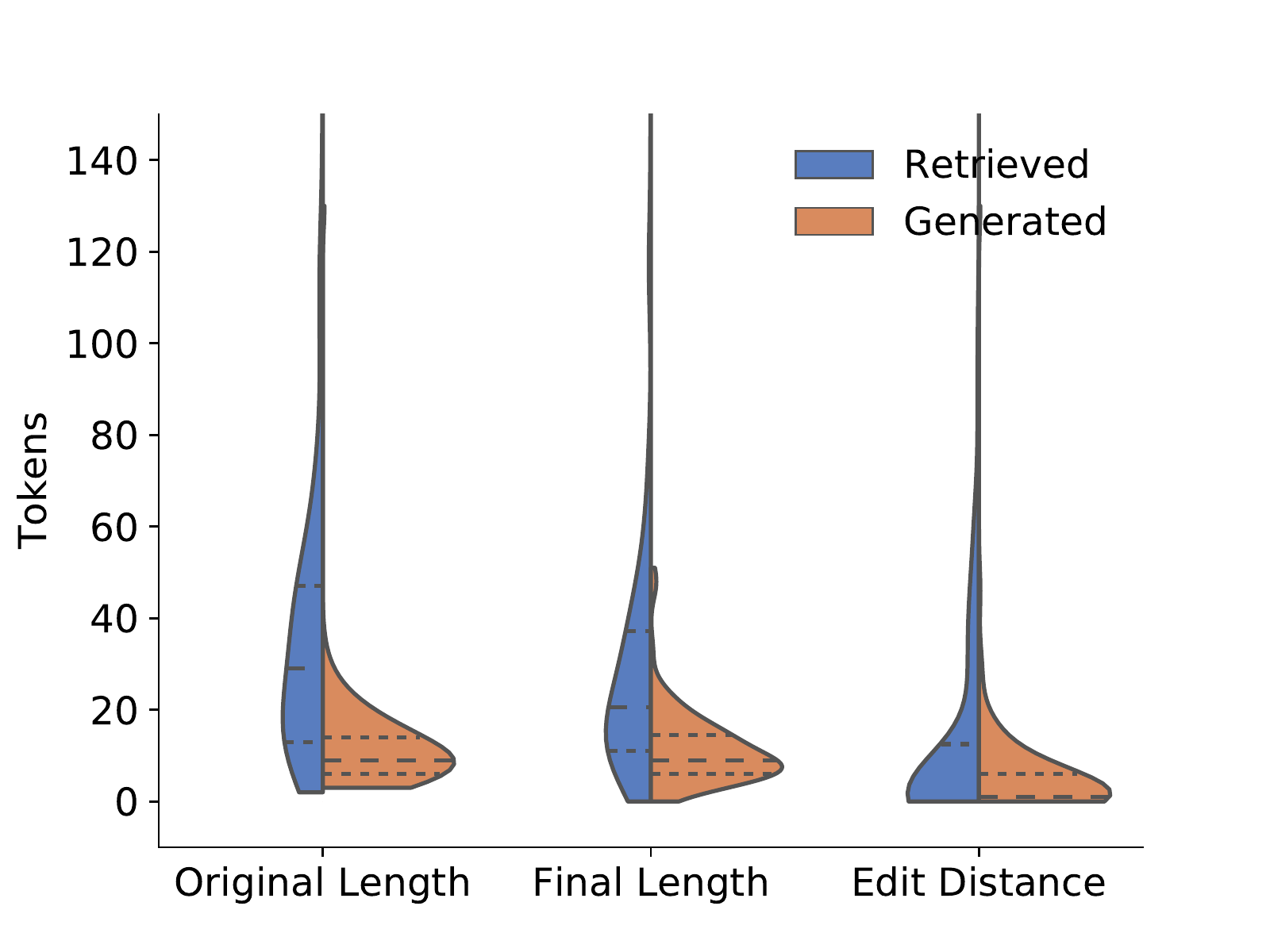}
    \caption{Split violin plots comparing the length (in tokens) of the code snippets chosen by the study participants across all successful queries, before and after potential edits in the IDE. The horizontal dotted lines represent 25\% and 75\% quartiles, and the dashed lines represent medians.
    }
    \label{fig:edit_comparison}
\end{figure}

Specifically, we first tokenize and tag each version of a code snippet using a Python tokenizer, and then compare the pairs of distributions of lengths before and after edits for code snippets originating from each of the two underlying models, generation and retrieval, using the non-parametric Wilcoxon signed-rank test;
in addition, as a measure of effect size we compute the median difference between members of the two groups, \ie the Hodges–Lehman estimator~\cite{hodges1963estimates}.
We also compute and report on the Levenshtein edit distance between the two versions, in terms of number of tokens.
Figure~\ref{fig:edit_comparison} visualizes these different distributions.

\looseness=-1
\mysec{Threats to Validity}
We note two potential threats to construct and external validity related to the analysis in this section. First, we have no way of enforcing that study participants contain their code edits related to a particular intent to the section of the source file specially delimited by code comments for this purpose. One may include unrelated edits in the same code region, or make related edits outside of the designated region. Therefore, our measurement of \textit{snippet length post edits} may not accurately reflect the construct of snippet length as related to a particular intent. To mitigate this, we gave clear instructions to participants at the beginning of the study and manually reviewed a small sample of the edited versions of a snippet, not discovering any obvious noise.
Second, not all study participants followed our instructions every time they used the plugin, and submitted their final (edited or not) version of the snippet back to our server. Only 303 out of the \numqueries successful queries recorded (76.3\%) had final code snippets uploaded back to our server. Since this was not a random sample, our findings on this sample may not generalize to the entire population of \numqueries successful queries. To assess the severity of this potential threat, we compared the distributions of plugin-returned code snippet lengths between all successful queries and just the 303 queries where study participants uploaded their edits onto our server; for both generated (Wilcoxon $p = 0.54$) and retrieved ($p = 0.93$) code snippets, we found the respective two distributions statistically indistinguishable, therefore we expect this to not be a sizable threat to validity.

\mysec{Results}
Comparing the two distributions of token lengths for acceptable code snippets from the code generation model before and after edits, we do not find any statistically significant differences in their mean ranks ($p = 0.345$).
The mean edit distance between the two versions of these snippets is 5.2 tokens (min 0, max 130, median 1).

In contrast, comparing the two distributions of token lengths for acceptable code snippets from the code retrieval engine before and after edits, we find a statistically significant difference in their mean ranks ($p = \num{1.195e-07}$).
The Hodges–Lehman median difference between the edited and unedited versions of these snippets is 18 tokens, with a 95\% confidence interval from 11 to 23 tokens.
The edit distance metric paints a similar picture---acceptable code snippets from the code retrieval engine, before and after edits, are at a mean edit distance of 13.2 tokens from each other (min 0, max 182, median 0).

We also note that code retrieval snippets tend to be longer than code generation ones both before ($p < \num{2.2e-16}$; median difference 18 tokens, with a 95\% confidence interval from 14 to Infinity) and after edits ($p = \num{2.657e-14}$; median difference 10 tokens, with a 95\% confidence interval from 7 to Infinity).
This may help explain why the retrieved snippets require more edits to correct the code to better suit the current programming code context, compared to the generated snippets.

Diving deeper into the edits to the plugin-supplied version of the different snippets, we compute the frequency distribution of tokens in both versions (plugin and final), normalized based on total token count in each corpus.
Table~\ref{tab:top-token-edits} highlights the tokens with the greatest increases and decreases in relative frequency during editing.
We observe that study participants seem to add common keywords such as ``in, for, if, with'', built-in names and functions such as ``key, print'', and common variable names such as ``line, filename'' to the generated/retrieved candidates.
Stated differently, in these cases the code snippets seem to miss substantive parts and relevant functionality, which also may be partly due to the lack of specificity described in the previous section.

In contrast, study participants seem to delete number and string literals from the code snippets. This may be explained by the fact that the tool used retrieved code snippets as they appeared on Stack Oveflow, and thus many retrieved code snippets contain additional boilerplate code required for initialization or setup, and hard-coded example inputs and outputs.
We also observe some commonly used variable names like ``df, plt'' that get deleted, suggesting that variable replacement is one of the common operations when reusing the code snippets.
An interesting observation here is that ``In'' and ``Out'' are getting deleted frequently. We find that it's mostly due to some of the code snippets retrieved from Stack Overflow being in the format of IPython REPL, which uses ``In'' and ``Out'' to separate the Python source code and execution outputs.
When integrating these snippets, the users will have to remove this superfluous text.
Figure~\ref{fig:snippet_edits} shows a representative example of such user edits after selecting a candidate snippet, which involves deleting IPython REPL contents, variable replacement and addition, as well as literal replacements.

{\footnotesize 
\begin{table}[t]
\centering
\caption{Most frequently added/deleted tokens after user edits to plugin-returned code snippets.}
\label{tab:top-token-edits}
\small
\small
\begin{tabular}{llllllll}
\toprule
\multicolumn{4}{c}{\emph{Addition}} & \multicolumn{4}{c}{\emph{Deletion}} \\ 
\cmidrule(lr){1-4} \cmidrule(lr){5-8}
\multicolumn{1}{l}{$\Delta$Freq.} & \multicolumn{1}{l}{Token} & \multicolumn{1}{l}{$\Delta$Freq.} & \multicolumn{1}{l}{Token} & \multicolumn{1}{l}{$\Delta$Freq.} & \multicolumn{1}{l}{Token}  & \multicolumn{1}{l}{$\Delta$Freq.} & \multicolumn{1}{l}{Token} \\ 
\cmidrule(lr){1-2} \cmidrule(lr){3-4} \cmidrule(lr){5-6} \cmidrule(lr){7-8}
0.0040                     & in                         & 0.0016                    & w                            & -0.0071                     & 2                            & -0.0016                     & In                           \\ 
0.0037                     & for                        & 0.0015                    & with                         & -0.0071                     & 1                            & -0.0016                     & 11                           \\ 
0.0030                     & line                       & 0.0015                    & \`{}\`{}                           & -0.0043                     & a                            & -0.0015                     & y                            \\ 
0.0024                     & file                       & 0.0015                    & days                         & -0.0038                     & 0                            & -0.0014                     & Seattle                      \\ 
0.0023                     & key                        & 0.0015                    & cur\_v                       & -0.0034                     & 3                            & -0.0014                     & 12                           \\ 
0.0023                     & os.path.join               & 0.0015                    & company\_info                & -0.0025                     & plt                          & -0.0013                     & 4                            \\ 
0.0021                     & dic                        & 0.0015                    & n                            & -0.0023                     & 50                           & -0.0013                     & iris                         \\ 
0.0021                     & filename                   & 0.0014                    & output                       & -0.0021                     & id\_generator                & -0.0013                     & string.digits                \\ 
0.0018                     & print                      & 0.0014                    & codecs.open                  & -0.0018                     & Out                          & -0.0013                     & 10                           \\ 
0.0017                     & if                         & 0.0014                    & v                            & -0.0017                     & df                           & -0.0013                     & matplotlib.pyplot            \\ 
\bottomrule
\end{tabular}
\end{table}}

\begin{figure}[ht]
\centering
{\footnotesize 
\begin{minipage}[t]{0.35\linewidth}
\begin{Verbatim}[frame=topline,numbers=left,label=Unedited,framesep=3mm]
In [479]: df
Out[479]: 
    ID  birthyear   weight
0   619040  1962    0.123123
1   600161  1963    0.981742
2   25602033    1963    1.312312
3   624870  1987    0.942120

In [480]: df["weight"].mean()
Out[480]: 0.83982437500000007
\end{Verbatim}
\end{minipage}\qquad\quad
\begin{minipage}[t]{0.43\linewidth}
\begin{Verbatim}[frame=topline,numbers=left,label=Edited,framesep=3mm]
car_prices = car_prices["price"].mean()
\end{Verbatim}
\end{minipage}
}
\caption{Representative example of user edits to a code snippet retrieved from Stack Overflow.}
\label{fig:snippet_edits}
\end{figure}

{\footnotesize 
\begin{table}[t]
\centering
\caption{Frequency changes of different token types after user edits to plugin-returned code snippets. Sorted in descending order, positive number represents addition and negative number represents deletion.}
\label{tab:top-token-types-edits}
\small
\begin{tabular}{llllllll}
\toprule
\multicolumn{1}{l}{$\Delta$Freq.} & \multicolumn{1}{l}{Type} & \multicolumn{1}{l}{$\Delta$Freq.} & \multicolumn{1}{l}{Type} & \multicolumn{1}{l}{$\Delta$Freq.} & \multicolumn{1}{l}{Type}  & \multicolumn{1}{l}{$\Delta$Freq.} & \multicolumn{1}{l}{Type} \\ 
\cmidrule(lr){1-2} \cmidrule(lr){3-4} \cmidrule(lr){5-6} \cmidrule(lr){7-8}
0.0138 & \texttt{NAME}  & 0.0053  & \texttt{DEDENT}  & 0.0004 & \texttt{COMMENT} & -0.0095  & \texttt{OP}  \\ 
0.0053  & \texttt{INDENT}   & 0.0022   & \texttt{STRING} & -0.0049 & \texttt{NEWLINE}   & -0.0248   & \texttt{NUMBER}      \\ 
\bottomrule
\end{tabular}
\end{table}}

Furthermore, following the previous observations on actual tokens, we are interested in how the frequency of different \emph{types} of tokens changes before and after users edit the plugin-returned code snippets.
We use the \texttt{tokenize}\footnote{\url{https://docs.python.org/3/library/tokenize.html}} Python 3 library to parse and tag the code snippets, and compare the frequency changes by token type, similar to the previous analysis.%
\footnote{3 of the retrieved snippets cannot be parsed and thus are omitted. See full explanation of different token types at \url{https://www.asmeurer.com/brown-water-python/tokens.html}. We also left out some uninteresting token types, such as \texttt{ENCODING}, \texttt{ENDMARKER}, \texttt{NL}.}
The results are shown in Table~\ref{tab:top-token-types-edits}.
We find that users add new \texttt{NAME} (identifiers, keywords) tokens the most, with the frequency of \texttt{STRING} (string literal) tokens slightly increased, and \texttt{COMMENT} (comment strings) tokens staying roughly the same after the edits. 
\texttt{NUMBER} (numeric literal) tokens are deleted the most, in line with the observation above, again suggesting that many plugin-returned snippets are not tailored to specific identifiers and parameters that the user desires.
Interestingly, we also see a slight decrease in frequency of \texttt{NEWLINE} tokens, representing a decrease in the number of logical lines of Python code after edits.
This suggests that the plugin-returned code snippets are not concise enough in some cases.

\section{\RQref{opinions}: User Perceptions of the NL2Code Plugin}
\label{sec:perceptions}

Our last research question gauges how study participants perceived working with the NL2Code plugin, their pain points, and their suggestions for improvement.

\mysec{Methodology}
As part of our post-test survey, we asked the participants open-ended questions about what worked well when using the plugin and, separately, what they think should be improved. In addition, we asked participants to rate their overall experience using the plugin on a Likert scale, ranging from 1 (very bad) to 5 (very good).
We then qualitatively coded the answers to open-ended questions to identify themes in the responses for the \numparticipants who completed all their assigned tasks.

\mysec{Threats to Validity}
We acknowledge usual threats to trustworthiness and transferability from qualitatively analyzing a relatively small set of open-ended survey data~\cite{nowell2017thematic}, as also discussed above.
In particular, we note that only one researcher was involved in coding.
To mitigate these threats, we release all verbatim survey responses as part of our replication package.


\mysec{Results}
Overall, study participants report having a neutral (15/31; 48.4\%) or at least somewhat positive (15/31; 48.4\%) experience using the NL2Code plugin, with only one participant rating their experience as somewhat negative.


Among the aspects the participants report as \textbf{positive}, we distill two main themes:

\paragraph{The plugin helps find code snippets the developer is aware of but cannot fully remember} (P1, P2, P8, P10, P11, P19, P20, P21, P22, P30, P31)
These tend to be small commands or less familiar API calls and API usage patterns, that users have seen before. Two participants summarize this well:
%
\begin{quote}
    ``On a few occasions, the plugin very conveniently gave me the snippet of code I was looking for, [which] was "on the tip of my tongue".'' \partic{10}
\end{quote}
\begin{quote}
    ``Sometimes I just cannot remember the exact code, but I remember the shape. I could select the correct one easily.'' \partic{2}
\end{quote}
Respondents expressed appreciation for both the generation and retrieval results, and there was little expression of preference for one method over the other, \textit{e.g.}:
\begin{quote}
    ``Even just having the snippets mined from Stack Overflow visible in the IDE was a good memory refresher / source of ideas'' \partic{10}
\end{quote}

\begin{quote}
    ``It was somewhat convenient to not have to switch tabs to Google things, ..., based on my memory, that most of the suggestions I got were from the internet anyway.'' \partic{5}
\end{quote}

\begin{quote}
    ``It has all resources needed at one place.'' \partic{6}
\end{quote}

\paragraph{Using an in-IDE plugin is less disruptive than using a web browser} (P1, P4, P5, P6, P7, P10, P18, P20, P24, P27)
Many of our respondents who were positive about the plugin reiterate expected context-switching benefits of not leaving the IDE while programming, \textit{e.g.}:
\begin{quote}
    ``I like that the plugin stops me having to go and search online for solutions. [...] It can be very easy to get distracted when searching for solutions online.'' \partic{20}
\end{quote}
\begin{quote}
    ``Compared with manual search, this is faster and less disruptive.'' \partic{1}
\end{quote}


\smallskip
Participants also describe many aspects of the plugin that \textbf{could be improved}.

\paragraph{The quality of code generation and retrieval results could be higher} (P3, P4, P5, P7, P9, P13, P14, P23, P27, P29,  P31)
Respondents mentioned that it was \inlineQuote{rare}{7} when they could directly use code from plugin, without modifications. In some cases, results from the plugin were \inlineQuote{not related to the search}{14},
and users 
\inlineQuote{didn't find what [they were] searching for}{31}.
As one respondent humbly summarized it:
\begin{quote}
``The model needs some improvements.'' \partic{4} 
\end{quote}

The insufficient quality of the plugin's results was especially felt as the tasks became more complex and involved APIs with complex usage patterns. One participant summarized this well:
\begin{quote}
``For easy tasks, like walking through a directory in the filesystem, the plugin saves me time because what I did previously was to go to Stack Overflow and copy the code. But for difficult tasks like data processing or ML, the plugin is not helpful. Most snippets are not useful and I have to go to the website of sklearn to read the full doc to understand what I should do.'' \partic{3}
\end{quote}

A particular related pain point is that the snippets from the code retrieval engine often contain spurious elements (as also noted above). In one participant's words:
\begin{quote}
    ``When inserting the code into my program, I would like to **not** copy the input/output examples, and I can't imagine ever wanting those in the program itself.'' \partic{5}
\end{quote}

\paragraph{Users could benefit from additional context} (P3, P5, P8, P18, P19, P20, P24, P26, P27)
Some respondents mention that it would be useful to include additional (links to) explanations and documentation alongside the returned code snippets so that the user could understand what the snippet is supposed to do, or even \inlineQuote{which of the suggestions is the correct one when you are not familiar with a module}{11}.
In two participants' words:
\begin{quote}
``It would be nice if the examples from the internet could contain the relevant context of the discussion (e.g., things to consider when using this suggestion), as well as the input/output examples.'' \partic{5}
\end{quote}
\begin{quote}
``I hope the generated code snippet can have more comments or usage [examples]. Otherwise I still need to search the web to understand what it is.'' \partic{3}
\end{quote}

\smallskip
A closely related theme is that \textit{using the plugin assumes one has a
{\smaller \fontfamily{cmss}\selectfont ``good background understanding of the underlying principles/modules/frameworks''}} {\smaller \partic{11}},
and they primarily need help with \inlineQuote{look[ing] up little syntax bits that you have forgotten}{11}. (P1, P11, P16, P25)
One participant was especially critical:
\begin{quote}
``For more complex problems, I think the plugin does not help at all, because the programmer needs to know the theoretical background.'' \partic{16}
\end{quote}


\paragraph{The plugin could benefit from additional context}
(P4, P9, P10, P17, P30)
Some participants suggest that the plugin could be ``smarter'' if it becomes more aware of the local context in the developer's IDE, \textit{e.g.}:
\begin{quote}
``Sometimes I want to generate an expression to be inserted somewhere, to be assigned to a variable, or to match the indentation level, without having to tell the plugin this explicitly. I didn't feel like the plugin was aware of context.'' \partic{10}
\end{quote}

\smallskip
Participants also comment on how \textit{the plugin's query syntax takes some getting used to} (P2, P12, P15), referring in particular to the way the code generation model expects queries to include variables, while the web search code retrieval engine allows users to only use keywords. For example:
\begin{quote}
``[It became] useful to me towards the end when I got the hang of it and could formulate questions in the correct way (which I feel is somewhat of a skill in itself)'' \partic{15}
\end{quote}
\begin{quote}
``It is not very natural for me to `instantiate' my questions, I mostly like to search [using] keywords or just a description of what I want to achieve.'' \partic{2}
\end{quote}

\paragraph{Querying the plugin could be interactive} (P11, P20, P30)
Finally, some participants suggest to make querying interactive, dialogue-based, rather than unidirectional. This could with refining queries until they are sufficiently well-specified, or to decompose complex functionality into smaller steps, \textit{e.g.}:
\begin{quote}
``A chatbot [...] could identify the rough area in which the user needs assistance, [and] could help narrow it down further, helping to pinpoint an exact solution.'' \partic{20}
\end{quote}

\section{Discussion and Implications}
\label{sec:discussionimplications}

Recent years have seen much progress from machine learning and software engineering researchers developing techniques to better assist programmers in their coding tasks, that exploit the advancements in (deep) learning technology and the availability of very large amounts of data from Big Code repositories like GitHub and Stack Overflow.
A particularly promising research direction in this space has been that addressing the decades-old problem of ``natural language programming''~\cite{dijkstra1979foolishness}, \ie having people instruct machines in the same (natural) language they communicate in with each other, which can be useful in many scenarios, as discussed in the Introduction.
However, while excited about this research direction and actively contributing to it ourselves, we are also questioning whether the most impact from such work can be had by focusing primarily on making technological advancements (\eg as we write this, a one-trillion parameter language model has just been announced~\cite{fedus2021switch}, as only the most current development in a very rapidly evolving field) without also carefully considering how such proposed solutions can fit within the software development workflow, through human-centered research.

In this spirit, we have presented the results of a controlled experiment with \numparticipants participants with diverse background and programming expertise, observed while completing a range of Python programming tasks with and without the help of a NL2Code IDE plugin. The plugin allows users to enter descriptions of intent in natural language, and have corresponding code snippets, ideally implementing said intent, automatically returned.
We designed the plugin with two research goals in mind. First, we sought to evaluate, to our knowledge for the first time using a human-centered approach, the performance of some NL2Code \textit{generation} model with state-of-the-art performance on a benchmark dataset, but unknown performance ``in the wild''.
Second, we sought to contrast the performance and user experience interacting with such a relatively sophisticated model to those of a relatively basic NL2Code \textit{retrieval} engine, that ``merely'' retrieves existing code snippets from Stack Overflow given natural language search queries.
This way, we could estimate not only how far we are from not having to write \textit{any} code while programming, but also how far we have come on this problem given the many recent advancements in learning and availability of datasets.

\mysec{Main Results}
Overall, our results are mixed.
First, after careful statistical analysis in \RQref{task-performance}, comparing tasks completed with and without using the NL2Code plugin (and either of its underlying code generation or retrieval systems), we found no statistically significant differences in task completion times or task correctness scores.

The results for \textbf{code metrics} (SLOC and CC) can be seen as mixed. One the one hand, the code containing automatically generated or retrieved fragments is not, on average, any more complex or any less maintainable than the code written manually, insofar as the CC and SLOC metrics can distinguish. One the other hand, one could have expected the opposite result, \ie that since NL2Code tools are typically trained on idiomatic code, using them should lead to ``better'', more idiomatic code overall, which might suggest lower SLOC and CC values, on average. 

Among the possible explanations for why we don't find supporting evidence for the ``better code'' hypothesis, two stand out: (i)~the two metrics are only crude approximations of the complex, multifaceted concept of code quality; and (ii)~even when writing code ``manually'', developers still consult the Web and Stack Overflow (\ie the same resources that these NL2Code tools were trained on) and copy-paste code therein. 
To better understand the interaction between using the plugin and using a traditional Web browser, we used the event logs from our instrumented environment and compared the distributions of in-browser Web searches between tasks where the \numparticipants study participants used the NL2Code plugin (median 3, mean 5, min 0, max 35 searches per user per task) and tasks where they did not (median 4, mean 7, min 0, max 48).
A mixed-effects regression model similar to the ones in Section~\ref{sec:results:task-performance}, controlling for individual self-reported experience and with random effects for {\small \texttt{user}} and {\small \texttt{task}}, reveals a statistically significant effect of using the plugin on the number of in-browser Web searches: on average, using the plugin is associated with 2.8 \textit{fewer} in-browser Web searches; however, this effect is smaller than the standard deviation of the random {\small \texttt{user}} intercept (\textasciitilde 4 in-browser Web searches). 
We conclude that developers still search the Web when using the plugin, even if slightly less than when not using the plugin.

Using a similar argument, the result for \textbf{task correctness scores} can be seen as mixed. Code containing automatically generated or retrieved snippets is not, on average, any less appropriate for a given task as per our rubric than code written manually. However, using the NL2Code plugin doesn't seem to help our study participants significantly improve their scores either, despite there being room for improvement. Even though across our sample the median score per task was 7 out of 10 when using the plugin and 6 when not using the plugin, the multivariate regression analysis did not find the difference to be statistically significant.

The result for \textbf{task completion times} can be seen as negative and, thus, is perhaps the most surprising of our results: on average, study participants do not complete their tasks statistically significantly faster when using the NL2Code plugin compared to when they are not using it. There are several possible explanations for this negative result. First, we acknowledge fundamental limitations of our study design, which we hope future researchers can improve on. In particular, our tasks, despite their diversity and, we believe, representativeness of real-world Python use, may not lend themselves sufficiently well to NL2Code queries and, therefore, study participants may not have sufficient opportunities to use, and benefit from, the plugin. Moreover, our study population (\numparticipants participants) may not be large enough for us to detect effects with small sizes, should they exist.

However, even with these limitations, considering also our results for \RQref{types-of-code} and \RQref{opinions} we argue that another explanation is plausible: \textit{our NL2Code plugin and its main underlying code generation technology, despite state-of-the-art (BLEU-score) performance on a benchmark dataset, is not developed enough to be markedly useful in practice just yet}. Our telemetry data (\RQref{types-of-code}) shows not only that study participants still carry out in-browser Web searches even though the NL2Code plugin was available, as discussed above, but also that the code snippets returned by the plugin, when used, undergo edits after insertion in the IDE, suggesting insufficient quality to begin with. Our qualitative survey data (\RQref{opinions}) paints a similar picture of overall insufficient quality of the NL2Code results.

\mysec{Implications}
While our study suggests that state-of-the-art learning-based natural language to code generation technology is ways away from being useful in practice, our results should be interpreted more optimistically. 

First, we argue that \textbf{the problem is worth working on}. In contemporary software development, which involves countless and constantly changing programming languages and APIs, natural language can be a useful medium to turn ideas into code, even for experienced programmers. A large fraction of our study participants commended NL2Code developer assistants for helping them remember the precise syntax or sequence of API calls and their arguments, required to implement some particular piece of functionality. When integrated into the development workflow, \eg through an IDE plugin, such systems can help developers focus by reducing the need for context switching, further improving their productivity. Our quantitative task performance results for the current version of this NL2Code plugin, while negative, do not imply that future, better performing such systems will also not be markedly useful in practice; the qualitative data from our our study participants already suggests otherwise, as does quantitative data from prior research on the usefulness of in-IDE code search plugins~\cite{Ponzanelli2014MiningST}.

Second, we argue that \textbf{this particular style of code \textit{generation} is worth working on}. Our analysis of input queries and resulting code snippets for \RQref{types-of-code} shows that the code generation model produces fundamentally different results than the (simple) code retrieval engine we used for comparison, and that study participants choose snippets returned by the code generation model almost as frequently as they do snippets from the code retrieval engine. In turn, this suggests that, at least within the scope of the current study, one type of model cannot be used as a substitute for the other. As discussed above, the code generation model does almost always produce different results than the code retrieval model. However, it was unclear from that analysis whether the generated code snippets reflect some fundamentally higher level of sophistication inherent to the code generation model, or whether the code retrieval engine we used for comparison is simply too naive. 

To further test this, we performed an additional analysis. 
Specifically, we looked up the chosen code generation snippets in the manually-labeled Stack Overflow dataset used for training the code generation model, to assess whether the model is simply memorizing the training inputs. Only 13 out of the 173 unique queries (\textasciitilde 7.5\%) had as the chosen code fragment snippets found verbatim in the model's training dataset.
Therefore, the evidence so far suggests that the code generation model does add some level of sophistication, and customization of results to the developers' intent (\eg composing function calls), compared to what \textit{any} code retrieval engine could.

Third, we provide the following \textbf{concrete future work recommendations} for researchers and toolsmiths in this area, informed by our results:
\begin{itemize}
\setlength\itemsep{4pt}

    \item  \textit{Combine code generation with code retrieval.} Our results suggest that some queries may be better answered through code retrieval techniques, and others through code generation. We recommend that future research continue to explore these types of approaches jointly, \eg using hybrid models~\cite{hashimoto2018retrieve,hayati-etal-2018-retrieval} that may be able to combine the best of both worlds.
    
    \item  \textit{Consider the user's local context as part of the input.} Our oracle comparison revealed that users' natural language queries can often be disambiguated by considering the local context provided by the source files they were working in at the time, which in turn could lead to better performance of the code generation model. There is already convincing evidence from prior work that considering a user's local context provides unique information about what code they might type next~\cite{tu2014localness}.
    In addition, some work on code retrieval has also considered how to incorporate context to improve retrieval results \cite{campbell2017nlp2code}; this may be similarly incorporated.
    
    \item  \textit{Consider the user's local context as part of the output.} Considering where in their local IDE users are when invoking an NL2Code assistant can also help with localizing the returned code snippets for that context. Some transformations are relatively simple, \eg pretty printing and indentation. Other transformations may require more advanced program analysis but are still well within reach of current technology, \eg renaming variables used in the returned snippet to match the local context (the Bing Developer Assistant code retrieval engine~\cite{wei2015building} already does this), or applying coding conventions~\cite{allamanis2014learning}.
    
    \item \textit{Provide more context for each returned snippet.} Our study shows that NL2Code generation or retrieval systems can be useful when users already know what the right answer is, but they need help retrieving it. At the same time, many of our study participants reported lacking sufficient background knowledge, be it domain-specific or API-specific, to recognize when a plugin-returned code snippet is the right one given their query, or what the snippet does in detail. Future research should consider incorporating more context and documentation together with the plugin's results, that allows users to better understand the code, \eg links to Stack Overflow, official documentation pages, explanations of domain-specific concepts, other API usage examples. 
    One example of this is the work of \citet{moreno2015can}, which retrieves usage examples that show how to use a specific method.

    \item  \textit{Provide a unified and intuitive query syntax.} We observed that users are not always formulating queries in the way that we would expect, perhaps because they are used to traditional search engines that are more robust to noisy inputs and designed for keyword-based search. The NL2Code generation model we experimented with in this study was trained on natural language queries that are not only complete English sentences, but also include references to variables or literals involved with an intent, specially delimited by dedicated syntax (grave accents). As our respondents commented in the post-test survey, getting used to formulating queries this way takes some practice. Future research should consider not only what is the most natural way for users to describe their intent using natural language, but also how to provide a unified query syntax for both code generation and code retrieval, to minimize confusion. Robust semantic parsing techniques~\cite{arthur2015semantic,radhakrishnan2020colloql} may also help with interpreting ill-specified user queries.
    
    \item  \textit{Provide dialogue-based query capability.} Dialogue-based querying could allow users to refine their natural language intents until they are sufficiently precise for the underlying models to confidently provide some results.
    Future systems may reference work on query reformulation in information retrieval, where the user queries are refined to improve retrieval results both for standard information retrieval~\cite{arens1996query} and code retrieval~\cite{,Haiduc2013AutomaticQR,Hill2014NLbasedQR}.
    In addition, in the NLP community there have been notable advancements recently in interactive semantic parsing~\cite{yao2019interactive,karamcheti-etal-2020-learning}, \ie soliciting user input when dealing with missing information or ambiguity while processing the initial natural language query, which could be of use as well.
    
    \item  \textit{Consider new paradigms of evaluation for code generation and retrieval systems.} Usage log data, such as the ones we collected here, is arguably very informative and useful for researchers looking to evaluate NL2Code systems. However, compared to automated metrics such as BLEU, such data is much less readily available. We argue that such data is worth collecting even if only in small quantities. For example, with little but high quality data, one could still train a reranker~\cite{yin-neubig-2019-reranking} to try to select the outputs that a human user selected; if the predictive power exceeds that of BLEU alone, then the trained reranker could be used to automatically evaluate the quality of the generated or retrieved code more realistically than by using BLEU.
    
\end{itemize}

\section{Related Work}
\label{sec:relatedwork}

Finally, we more extensively discuss how this work fits in the landscape of the many other related works in the area.

\subsection{NL2Code Generation}

While we took a particular approach to code generation, there are a wide variety of other options. 
Researchers have proposed that natural language dialogue could be a 
new form of human-computer interaction since nearly the 
advent of modern computers~\cite{heidorn1976automatic,Ginsparg1978NaturalLP,dijkstra1979foolishness,mihalcea2006nlp}.
The bulk of prior work either targeted domain-specific languages (DSLs),
or focused on task-specific code generation for general-purpose languages, 
where more progress could be made given the relatively constrained vocabulary 
and output code space.
Examples include generating formatted input file parsers~\cite{Lei2013FromNL}; 
structured, idiomatic sequences of API calls~\cite{Raghothaman2016SWIMSW}; 
regular expressions~\cite{Kushman2013UsingSU,Manshadi2013IntegratingPB,Parisotto2017NeuroSymbolicPS}; 
string manipulation DSL programs~\cite{Raza2015CompositionalPS}; 
card implementations for trading card games~\cite{DBLP:conf/acl/LingBGHKWS16};
and solutions to the simplest of programming competition-style problems~\cite{Balog2017DeepCoderLT}.

With the recent boom of neural networks and deep learning in 
natural language processing, generating arbitrary code in a general-purpose language~\cite{yin2017syntactic,yin2018tranx} are becoming more feasible. 
Some have been trained on both official 
API documentation and Stack Overflow questions and answers~\cite{xu2020incorporating}.
There are also similar systems%
\footnote{This is, of course, among the many other use cases for neural network models of code and natural language such as code summarization~\cite{Iyer2016SummarizingSC,Yao2018StaQCAS}, or embedding models that represent programming languages together with natural languages~\cite{Feng2020CodeBERTAP}. \citet{allamanis2018survey} provide a comprehensive survey of the use cases of machine learning models in this area.}
able to generate class member functions 
given natural language descriptions of intent and the programmatic 
context provided by the rest of the class~\cite{Iyer2018MappingLT},
and to generate the API call sequence in a Jupyter Notebook code cell
given the natural language and code history up to that particular  
cell~\cite{Agashe2019JuICeAL}.

\subsection{NL2Code Retrieval}

Code retrieval has similarly seen a wide variety of approaches.
The simplest way to perform retrieval is to start with existing information retrieval models designed for natural language search, and adapt them specifically for the source code domain through query reformulation or other methods~\cite{Haiduc2013AutomaticQR,Hill2014NLbasedQR,Keivanloo2014SpottingWC,wei2015building,Lu2015QueryEV,Vinayakarao2017ANNEIS}.
Other research works utilize deep learning models \cite{Allamanis2015BimodalMO,Iyer2016SummarizingSC,gu2018deep,husain2019codesearchnet} to train a relevance model between natural language queries and corresponding code snippets.
It is also possible to exploit code annotations to generate additional information to help improve code retrieval performance \cite{Yao2019CoaCorCA} or extracted abstract programming patterns and associated natural language keywords for more content-based code search~\cite{Keivanloo2014SpottingWC}.
Many of the models achieve good performance on human annotated relevance benchmark datasets between natural language and code snippets.
Practically, however, many developers simply rely on generic natural-language search engines like Google to find appropriate code snippets by first locating pages that contain code snippets through natural language queries~\cite{sadowski2015developers} on programming QA websites like Stack Overflow.

\subsection{Evaluation of NL2Code Methods}
In order to evaluate whether NL2Code methods are succeeding, the most common way is to create a ``reference'' program that indeed implements the desired functionality, and measure the similarity of the generated program to this reference program.
Because deciding whether two programs are equivalent is, in the general case, undecidable~\cite{rice1953classes}, alternative means are necessary.
For code generation in limited domains, this is often done by creating a small number of input-output examples and making sure that the generated program returns the same values as the reference program over these tests~\cite{zelle1996learning,berant2013semantic,zettlemoyer2007online,wang2015building,yao2014information,zhong2017seq2sql,zavershynskyi2018naps,Kulal2019SPoCSP,Zhong2020SemanticSF}.
However, when scaling to broader domains, creating a thorough and comprehensive suite of test cases over programs that have a wide variety of assumptions about the input and output data formats is not trivial.

As a result, much research work on code generation and retrieval take a different tack.
Specifically, many code generation methods \cite{yin2017syntactic,Iyer2018MappingLT,Agashe2019JuICeAL,xu2020incorporating} aim to directly compare generated code snippets against ground truth snippets, using token sequence comparison metrics borrowed from machine translation tasks, such as BLEU score~\cite{papineni-etal-2002-bleu}.
However, many code snippets are equivalent in functionality but differ quite largely in terms of token sequences, or differ only slightly in token sequence but greatly in functionality, and thus BLEU is an imperfect metric of correctness of a source code snippet~\cite{tran2019does}.

Code retrieval, on the other hand, is the task of retrieving relevant code given a natural language query, that is related to other information retrieval tasks. 
Since code retrieval is often used to search for vague concepts and ideas, human-annotated relevance annotations are needed for evaluation.
The common methods used in research work~\cite{gu2018deep,Yao2018StaQCAS,husain2019codesearchnet} compare the retrieved code snippet candidates given a natural language query, with a human annotated list of code snippet relevance, using common automatic information retrieval metrics like NDCG, MRR, etc.~\cite{manning2008introduction}
The drawback of this evaluation method is that the cost of retrieval relevance annotation is high, and often requires experts in the specific area.
Also, since the candidate lists are usually long, only a few unique natural language queries could be annotated.
For example, one of the most recent large scale code search challenge  CodeSearchNet~\cite{husain2019codesearchnet} contains only 99 unique natural language queries, along with their corresponding code snippet relevance expert  annotations, leading to smaller coverage of real world development scenarios in evaluation.

Regardless of the automatic metrics above, in the end our final goal is to help developers in their task of writing code.
This paper fills the gap of the fundamental question of whether these methods will be useful within the developer workflow.

\subsection{In-IDE Plugins}
Similarly, many research works on deploying plugins inside IDEs to help developers have been performed.
Both \citet{ponzanelli2013seahawk} and \citet{Ponzanelli2014MiningST} focus on reducing context switching in IDE by incorporating Stack Overflow, by using the context in the IDE to automatically retrieve pertinent discussions from Stack Overflow.
\citet{Subramanian2014LiveAD} proposes a plugin to enhance traditional API documentation with up-to-date source code examples.
\citet{Rahman2014SurfClipseCM} and \citet{Liu2016EXPSOLRO} designs the plugin to help developers find solutions on the Internet to program exceptions and errors.
Following the similar route, \citet{Brandt2009TwoSO} studies  opportunistic programming where programmers leverage online resources with a range of intentions, including the assistance that could be accessed from inside the IDE.

Besides plugin developed to reduce context-switching to other resources in developer workflows, \citet{Amann2016FeedBaGAI} focus on collecting data of various developer activities from inside the IDE that fuel empirical research on the area~\cite{proksch2018enriched}.

This paper proposes an in-IDE plugin that incorporates code generation in addition to code retrieval to test the user experience in the real development workflow. 
In the meantime it also collects fine-grained user activities interacting with the plugin as well as editing the code snippet candidates, to provide public data for future work.

\subsection{End-User Development}
The direction of exploring using natural language intents to generate code snippets is closely related to end-user development~\cite{Lieberman2006EndUserDA}, which allows end-users (people who are not professional software developers) to program computers.
\citet{Cypher1993WatchWI} is among the first work that enables end-user to program by demonstration.

Traditionally, programming has been performed by software developers who write code directly in programming languages for the majority of functionality they wish to implement.
However, acquiring the requisite knowledge to perform this task requires time-consuming training and practice, and even for skilled programmers, writing programs requires a great amount of time and effort.
To this end, there have been many recent developments on no-code or low-code software development platforms that allow both programmers and non-programmers to develop in modalities of interaction other than code~\cite{Sahay2020SupportingTU}.
Some examples include visual programming languages such as Scratch~\cite{Maloney2010TheSP} that offers a building-block style graphical user interface to implement logic.
In specific domains such as user interface design and prototyping, recent advances in deep learning models also enable developers to sketch the user interface visually and then automatically generates user interface code with the sketch~\cite{Beltramelli2018pix2codeGC}, or using existing screenshots~\cite{Nguyen2015ReverseEM}.

Besides visual no-code or low-code programming interfaces, there has also been much progress on program synthesis~\cite{Basin2004SynthesisOP,Feser2015SynthesizingDS,Feng2018ProgramSU,Bodk2008ProgramSB}, which uses input-output examples, logic sketches, etc. to automatically generate functions, with some recent advances that use machine learning models~\cite{Balog2017DeepCoderLT,Chen2019ExecutionGuidedNP,Ellis2019WriteEA,Shin2019ProgramSA}. 
Some work also generate programs from easier-to-write pseudo-code~\cite{Kulal2019SPoCSP,Zhong2020SemanticSF}.

There are other work in the area.
\citet{Chasins2015BrowserRA,Barman2016RingerWA,Chasins2018RousillonSD} make web automation accessible to non-coders through programming by demonstration, while \citet{Li2017SUGILITECM,Li2018APPINITEAM,Li2019PUMICEAM} automates mobile applications with multimodal inputs including demonstration and natural language intents.
\citet{Head2017WritingRC} combines teacher expertise with data-driven program synthesis techniques to learn bug-fixing code transformations in classroom scenarios.
\citet{Head2018InteractiveEO} helps users extract executable, simplified code from existing code.
\citet{Ko2004DesigningTW,Ko2008DebuggingR} provides a debugging interface for asking questions about program behavior.
\citet{Myers2016ImprovingAU} discusses API designers should consider usability as a step towards enabling end-user programming.
\citet{Kery2017ExploringEP,Kery2017VarioliteSE} enable data scientists to explore data easily with exploratory programming. 
Our paper's plugin of using both state-of-the-art code generation and code retrieval to provide more natural programming experience to developers, with the potential future of enabling end-user programming, is related to \citet{Myers2004NaturalPL} that envisions natural language programming.

\subsection{Code Completion}
Many developers use Integrated Development Environments (IDEs) as a convenient solution to help with many aspects during development.
Most importantly, many developers actively rely on intelligent code-completion aid like IntelliSense\footnote{\url{https://docs.microsoft.com/en-us/visualstudio/ide/using-intellisense}}  for Visual Studio~\cite{amann2016study,proksch2018enriched} to help developers learn more about the code, keep track of the parameters, and add calls to properties and methods with only a few keystrokes. 
Many of intelligent code-completion tools also consider the current code context where the developer is editing.
With the recent advances in machine learning and deep learning, example tools like IntelliCode\footnote{\url{https://visualstudio.microsoft.com/services/intellicode}} for Visual Studio, Codota\footnote{\url{https://www.codota.com/}} and TabNine\footnote{https://www.tabnine.com/} present AI-assisted code-suggestion and code-completion based on the current source code context, learned from abundant amounts of projects over the Internet.
The scope of our paper is to investigate generating or retrieving code using natural language queries, rather than based on the context of the current source code.

\section{Conclusion}
In this paper, we performed an extensive user study of in-IDE code generation and retrieval, developing an experimental harness and framework for analysis.
This demonstrated challenges and limitations in the current state of both code generation and code retrieval; results were mixed with regards to the impact on the developer workflow, including time efficiency, code correctness and code quality.
However, there was also promise: developers subjectively enjoyed the experience of using in-IDE developer management tools, and provided several concrete areas for improvement. 
We believe that these results will spur future, targeted development in productive directions for code generation and retrieval models.

\begin{acks}
This research was supported by NSF Award No. 1815287 ``Open-domain, Data-driven Code Synthesis from Natural Language.''
We thank William Qian, who was involved in development of an early version of the plugin.
We thank all participants taken part in the user study experiments for their effort on completing the tasks testing the intelligent programming interface.
We would like to give special thanks to Ziyu Yao, and NeuLab members Shuyan Zhou, Zecong Hu, among others for the early testing of the plugin and the user study and their valuable feedback.
We also thank anonymous reviewers for their comments on revising this paper.
\end{acks}

\bibliographystyle{ACM-Reference-Format}
\bibliography{sample-base,ref}


\begin{thebibliography}{130}


\ifx \showCODEN    \undefined \def \showCODEN     #1{\unskip}     \fi
\ifx \showDOI      \undefined \def \showDOI       #1{#1}\fi
\ifx \showISBNx    \undefined \def \showISBNx     #1{\unskip}     \fi
\ifx \showISBNxiii \undefined \def \showISBNxiii  #1{\unskip}     \fi
\ifx \showISSN     \undefined \def \showISSN      #1{\unskip}     \fi
\ifx \showLCCN     \undefined \def \showLCCN      #1{\unskip}     \fi
\ifx \shownote     \undefined \def \shownote      #1{#1}          \fi
\ifx \showarticletitle \undefined \def \showarticletitle #1{#1}   \fi
\ifx \showURL      \undefined \def \showURL       {\relax}        \fi
\providecommand\bibfield[2]{#2}
\providecommand\bibinfo[2]{#2}
\providecommand\natexlab[1]{#1}
\providecommand\showeprint[2][]{arXiv:#2}

\bibitem[\protect\citeauthoryear{Agashe, Iyer, and Zettlemoyer}{Agashe
  et~al\mbox{.}}{2019}]%
        {Agashe2019JuICeAL}
\bibfield{author}{\bibinfo{person}{R. Agashe}, \bibinfo{person}{Srini Iyer},
  {and} \bibinfo{person}{Luke Zettlemoyer}.} \bibinfo{year}{2019}\natexlab{}.
\newblock \showarticletitle{JuICe: A Large Scale Distantly Supervised Dataset
  for Open Domain Context-based Code Generation}. In
  \bibinfo{booktitle}{\emph{2019 Conference on Empirical Methods in Natural
  Language Processing and 9th International Joint Conference on Natural
  Language Processing (EMNLP/IJCNLP)}}.
\newblock


\bibitem[\protect\citeauthoryear{Allamanis, Barr, Bird, and Sutton}{Allamanis
  et~al\mbox{.}}{2014}]%
        {allamanis2014learning}
\bibfield{author}{\bibinfo{person}{Miltiadis Allamanis},
  \bibinfo{person}{Earl~T Barr}, \bibinfo{person}{Christian Bird}, {and}
  \bibinfo{person}{Charles Sutton}.} \bibinfo{year}{2014}\natexlab{}.
\newblock \showarticletitle{Learning natural coding conventions}. In
  \bibinfo{booktitle}{\emph{International Symposium on Foundations of Software
  Engineering (ESEC/FSE)}}. \bibinfo{pages}{281--293}.
\newblock


\bibitem[\protect\citeauthoryear{Allamanis, Barr, Devanbu, and
  Sutton}{Allamanis et~al\mbox{.}}{2018}]%
        {allamanis2018survey}
\bibfield{author}{\bibinfo{person}{Miltiadis Allamanis},
  \bibinfo{person}{Earl~T Barr}, \bibinfo{person}{Premkumar Devanbu}, {and}
  \bibinfo{person}{Charles Sutton}.} \bibinfo{year}{2018}\natexlab{}.
\newblock \showarticletitle{A survey of machine learning for big code and
  naturalness}.
\newblock \bibinfo{journal}{\emph{ACM Computing Surveys (CSUR)}}
  \bibinfo{volume}{51}, \bibinfo{number}{4} (\bibinfo{year}{2018}),
  \bibinfo{pages}{1--37}.
\newblock


\bibitem[\protect\citeauthoryear{Allamanis, Tarlow, Gordon, and Wei}{Allamanis
  et~al\mbox{.}}{2015}]%
        {Allamanis2015BimodalMO}
\bibfield{author}{\bibinfo{person}{Miltiadis Allamanis},
  \bibinfo{person}{Daniel Tarlow}, \bibinfo{person}{A. Gordon}, {and}
  \bibinfo{person}{Y. Wei}.} \bibinfo{year}{2015}\natexlab{}.
\newblock \showarticletitle{Bimodal Modelling of Source Code and Natural
  Language}. In \bibinfo{booktitle}{\emph{The 32nd International Conference on
  Machine Learning (ICML)}}.
\newblock


\bibitem[\protect\citeauthoryear{Amann, Proksch, and Nadi}{Amann
  et~al\mbox{.}}{2016a}]%
        {Amann2016FeedBaGAI}
\bibfield{author}{\bibinfo{person}{S. Amann}, \bibinfo{person}{Sebastian
  Proksch}, {and} \bibinfo{person}{S. Nadi}.} \bibinfo{year}{2016}\natexlab{a}.
\newblock \showarticletitle{FeedBaG: An interaction tracker for Visual Studio}.
\newblock \bibinfo{journal}{\emph{International Conference on Program
  Comprehension (ICPC)}} (\bibinfo{year}{2016}), \bibinfo{pages}{1--3}.
\newblock


\bibitem[\protect\citeauthoryear{Amann, Proksch, Nadi, and Mezini}{Amann
  et~al\mbox{.}}{2016b}]%
        {amann2016study}
\bibfield{author}{\bibinfo{person}{Sven Amann}, \bibinfo{person}{Sebastian
  Proksch}, \bibinfo{person}{Sarah Nadi}, {and} \bibinfo{person}{Mira Mezini}.}
  \bibinfo{year}{2016}\natexlab{b}.
\newblock \showarticletitle{A study of visual studio usage in practice}. In
  \bibinfo{booktitle}{\emph{2016 IEEE 23rd International Conference on Software
  Analysis, Evolution, and Reengineering (SANER)}}, Vol.~\bibinfo{volume}{1}.
  IEEE, \bibinfo{pages}{124--134}.
\newblock


\bibitem[\protect\citeauthoryear{Arens, Knoblock, and Shen}{Arens
  et~al\mbox{.}}{1996}]%
        {arens1996query}
\bibfield{author}{\bibinfo{person}{Yigal Arens}, \bibinfo{person}{Craig~A
  Knoblock}, {and} \bibinfo{person}{Wei-Min Shen}.}
  \bibinfo{year}{1996}\natexlab{}.
\newblock \showarticletitle{Query reformulation for dynamic information
  integration}.
\newblock \bibinfo{journal}{\emph{Journal of Intelligent Information Systems}}
  \bibinfo{volume}{6}, \bibinfo{number}{2-3} (\bibinfo{year}{1996}),
  \bibinfo{pages}{99--130}.
\newblock


\bibitem[\protect\citeauthoryear{Arthur, Neubig, Sakti, Toda, and
  Nakamura}{Arthur et~al\mbox{.}}{2015}]%
        {arthur2015semantic}
\bibfield{author}{\bibinfo{person}{Philip Arthur}, \bibinfo{person}{Graham
  Neubig}, \bibinfo{person}{Sakriani Sakti}, \bibinfo{person}{Tomoki Toda},
  {and} \bibinfo{person}{Satoshi Nakamura}.} \bibinfo{year}{2015}\natexlab{}.
\newblock \showarticletitle{Semantic parsing of ambiguous input through
  paraphrasing and verification}.
\newblock \bibinfo{journal}{\emph{Transactions of the Association for
  Computational Linguistics (TACL)}}  \bibinfo{volume}{3}
  (\bibinfo{year}{2015}), \bibinfo{pages}{571--584}.
\newblock


\bibitem[\protect\citeauthoryear{Bacchelli, Ponzanelli, and Lanza}{Bacchelli
  et~al\mbox{.}}{2012}]%
        {bacchelli2012harnessing}
\bibfield{author}{\bibinfo{person}{Alberto Bacchelli}, \bibinfo{person}{Luca
  Ponzanelli}, {and} \bibinfo{person}{Michele Lanza}.}
  \bibinfo{year}{2012}\natexlab{}.
\newblock \showarticletitle{Harnessing Stack Overflow for the IDE}. In
  \bibinfo{booktitle}{\emph{International Workshop on Recommendation Systems
  for Software Engineering (RSSE)}}. IEEE, \bibinfo{pages}{26--30}.
\newblock


\bibitem[\protect\citeauthoryear{Balog, Gaunt, Brockschmidt, Nowozin, and
  Tarlow}{Balog et~al\mbox{.}}{2017}]%
        {Balog2017DeepCoderLT}
\bibfield{author}{\bibinfo{person}{Matej Balog}, \bibinfo{person}{Alexander~L.
  Gaunt}, \bibinfo{person}{Marc Brockschmidt}, \bibinfo{person}{Sebastian
  Nowozin}, {and} \bibinfo{person}{Daniel Tarlow}.}
  \bibinfo{year}{2017}\natexlab{}.
\newblock \showarticletitle{DeepCoder: Learning to Write Programs}.
\newblock \bibinfo{journal}{\emph{5th International Conference on Learning
  Representations (ICLR)}} (\bibinfo{year}{2017}).
\newblock


\bibitem[\protect\citeauthoryear{Barman, Chasins, Bod{\'i}k, and
  Gulwani}{Barman et~al\mbox{.}}{2016}]%
        {Barman2016RingerWA}
\bibfield{author}{\bibinfo{person}{S. Barman}, \bibinfo{person}{Sarah~E.
  Chasins}, \bibinfo{person}{Rastislav Bod{\'i}k}, {and} \bibinfo{person}{Sumit
  Gulwani}.} \bibinfo{year}{2016}\natexlab{}.
\newblock \showarticletitle{Ringer: web automation by demonstration}.
\newblock \bibinfo{journal}{\emph{Proceedings of the 2016 ACM SIGPLAN
  International Conference on Object-Oriented Programming, Systems, Languages,
  and Applications}} (\bibinfo{year}{2016}).
\newblock


\bibitem[\protect\citeauthoryear{Basin, Deville, Flener, Hamfelt, and
  Nilsson}{Basin et~al\mbox{.}}{2004}]%
        {Basin2004SynthesisOP}
\bibfield{author}{\bibinfo{person}{D. Basin}, \bibinfo{person}{Y. Deville},
  \bibinfo{person}{P. Flener}, \bibinfo{person}{A. Hamfelt}, {and}
  \bibinfo{person}{J{\o}rgen~Fischer Nilsson}.}
  \bibinfo{year}{2004}\natexlab{}.
\newblock \showarticletitle{Synthesis of Programs in Computational Logic}. In
  \bibinfo{booktitle}{\emph{Program Development in Computational Logic}}.
\newblock


\bibitem[\protect\citeauthoryear{Bell, Fairbrother, and Jones}{Bell
  et~al\mbox{.}}{2019}]%
        {bell2019fixed}
\bibfield{author}{\bibinfo{person}{Andrew Bell}, \bibinfo{person}{Malcolm
  Fairbrother}, {and} \bibinfo{person}{Kelvyn Jones}.}
  \bibinfo{year}{2019}\natexlab{}.
\newblock \showarticletitle{Fixed and random effects models: making an informed
  choice}.
\newblock \bibinfo{journal}{\emph{Quality \& Quantity}} \bibinfo{volume}{53},
  \bibinfo{number}{2} (\bibinfo{year}{2019}), \bibinfo{pages}{1051--1074}.
\newblock


\bibitem[\protect\citeauthoryear{Beltramelli}{Beltramelli}{2018}]%
        {Beltramelli2018pix2codeGC}
\bibfield{author}{\bibinfo{person}{Tony Beltramelli}.}
  \bibinfo{year}{2018}\natexlab{}.
\newblock \showarticletitle{pix2code: Generating Code from a Graphical User
  Interface Screenshot}. In \bibinfo{booktitle}{\emph{Proceedings of the {ACM}
  {SIGCHI} Symposium on Engineering Interactive Computing Systems, {EICS} 2018,
  Paris, France, June 19-22, 2018}}. \bibinfo{publisher}{{ACM}},
  \bibinfo{pages}{3:1--3:6}.
\newblock
\urldef\tempurl%
\url{https://doi.org/10.1145/3220134.3220135}
\showDOI{\tempurl}


\bibitem[\protect\citeauthoryear{Berant, Chou, Frostig, and Liang}{Berant
  et~al\mbox{.}}{2013}]%
        {berant2013semantic}
\bibfield{author}{\bibinfo{person}{Jonathan Berant}, \bibinfo{person}{Andrew
  Chou}, \bibinfo{person}{Roy Frostig}, {and} \bibinfo{person}{Percy Liang}.}
  \bibinfo{year}{2013}\natexlab{}.
\newblock \showarticletitle{Semantic parsing on freebase from question-answer
  pairs}. In \bibinfo{booktitle}{\emph{Proceedings of the 2013 conference on
  empirical methods in natural language processing (EMNLP)}}.
  \bibinfo{pages}{1533--1544}.
\newblock


\bibitem[\protect\citeauthoryear{Brandt, Guo, Lewenstein, Dontcheva, and
  Klemmer}{Brandt et~al\mbox{.}}{2009}]%
        {Brandt2009TwoSO}
\bibfield{author}{\bibinfo{person}{J. Brandt}, \bibinfo{person}{P. Guo},
  \bibinfo{person}{J. Lewenstein}, \bibinfo{person}{Mira Dontcheva}, {and}
  \bibinfo{person}{Scott~R. Klemmer}.} \bibinfo{year}{2009}\natexlab{}.
\newblock \showarticletitle{Two studies of opportunistic programming:
  interleaving web foraging, learning, and writing code}. In
  \bibinfo{booktitle}{\emph{Proceedings of the SIGCHI Conference on Human
  Factors in Computing Systems (CHI)}}.
\newblock


\bibitem[\protect\citeauthoryear{Campbell and Treude}{Campbell and
  Treude}{2017}]%
        {campbell2017nlp2code}
\bibfield{author}{\bibinfo{person}{Brock~Angus Campbell} {and}
  \bibinfo{person}{Christoph Treude}.} \bibinfo{year}{2017}\natexlab{}.
\newblock \showarticletitle{NLP2Code: Code snippet content assist via natural
  language tasks}. In \bibinfo{booktitle}{\emph{International Conference on
  Software Maintenance and Evolution (ICSME)}}. IEEE,
  \bibinfo{pages}{628--632}.
\newblock


\bibitem[\protect\citeauthoryear{Catet{\'e} and Barnes}{Catet{\'e} and
  Barnes}{2017}]%
        {Catet2017ApplicationOT}
\bibfield{author}{\bibinfo{person}{Veronica Catet{\'e}} {and}
  \bibinfo{person}{T. Barnes}.} \bibinfo{year}{2017}\natexlab{}.
\newblock \showarticletitle{Application of the Delphi Method in Computer
  Science Principles Rubric Creation}.
\newblock \bibinfo{journal}{\emph{Proceedings of the 2017 ACM Conference on
  Innovation and Technology in Computer Science Education}}
  (\bibinfo{year}{2017}).
\newblock


\bibitem[\protect\citeauthoryear{Chasins, Barman, Bod{\'i}k, and
  Gulwani}{Chasins et~al\mbox{.}}{2015}]%
        {Chasins2015BrowserRA}
\bibfield{author}{\bibinfo{person}{Sarah~E. Chasins}, \bibinfo{person}{S.
  Barman}, \bibinfo{person}{Rastislav Bod{\'i}k}, {and} \bibinfo{person}{Sumit
  Gulwani}.} \bibinfo{year}{2015}\natexlab{}.
\newblock \showarticletitle{Browser Record and Replay as a Building Block for
  End-User Web Automation Tools}.
\newblock \bibinfo{journal}{\emph{Proceedings of the 24th International
  Conference on World Wide Web (WWW)}} (\bibinfo{year}{2015}).
\newblock


\bibitem[\protect\citeauthoryear{Chasins, Mueller, and Bod{\'i}k}{Chasins
  et~al\mbox{.}}{2018}]%
        {Chasins2018RousillonSD}
\bibfield{author}{\bibinfo{person}{Sarah~E. Chasins}, \bibinfo{person}{Maria
  Mueller}, {and} \bibinfo{person}{Rastislav Bod{\'i}k}.}
  \bibinfo{year}{2018}\natexlab{}.
\newblock \showarticletitle{Rousillon: Scraping Distributed Hierarchical Web
  Data}.
\newblock \bibinfo{journal}{\emph{Proceedings of the 31st Annual ACM Symposium
  on User Interface Software and Technology (UIST)}} (\bibinfo{year}{2018}).
\newblock


\bibitem[\protect\citeauthoryear{Chen, Liu, and Song}{Chen
  et~al\mbox{.}}{2019}]%
        {Chen2019ExecutionGuidedNP}
\bibfield{author}{\bibinfo{person}{X. Chen}, \bibinfo{person}{C. Liu}, {and}
  \bibinfo{person}{D. Song}.} \bibinfo{year}{2019}\natexlab{}.
\newblock \showarticletitle{Execution-Guided Neural Program Synthesis}. In
  \bibinfo{booktitle}{\emph{7th International Conference on Learning
  Representations (ICLR)}}.
\newblock


\bibitem[\protect\citeauthoryear{Cohen}{Cohen}{2003}]%
        {cohen2003applied}
\bibfield{author}{\bibinfo{person}{J. Cohen}.} \bibinfo{year}{2003}\natexlab{}.
\newblock \bibinfo{booktitle}{\emph{{Applied multiple regression/correlation
  analysis for the behavioral sciences}}}.
\newblock \bibinfo{publisher}{Lawrence Erlbaum}.
\newblock
\showISBNx{0805822232}


\bibitem[\protect\citeauthoryear{Cram{\'e}r}{Cram{\'e}r}{1999}]%
        {cramer1999mathematical}
\bibfield{author}{\bibinfo{person}{Harald Cram{\'e}r}.}
  \bibinfo{year}{1999}\natexlab{}.
\newblock \bibinfo{booktitle}{\emph{Mathematical methods of statistics}}.
  Vol.~\bibinfo{volume}{43}.
\newblock \bibinfo{publisher}{Princeton University Press}.
\newblock


\bibitem[\protect\citeauthoryear{Cypher, Halbert, Kurlander, Lieberman,
  Maulsby, Myers, and Turransky}{Cypher et~al\mbox{.}}{1993}]%
        {Cypher1993WatchWI}
\bibfield{author}{\bibinfo{person}{A. Cypher}, \bibinfo{person}{Daniel~C.
  Halbert}, \bibinfo{person}{D. Kurlander}, \bibinfo{person}{H. Lieberman},
  \bibinfo{person}{D. Maulsby}, \bibinfo{person}{B. Myers}, {and}
  \bibinfo{person}{Alan Turransky}.} \bibinfo{year}{1993}\natexlab{}.
\newblock \showarticletitle{Watch what I do: programming by demonstration}.
\newblock


\bibitem[\protect\citeauthoryear{Dawood, Buragga, Khan, and Zaman}{Dawood
  et~al\mbox{.}}{2013}]%
        {Dawood2013RubricBA}
\bibfield{author}{\bibinfo{person}{M. Dawood}, \bibinfo{person}{Khalid~A.
  Buragga}, \bibinfo{person}{Abdul~Raouf Khan}, {and} \bibinfo{person}{Noor
  Zaman}.} \bibinfo{year}{2013}\natexlab{}.
\newblock \showarticletitle{Rubric based assessment plan implementation for
  Computer Science program: A practical approach}.
\newblock \bibinfo{journal}{\emph{Proceedings of 2013 IEEE International
  Conference on Teaching, Assessment and Learning for Engineering (TALE)}}
  (\bibinfo{year}{2013}), \bibinfo{pages}{551--555}.
\newblock


\bibitem[\protect\citeauthoryear{Dijkstra}{Dijkstra}{1979}]%
        {dijkstra1979foolishness}
\bibfield{author}{\bibinfo{person}{Edsger~W Dijkstra}.}
  \bibinfo{year}{1979}\natexlab{}.
\newblock \showarticletitle{On the foolishness of ``natural language
  programming''}.
\newblock In \bibinfo{booktitle}{\emph{Program Construction}}.
  \bibinfo{publisher}{Springer}, \bibinfo{pages}{51--53}.
\newblock


\bibitem[\protect\citeauthoryear{Ellis, Nye, Pu, Sosa, Tenenbaum, and
  Solar-Lezama}{Ellis et~al\mbox{.}}{2019}]%
        {Ellis2019WriteEA}
\bibfield{author}{\bibinfo{person}{K. Ellis}, \bibinfo{person}{Maxwell Nye},
  \bibinfo{person}{Y. Pu}, \bibinfo{person}{Felix Sosa}, \bibinfo{person}{J.
  Tenenbaum}, {and} \bibinfo{person}{Armando Solar-Lezama}.}
  \bibinfo{year}{2019}\natexlab{}.
\newblock \showarticletitle{Write, Execute, Assess: Program Synthesis with a
  REPL}. In \bibinfo{booktitle}{\emph{33rd Conference on Neural Information
  Processing Systems (NeurIPS)}}.
\newblock


\bibitem[\protect\citeauthoryear{Fedus, Zoph, and Shazeer}{Fedus
  et~al\mbox{.}}{2021}]%
        {fedus2021switch}
\bibfield{author}{\bibinfo{person}{William Fedus}, \bibinfo{person}{Barret
  Zoph}, {and} \bibinfo{person}{Noam Shazeer}.}
  \bibinfo{year}{2021}\natexlab{}.
\newblock \showarticletitle{Switch Transformers: Scaling to Trillion Parameter
  Models with Simple and Efficient Sparsity}.
\newblock \bibinfo{journal}{\emph{arXiv preprint arXiv:2101.03961}}
  (\bibinfo{year}{2021}).
\newblock


\bibitem[\protect\citeauthoryear{Feng, Martins, Bastani, and Dillig}{Feng
  et~al\mbox{.}}{2018}]%
        {Feng2018ProgramSU}
\bibfield{author}{\bibinfo{person}{Y. Feng}, \bibinfo{person}{R. Martins},
  \bibinfo{person}{Osbert Bastani}, {and} \bibinfo{person}{Isil Dillig}.}
  \bibinfo{year}{2018}\natexlab{}.
\newblock \showarticletitle{Program synthesis using conflict-driven learning}.
\newblock \bibinfo{journal}{\emph{Proceedings of the 39th ACM SIGPLAN
  Conference on Programming Language Design and Implementation}}
  (\bibinfo{year}{2018}).
\newblock


\bibitem[\protect\citeauthoryear{Feng, Guo, Tang, Duan, Feng, Gong, Shou, Qin,
  Liu, Jiang, and Zhou}{Feng et~al\mbox{.}}{2020}]%
        {Feng2020CodeBERTAP}
\bibfield{author}{\bibinfo{person}{Zhangyin Feng}, \bibinfo{person}{Daya Guo},
  \bibinfo{person}{Duyu Tang}, \bibinfo{person}{N. Duan}, \bibinfo{person}{X.
  Feng}, \bibinfo{person}{Ming Gong}, \bibinfo{person}{Linjun Shou},
  \bibinfo{person}{B. Qin}, \bibinfo{person}{Ting Liu}, \bibinfo{person}{Daxin
  Jiang}, {and} \bibinfo{person}{M. Zhou}.} \bibinfo{year}{2020}\natexlab{}.
\newblock \showarticletitle{CodeBERT: A Pre-Trained Model for Programming and
  Natural Languages}. In \bibinfo{booktitle}{\emph{2020 Conference on Empirical
  Methods in Natural Language Processing (EMNLP)}}.
\newblock


\bibitem[\protect\citeauthoryear{Feser, Chaudhuri, and Dillig}{Feser
  et~al\mbox{.}}{2015}]%
        {Feser2015SynthesizingDS}
\bibfield{author}{\bibinfo{person}{John~K. Feser}, \bibinfo{person}{S.
  Chaudhuri}, {and} \bibinfo{person}{Isil Dillig}.}
  \bibinfo{year}{2015}\natexlab{}.
\newblock \showarticletitle{Synthesizing data structure transformations from
  input-output examples}. In \bibinfo{booktitle}{\emph{36th annual ACM SIGPLAN
  conference on Programming Language Design and Implementation (PLDI)}}.
\newblock


\bibitem[\protect\citeauthoryear{Franks, Tu, Devanbu, and Hellendoorn}{Franks
  et~al\mbox{.}}{2015}]%
        {franks2015cacheca}
\bibfield{author}{\bibinfo{person}{Christine Franks}, \bibinfo{person}{Zhaopeng
  Tu}, \bibinfo{person}{Premkumar Devanbu}, {and} \bibinfo{person}{Vincent
  Hellendoorn}.} \bibinfo{year}{2015}\natexlab{}.
\newblock \showarticletitle{{CACHECA}: A cache language model based code
  suggestion tool}. In \bibinfo{booktitle}{\emph{International Conference on
  Software Engineering (ICSE)}}, Vol.~\bibinfo{volume}{2}. IEEE,
  \bibinfo{pages}{705--708}.
\newblock


\bibitem[\protect\citeauthoryear{Fraser, Staats, McMinn, Arcuri, and
  Padberg}{Fraser et~al\mbox{.}}{2015}]%
        {fraser2015does}
\bibfield{author}{\bibinfo{person}{Gordon Fraser}, \bibinfo{person}{Matt
  Staats}, \bibinfo{person}{Phil McMinn}, \bibinfo{person}{Andrea Arcuri},
  {and} \bibinfo{person}{Frank Padberg}.} \bibinfo{year}{2015}\natexlab{}.
\newblock \showarticletitle{Does automated unit test generation really help
  software testers? a controlled empirical study}.
\newblock \bibinfo{journal}{\emph{ACM Transactions on Software Engineering and
  Methodology (TOSEM)}} \bibinfo{volume}{24}, \bibinfo{number}{4}
  (\bibinfo{year}{2015}), \bibinfo{pages}{1--49}.
\newblock


\bibitem[\protect\citeauthoryear{Gelman and Hill}{Gelman and Hill}{2006}]%
        {gelman2006data}
\bibfield{author}{\bibinfo{person}{Andrew Gelman} {and}
  \bibinfo{person}{Jennifer Hill}.} \bibinfo{year}{2006}\natexlab{}.
\newblock \bibinfo{booktitle}{\emph{Data analysis using regression and
  multilevel/hierarchical models}}.
\newblock \bibinfo{publisher}{Cambridge University Press}.
\newblock


\bibitem[\protect\citeauthoryear{Ginsparg}{Ginsparg}{1978}]%
        {Ginsparg1978NaturalLP}
\bibfield{author}{\bibinfo{person}{J. Ginsparg}.}
  \bibinfo{year}{1978}\natexlab{}.
\newblock \showarticletitle{Natural Language Processing in an Automatic
  Programming Domain}.
\newblock


\bibitem[\protect\citeauthoryear{Grover, Basu, and Schank}{Grover
  et~al\mbox{.}}{2018}]%
        {Grover2018WhatWC}
\bibfield{author}{\bibinfo{person}{Shuchi Grover}, \bibinfo{person}{S. Basu},
  {and} \bibinfo{person}{Patricia~K. Schank}.} \bibinfo{year}{2018}\natexlab{}.
\newblock \showarticletitle{What We Can Learn About Student Learning From
  Open-Ended Programming Projects in Middle School Computer Science}.
\newblock \bibinfo{journal}{\emph{Proceedings of the 49th ACM Technical
  Symposium on Computer Science Education}} (\bibinfo{year}{2018}).
\newblock


\bibitem[\protect\citeauthoryear{Gu, Zhang, and Kim}{Gu et~al\mbox{.}}{2018}]%
        {gu2018deep}
\bibfield{author}{\bibinfo{person}{Xiaodong Gu}, \bibinfo{person}{Hongyu
  Zhang}, {and} \bibinfo{person}{Sunghun Kim}.}
  \bibinfo{year}{2018}\natexlab{}.
\newblock \showarticletitle{Deep code search}. In
  \bibinfo{booktitle}{\emph{2018 IEEE/ACM 40th International Conference on
  Software Engineering (ICSE)}}. IEEE, \bibinfo{pages}{933--944}.
\newblock


\bibitem[\protect\citeauthoryear{Gulwani}{Gulwani}{2011}]%
        {gulwani2011automating}
\bibfield{author}{\bibinfo{person}{Sumit Gulwani}.}
  \bibinfo{year}{2011}\natexlab{}.
\newblock \showarticletitle{Automating string processing in spreadsheets using
  input-output examples}.
\newblock \bibinfo{journal}{\emph{ACM SIGPLAN Notices}} \bibinfo{volume}{46},
  \bibinfo{number}{1} (\bibinfo{year}{2011}), \bibinfo{pages}{317--330}.
\newblock


\bibitem[\protect\citeauthoryear{Haiduc, Bavota, Marcus, Oliveto, Lucia, and
  Menzies}{Haiduc et~al\mbox{.}}{2013}]%
        {Haiduc2013AutomaticQR}
\bibfield{author}{\bibinfo{person}{Sonia Haiduc}, \bibinfo{person}{G. Bavota},
  \bibinfo{person}{A. Marcus}, \bibinfo{person}{R. Oliveto},
  \bibinfo{person}{A. Lucia}, {and} \bibinfo{person}{T. Menzies}.}
  \bibinfo{year}{2013}\natexlab{}.
\newblock \showarticletitle{Automatic query reformulations for text retrieval
  in software engineering}.
\newblock \bibinfo{journal}{\emph{2013 35th International Conference on
  Software Engineering (ICSE)}} (\bibinfo{year}{2013}),
  \bibinfo{pages}{842--851}.
\newblock


\bibitem[\protect\citeauthoryear{Hashimoto, Guu, Oren, and Liang}{Hashimoto
  et~al\mbox{.}}{2018}]%
        {hashimoto2018retrieve}
\bibfield{author}{\bibinfo{person}{Tatsunori~B Hashimoto},
  \bibinfo{person}{Kelvin Guu}, \bibinfo{person}{Yonatan Oren}, {and}
  \bibinfo{person}{Percy~S Liang}.} \bibinfo{year}{2018}\natexlab{}.
\newblock \showarticletitle{A retrieve-and-edit framework for predicting
  structured outputs}.
\newblock \bibinfo{journal}{\emph{Advances in Neural Information Processing
  Systems (NeurIPS)}}  \bibinfo{volume}{31} (\bibinfo{year}{2018}),
  \bibinfo{pages}{10052--10062}.
\newblock


\bibitem[\protect\citeauthoryear{Hayati, Olivier, Avvaru, Yin, Tomasic, and
  Neubig}{Hayati et~al\mbox{.}}{2018}]%
        {hayati-etal-2018-retrieval}
\bibfield{author}{\bibinfo{person}{Shirley~Anugrah Hayati},
  \bibinfo{person}{Raphael Olivier}, \bibinfo{person}{Pravalika Avvaru},
  \bibinfo{person}{Pengcheng Yin}, \bibinfo{person}{Anthony Tomasic}, {and}
  \bibinfo{person}{Graham Neubig}.} \bibinfo{year}{2018}\natexlab{}.
\newblock \showarticletitle{Retrieval-Based Neural Code Generation}. In
  \bibinfo{booktitle}{\emph{Proceedings of the 2018 Conference on Empirical
  Methods in Natural Language Processing (EMNLP)}}.
  \bibinfo{publisher}{Association for Computational Linguistics},
  \bibinfo{address}{Brussels, Belgium}, \bibinfo{pages}{925--930}.
\newblock
\urldef\tempurl%
\url{https://doi.org/10.18653/v1/D18-1111}
\showDOI{\tempurl}


\bibitem[\protect\citeauthoryear{Head, Glassman, Hartmann, and Hearst}{Head
  et~al\mbox{.}}{2018}]%
        {Head2018InteractiveEO}
\bibfield{author}{\bibinfo{person}{Andrew Head}, \bibinfo{person}{Elena~Leah
  Glassman}, \bibinfo{person}{B. Hartmann}, {and} \bibinfo{person}{Marti~A.
  Hearst}.} \bibinfo{year}{2018}\natexlab{}.
\newblock \showarticletitle{Interactive Extraction of Examples from Existing
  Code}.
\newblock \bibinfo{journal}{\emph{Proceedings of the 2018 CHI Conference on
  Human Factors in Computing Systems}} (\bibinfo{year}{2018}).
\newblock


\bibitem[\protect\citeauthoryear{Head, Glassman, Soares, Suzuki, Figueredo,
  D'Antoni, and Hartmann}{Head et~al\mbox{.}}{2017}]%
        {Head2017WritingRC}
\bibfield{author}{\bibinfo{person}{Andrew Head}, \bibinfo{person}{Elena~Leah
  Glassman}, \bibinfo{person}{Gustavo Soares}, \bibinfo{person}{R. Suzuki},
  \bibinfo{person}{Lucas Figueredo}, \bibinfo{person}{L. D'Antoni}, {and}
  \bibinfo{person}{B. Hartmann}.} \bibinfo{year}{2017}\natexlab{}.
\newblock \showarticletitle{Writing Reusable Code Feedback at Scale with
  Mixed-Initiative Program Synthesis}.
\newblock \bibinfo{journal}{\emph{Proceedings of the Fourth (2017) ACM
  Conference on Learning @ Scale}} (\bibinfo{year}{2017}).
\newblock


\bibitem[\protect\citeauthoryear{Heidorn}{Heidorn}{1976}]%
        {heidorn1976automatic}
\bibfield{author}{\bibinfo{person}{George~E. Heidorn}.}
  \bibinfo{year}{1976}\natexlab{}.
\newblock \showarticletitle{Automatic programming through natural language
  dialogue: A survey}.
\newblock \bibinfo{journal}{\emph{IBM Journal of research and development}}
  \bibinfo{volume}{20}, \bibinfo{number}{4} (\bibinfo{year}{1976}),
  \bibinfo{pages}{302--313}.
\newblock


\bibitem[\protect\citeauthoryear{Hill, Roldan-Vega, Fails, and Mallet}{Hill
  et~al\mbox{.}}{2014}]%
        {Hill2014NLbasedQR}
\bibfield{author}{\bibinfo{person}{E. Hill}, \bibinfo{person}{Manuel
  Roldan-Vega}, \bibinfo{person}{J. Fails}, {and} \bibinfo{person}{Greg
  Mallet}.} \bibinfo{year}{2014}\natexlab{}.
\newblock \showarticletitle{NL-based query refinement and contextualized code
  search results: A user study}.
\newblock \bibinfo{journal}{\emph{2014 Software Evolution Week - IEEE
  Conference on Software Maintenance, Reengineering, and Reverse Engineering
  (CSMR-WCRE)}} (\bibinfo{year}{2014}), \bibinfo{pages}{34--43}.
\newblock


\bibitem[\protect\citeauthoryear{Hodges~Jr and Lehmann}{Hodges~Jr and
  Lehmann}{1963}]%
        {hodges1963estimates}
\bibfield{author}{\bibinfo{person}{Joseph~L Hodges~Jr} {and}
  \bibinfo{person}{Erich~L Lehmann}.} \bibinfo{year}{1963}\natexlab{}.
\newblock \showarticletitle{Estimates of location based on rank tests}.
\newblock \bibinfo{journal}{\emph{The Annals of Mathematical Statistics}}
  (\bibinfo{year}{1963}), \bibinfo{pages}{598--611}.
\newblock


\bibitem[\protect\citeauthoryear{Husain, Wu, Gazit, Allamanis, and
  Brockschmidt}{Husain et~al\mbox{.}}{2019}]%
        {husain2019codesearchnet}
\bibfield{author}{\bibinfo{person}{Hamel Husain}, \bibinfo{person}{Ho-Hsiang
  Wu}, \bibinfo{person}{Tiferet Gazit}, \bibinfo{person}{Miltiadis Allamanis},
  {and} \bibinfo{person}{Marc Brockschmidt}.} \bibinfo{year}{2019}\natexlab{}.
\newblock \showarticletitle{Codesearchnet challenge: Evaluating the state of
  semantic code search}.
\newblock \bibinfo{journal}{\emph{arXiv preprint arXiv:1909.09436}}
  (\bibinfo{year}{2019}).
\newblock


\bibitem[\protect\citeauthoryear{Iyer, Konstas, Cheung, and Zettlemoyer}{Iyer
  et~al\mbox{.}}{2016}]%
        {Iyer2016SummarizingSC}
\bibfield{author}{\bibinfo{person}{Srini Iyer}, \bibinfo{person}{Ioannis
  Konstas}, \bibinfo{person}{A. Cheung}, {and} \bibinfo{person}{Luke
  Zettlemoyer}.} \bibinfo{year}{2016}\natexlab{}.
\newblock \showarticletitle{Summarizing Source Code using a Neural Attention
  Model}. In \bibinfo{booktitle}{\emph{Proceedings of the 54th Annual Meeting
  of the Association for Computational Linguistics (ACL)}}.
\newblock


\bibitem[\protect\citeauthoryear{Iyer, Konstas, Cheung, and Zettlemoyer}{Iyer
  et~al\mbox{.}}{2018}]%
        {Iyer2018MappingLT}
\bibfield{author}{\bibinfo{person}{Srini Iyer}, \bibinfo{person}{Ioannis
  Konstas}, \bibinfo{person}{A. Cheung}, {and} \bibinfo{person}{Luke
  Zettlemoyer}.} \bibinfo{year}{2018}\natexlab{}.
\newblock \showarticletitle{Mapping Language to Code in Programmatic Context}.
  In \bibinfo{booktitle}{\emph{2018 Conference on Empirical Methods in Natural
  Language Processing (EMNLP)}}.
\newblock


\bibitem[\protect\citeauthoryear{Johnson}{Johnson}{2014}]%
        {johnson2014extension}
\bibfield{author}{\bibinfo{person}{Paul~CD Johnson}.}
  \bibinfo{year}{2014}\natexlab{}.
\newblock \showarticletitle{Extension of Nakagawa \& Schielzeth's $R^2_{GLMM}$
  to random slopes models}.
\newblock \bibinfo{journal}{\emph{Methods in Ecology and Evolution}}
  \bibinfo{volume}{5}, \bibinfo{number}{9} (\bibinfo{year}{2014}),
  \bibinfo{pages}{944--946}.
\newblock


\bibitem[\protect\citeauthoryear{Karamcheti, Sadigh, and Liang}{Karamcheti
  et~al\mbox{.}}{2020}]%
        {karamcheti-etal-2020-learning}
\bibfield{author}{\bibinfo{person}{Siddharth Karamcheti},
  \bibinfo{person}{Dorsa Sadigh}, {and} \bibinfo{person}{Percy Liang}.}
  \bibinfo{year}{2020}\natexlab{}.
\newblock \showarticletitle{Learning Adaptive Language Interfaces through
  Decomposition}. In \bibinfo{booktitle}{\emph{Proceedings of the First
  Workshop on Interactive and Executable Semantic Parsing}}.
  \bibinfo{publisher}{Association for Computational Linguistics},
  \bibinfo{address}{Online}, \bibinfo{pages}{23--33}.
\newblock
\urldef\tempurl%
\url{https://doi.org/10.18653/v1/2020.intexsempar-1.4}
\showDOI{\tempurl}


\bibitem[\protect\citeauthoryear{Keivanloo, Rilling, and Zou}{Keivanloo
  et~al\mbox{.}}{2014}]%
        {Keivanloo2014SpottingWC}
\bibfield{author}{\bibinfo{person}{I. Keivanloo}, \bibinfo{person}{J. Rilling},
  {and} \bibinfo{person}{Ying Zou}.} \bibinfo{year}{2014}\natexlab{}.
\newblock \showarticletitle{Spotting working code examples}. In
  \bibinfo{booktitle}{\emph{36th International Conference on Software
  Engineering (ICSE)}}.
\newblock


\bibitem[\protect\citeauthoryear{Kery, Horvath, and Myers}{Kery
  et~al\mbox{.}}{2017}]%
        {Kery2017VarioliteSE}
\bibfield{author}{\bibinfo{person}{Mary~Beth Kery}, \bibinfo{person}{Amber
  Horvath}, {and} \bibinfo{person}{B. Myers}.} \bibinfo{year}{2017}\natexlab{}.
\newblock \showarticletitle{Variolite: Supporting Exploratory Programming by
  Data Scientists}.
\newblock \bibinfo{journal}{\emph{Proceedings of the 2017 CHI Conference on
  Human Factors in Computing Systems (CHI)}} (\bibinfo{year}{2017}).
\newblock


\bibitem[\protect\citeauthoryear{Kery and Myers}{Kery and Myers}{2017}]%
        {Kery2017ExploringEP}
\bibfield{author}{\bibinfo{person}{Mary~Beth Kery} {and} \bibinfo{person}{B.
  Myers}.} \bibinfo{year}{2017}\natexlab{}.
\newblock \showarticletitle{Exploring exploratory programming}.
\newblock \bibinfo{journal}{\emph{2017 IEEE Symposium on Visual Languages and
  Human-Centric Computing (VL/HCC)}} (\bibinfo{year}{2017}),
  \bibinfo{pages}{25--29}.
\newblock


\bibitem[\protect\citeauthoryear{Ko and Myers}{Ko and Myers}{2004}]%
        {Ko2004DesigningTW}
\bibfield{author}{\bibinfo{person}{A. Ko} {and} \bibinfo{person}{B. Myers}.}
  \bibinfo{year}{2004}\natexlab{}.
\newblock \showarticletitle{Designing the whyline: a debugging interface for
  asking questions about program behavior}. In \bibinfo{booktitle}{\emph{CHI
  2004 Conference on Human Factors in Computing Systems (CHI)}}.
\newblock


\bibitem[\protect\citeauthoryear{Ko and Myers}{Ko and Myers}{2008}]%
        {Ko2008DebuggingR}
\bibfield{author}{\bibinfo{person}{A. Ko} {and} \bibinfo{person}{B. Myers}.}
  \bibinfo{year}{2008}\natexlab{}.
\newblock \showarticletitle{Debugging reinvented}.
\newblock \bibinfo{journal}{\emph{2008 ACM/IEEE 30th International Conference
  on Software Engineering (ICSE)}} (\bibinfo{year}{2008}),
  \bibinfo{pages}{301--310}.
\newblock


\bibitem[\protect\citeauthoryear{Ko, Myers, and Aung}{Ko et~al\mbox{.}}{2004}]%
        {ko2004six}
\bibfield{author}{\bibinfo{person}{Amy Ko}, \bibinfo{person}{Brad~A Myers},
  {and} \bibinfo{person}{Htet~Htet Aung}.} \bibinfo{year}{2004}\natexlab{}.
\newblock \showarticletitle{Six learning barriers in end-user programming
  systems}. In \bibinfo{booktitle}{\emph{IEEE Symposium on Visual Languages and
  Human-Centric Computing (VL/HCC)}}. IEEE, \bibinfo{pages}{199--206}.
\newblock


\bibitem[\protect\citeauthoryear{Kock and Lynn}{Kock and Lynn}{2012}]%
        {kock2012lateral}
\bibfield{author}{\bibinfo{person}{Ned Kock} {and} \bibinfo{person}{Gary
  Lynn}.} \bibinfo{year}{2012}\natexlab{}.
\newblock \showarticletitle{Lateral collinearity and misleading results in
  variance-based SEM: An illustration and recommendations}.
\newblock \bibinfo{journal}{\emph{Journal of the Association for information
  Systems}} \bibinfo{volume}{13}, \bibinfo{number}{7} (\bibinfo{year}{2012}).
\newblock


\bibitem[\protect\citeauthoryear{Kulal, Pasupat, Chandra, Lee, Padon, Aiken,
  and Liang}{Kulal et~al\mbox{.}}{2019}]%
        {Kulal2019SPoCSP}
\bibfield{author}{\bibinfo{person}{S. Kulal}, \bibinfo{person}{Panupong
  Pasupat}, \bibinfo{person}{K. Chandra}, \bibinfo{person}{Mina Lee},
  \bibinfo{person}{Oded Padon}, \bibinfo{person}{A. Aiken}, {and}
  \bibinfo{person}{Percy Liang}.} \bibinfo{year}{2019}\natexlab{}.
\newblock \showarticletitle{SPoC: Search-based Pseudocode to Code}. In
  \bibinfo{booktitle}{\emph{33rd Conference on Neural Information Processing
  Systems (NeurIPS)}}.
\newblock


\bibitem[\protect\citeauthoryear{Kushman and Barzilay}{Kushman and
  Barzilay}{2013}]%
        {Kushman2013UsingSU}
\bibfield{author}{\bibinfo{person}{Nate Kushman} {and} \bibinfo{person}{R.
  Barzilay}.} \bibinfo{year}{2013}\natexlab{}.
\newblock \showarticletitle{Using Semantic Unification to Generate Regular
  Expressions from Natural Language}. In \bibinfo{booktitle}{\emph{The 2013
  Conference of the North American Chapter of the Association for Computational
  Linguistics: Human Language Technologies (HLT-NAACL)}}.
\newblock


\bibitem[\protect\citeauthoryear{Landman, Serebrenik, Bouwers, and
  Vinju}{Landman et~al\mbox{.}}{2016}]%
        {landman2016empirical}
\bibfield{author}{\bibinfo{person}{Davy Landman}, \bibinfo{person}{Alexander
  Serebrenik}, \bibinfo{person}{Eric Bouwers}, {and} \bibinfo{person}{Jurgen~J
  Vinju}.} \bibinfo{year}{2016}\natexlab{}.
\newblock \showarticletitle{Empirical analysis of the relationship between CC
  and SLOC in a large corpus of Java methods and C functions}.
\newblock \bibinfo{journal}{\emph{Journal of Software: Evolution and Process}}
  \bibinfo{volume}{28}, \bibinfo{number}{7} (\bibinfo{year}{2016}),
  \bibinfo{pages}{589--618}.
\newblock


\bibitem[\protect\citeauthoryear{Le and Gulwani}{Le and Gulwani}{2014}]%
        {le2014flashextract}
\bibfield{author}{\bibinfo{person}{Vu Le} {and} \bibinfo{person}{Sumit
  Gulwani}.} \bibinfo{year}{2014}\natexlab{}.
\newblock \showarticletitle{FlashExtract: a framework for data extraction by
  examples}.
\newblock \bibinfo{journal}{\emph{ACM SIGPLAN Notices}} \bibinfo{volume}{49},
  \bibinfo{number}{6} (\bibinfo{year}{2014}), \bibinfo{pages}{542--553}.
\newblock


\bibitem[\protect\citeauthoryear{Lei, Long, Barzilay, and Rinard}{Lei
  et~al\mbox{.}}{2013}]%
        {Lei2013FromNL}
\bibfield{author}{\bibinfo{person}{Tao Lei}, \bibinfo{person}{F. Long},
  \bibinfo{person}{R. Barzilay}, {and} \bibinfo{person}{M. Rinard}.}
  \bibinfo{year}{2013}\natexlab{}.
\newblock \showarticletitle{From Natural Language Specifications to Program
  Input Parsers}. In \bibinfo{booktitle}{\emph{The 51st Annual Meeting of the
  Association for Computational Linguistics (ACL)}}.
\newblock


\bibitem[\protect\citeauthoryear{Li, Azaria, and Myers}{Li
  et~al\mbox{.}}{2017}]%
        {Li2017SUGILITECM}
\bibfield{author}{\bibinfo{person}{Toby Jia-Jun Li}, \bibinfo{person}{Amos
  Azaria}, {and} \bibinfo{person}{B. Myers}.} \bibinfo{year}{2017}\natexlab{}.
\newblock \showarticletitle{SUGILITE: Creating Multimodal Smartphone Automation
  by Demonstration}.
\newblock \bibinfo{journal}{\emph{Proceedings of the 2017 CHI Conference on
  Human Factors in Computing Systems (CHI)}} (\bibinfo{year}{2017}).
\newblock


\bibitem[\protect\citeauthoryear{Li, Labutov, Li, Zhang, Shi, Ding, Mitchell,
  and Myers}{Li et~al\mbox{.}}{2018}]%
        {Li2018APPINITEAM}
\bibfield{author}{\bibinfo{person}{Toby Jia-Jun Li}, \bibinfo{person}{I.
  Labutov}, \bibinfo{person}{X. Li}, \bibinfo{person}{X. Zhang},
  \bibinfo{person}{W. Shi}, \bibinfo{person}{Wanling Ding},
  \bibinfo{person}{Tom~Michael Mitchell}, {and} \bibinfo{person}{B. Myers}.}
  \bibinfo{year}{2018}\natexlab{}.
\newblock \showarticletitle{APPINITE: A Multi-Modal Interface for Specifying
  Data Descriptions in Programming by Demonstration Using Natural Language
  Instructions}.
\newblock \bibinfo{journal}{\emph{2018 IEEE Symposium on Visual Languages and
  Human-Centric Computing (VL/HCC)}} (\bibinfo{year}{2018}),
  \bibinfo{pages}{105--114}.
\newblock


\bibitem[\protect\citeauthoryear{Li, Radensky, Jia, Singarajah, Mitchell, and
  Myers}{Li et~al\mbox{.}}{2019}]%
        {Li2019PUMICEAM}
\bibfield{author}{\bibinfo{person}{Toby Jia-Jun Li}, \bibinfo{person}{Marissa
  Radensky}, \bibinfo{person}{J. Jia}, \bibinfo{person}{Kirielle Singarajah},
  \bibinfo{person}{Tom~Michael Mitchell}, {and} \bibinfo{person}{B. Myers}.}
  \bibinfo{year}{2019}\natexlab{}.
\newblock \showarticletitle{PUMICE: A Multi-Modal Agent that Learns Concepts
  and Conditionals from Natural Language and Demonstrations}.
\newblock \bibinfo{journal}{\emph{Proceedings of the 32nd Annual ACM Symposium
  on User Interface Software and Technology (UIST)}} (\bibinfo{year}{2019}).
\newblock


\bibitem[\protect\citeauthoryear{Lieberman, Patern{\`o}, Klann, and
  Wulf}{Lieberman et~al\mbox{.}}{2006}]%
        {Lieberman2006EndUserDA}
\bibfield{author}{\bibinfo{person}{H. Lieberman}, \bibinfo{person}{F.
  Patern{\`o}}, \bibinfo{person}{Markus Klann}, {and} \bibinfo{person}{V.
  Wulf}.} \bibinfo{year}{2006}\natexlab{}.
\newblock \showarticletitle{End-User Development: An Emerging Paradigm}. In
  \bibinfo{booktitle}{\emph{End User Development}}.
\newblock


\bibitem[\protect\citeauthoryear{Ling, Blunsom, Grefenstette, Hermann,
  Kocisk{\'{y}}, Wang, and Senior}{Ling et~al\mbox{.}}{2016}]%
        {DBLP:conf/acl/LingBGHKWS16}
\bibfield{author}{\bibinfo{person}{Wang Ling}, \bibinfo{person}{Phil Blunsom},
  \bibinfo{person}{Edward Grefenstette}, \bibinfo{person}{Karl~Moritz Hermann},
  \bibinfo{person}{Tom{\'{a}}s Kocisk{\'{y}}}, \bibinfo{person}{Fumin Wang},
  {and} \bibinfo{person}{Andrew~W. Senior}.} \bibinfo{year}{2016}\natexlab{}.
\newblock \showarticletitle{Latent Predictor Networks for Code Generation}. In
  \bibinfo{booktitle}{\emph{Proceedings of the 54th Annual Meeting of the
  Association for Computational Linguistics (ACL)}}. \bibinfo{publisher}{The
  Association for Computer Linguistics}.
\newblock
\urldef\tempurl%
\url{https://doi.org/10.18653/v1/p16-1057}
\showDOI{\tempurl}


\bibitem[\protect\citeauthoryear{Liu, Xia, Lo, Gao, Yang, and Grundy}{Liu
  et~al\mbox{.}}{2020}]%
        {Liu2020OpportunitiesAC}
\bibfield{author}{\bibinfo{person}{C. Liu}, \bibinfo{person}{Xin Xia},
  \bibinfo{person}{David Lo}, \bibinfo{person}{Cuiyun Gao},
  \bibinfo{person}{Xiaohu Yang}, {and} \bibinfo{person}{J. Grundy}.}
  \bibinfo{year}{2020}\natexlab{}.
\newblock \showarticletitle{Opportunities and Challenges in Code Search Tools}.
\newblock \bibinfo{journal}{\emph{ArXiv}}  \bibinfo{volume}{abs/2011.02297}
  (\bibinfo{year}{2020}).
\newblock


\bibitem[\protect\citeauthoryear{Liu, Shen, Zhong, and Zhu}{Liu
  et~al\mbox{.}}{2016}]%
        {Liu2016EXPSOLRO}
\bibfield{author}{\bibinfo{person}{X. Liu}, \bibinfo{person}{Beijun Shen},
  \bibinfo{person}{H. Zhong}, {and} \bibinfo{person}{Jiangang Zhu}.}
  \bibinfo{year}{2016}\natexlab{}.
\newblock \showarticletitle{EXPSOL: Recommending Online Threads for
  Exception-Related Bug Reports}.
\newblock \bibinfo{journal}{\emph{2016 23rd Asia-Pacific Software Engineering
  Conference (APSEC)}} (\bibinfo{year}{2016}), \bibinfo{pages}{25--32}.
\newblock


\bibitem[\protect\citeauthoryear{Lu, Sun, Wang, Lo, and Duan}{Lu
  et~al\mbox{.}}{2015}]%
        {Lu2015QueryEV}
\bibfield{author}{\bibinfo{person}{Meili Lu}, \bibinfo{person}{Xiaobing Sun},
  \bibinfo{person}{S. Wang}, \bibinfo{person}{D. Lo}, {and}
  \bibinfo{person}{Yucong Duan}.} \bibinfo{year}{2015}\natexlab{}.
\newblock \showarticletitle{Query expansion via WordNet for effective code
  search}. In \bibinfo{booktitle}{\emph{International Conference on Software
  Analysis, Evolution, and Reengineering (SANER)}}. \bibinfo{publisher}{IEEE},
  \bibinfo{pages}{545--549}.
\newblock


\bibitem[\protect\citeauthoryear{Maloney, Resnick, Rusk, Silverman, and
  Eastmond}{Maloney et~al\mbox{.}}{2010}]%
        {Maloney2010TheSP}
\bibfield{author}{\bibinfo{person}{J. Maloney}, \bibinfo{person}{M. Resnick},
  \bibinfo{person}{N. Rusk}, \bibinfo{person}{B. Silverman}, {and}
  \bibinfo{person}{Evelyn Eastmond}.} \bibinfo{year}{2010}\natexlab{}.
\newblock \showarticletitle{The Scratch Programming Language and Environment}.
\newblock \bibinfo{journal}{\emph{ACM Trans. Comput. Educ.}}
  \bibinfo{volume}{10} (\bibinfo{year}{2010}), \bibinfo{pages}{16:1--16:15}.
\newblock


\bibitem[\protect\citeauthoryear{Manning, Sch{\"u}tze, and Raghavan}{Manning
  et~al\mbox{.}}{2008}]%
        {manning2008introduction}
\bibfield{author}{\bibinfo{person}{Christopher~D Manning},
  \bibinfo{person}{Hinrich Sch{\"u}tze}, {and} \bibinfo{person}{Prabhakar
  Raghavan}.} \bibinfo{year}{2008}\natexlab{}.
\newblock \bibinfo{booktitle}{\emph{Introduction to information retrieval}}.
\newblock \bibinfo{publisher}{Cambridge university press}.
\newblock


\bibitem[\protect\citeauthoryear{Manshadi, Gildea, and Allen}{Manshadi
  et~al\mbox{.}}{2013}]%
        {Manshadi2013IntegratingPB}
\bibfield{author}{\bibinfo{person}{Mehdi Manshadi}, \bibinfo{person}{Daniel
  Gildea}, {and} \bibinfo{person}{James~F. Allen}.}
  \bibinfo{year}{2013}\natexlab{}.
\newblock \showarticletitle{Integrating Programming by Example and Natural
  Language Programming}. In \bibinfo{booktitle}{\emph{Proceedings of the AAAI
  Conference on Artificial Intelligence (AAAI)}}.
\newblock


\bibitem[\protect\citeauthoryear{McCabe}{McCabe}{1976}]%
        {McCabe1976ACM}
\bibfield{author}{\bibinfo{person}{T. McCabe}.}
  \bibinfo{year}{1976}\natexlab{}.
\newblock \showarticletitle{A Complexity Measure}.
\newblock \bibinfo{journal}{\emph{IEEE Transactions on Software Engineering}}
  \bibinfo{volume}{SE-2} (\bibinfo{year}{1976}), \bibinfo{pages}{308--320}.
\newblock


\bibitem[\protect\citeauthoryear{Mihalcea, Liu, and Lieberman}{Mihalcea
  et~al\mbox{.}}{2006}]%
        {mihalcea2006nlp}
\bibfield{author}{\bibinfo{person}{Rada Mihalcea}, \bibinfo{person}{Hugo Liu},
  {and} \bibinfo{person}{Henry Lieberman}.} \bibinfo{year}{2006}\natexlab{}.
\newblock \showarticletitle{NLP (natural language processing) for NLP (natural
  language programming)}. In \bibinfo{booktitle}{\emph{International Conference
  on Intelligent Text Processing and Computational Linguistics}}. Springer,
  \bibinfo{pages}{319--330}.
\newblock


\bibitem[\protect\citeauthoryear{Mohagheghi and Conradi}{Mohagheghi and
  Conradi}{2007}]%
        {mohagheghi2007quality}
\bibfield{author}{\bibinfo{person}{Parastoo Mohagheghi} {and}
  \bibinfo{person}{Reidar Conradi}.} \bibinfo{year}{2007}\natexlab{}.
\newblock \showarticletitle{Quality, productivity and economic benefits of
  software reuse: a review of industrial studies}.
\newblock \bibinfo{journal}{\emph{Empirical Software Engineering}}
  \bibinfo{volume}{12}, \bibinfo{number}{5} (\bibinfo{year}{2007}),
  \bibinfo{pages}{471--516}.
\newblock


\bibitem[\protect\citeauthoryear{Moreno, Bavota, Di~Penta, Oliveto, and
  Marcus}{Moreno et~al\mbox{.}}{2015}]%
        {moreno2015can}
\bibfield{author}{\bibinfo{person}{Laura Moreno}, \bibinfo{person}{Gabriele
  Bavota}, \bibinfo{person}{Massimiliano Di~Penta}, \bibinfo{person}{Rocco
  Oliveto}, {and} \bibinfo{person}{Andrian Marcus}.}
  \bibinfo{year}{2015}\natexlab{}.
\newblock \showarticletitle{How can I use this method?}. In
  \bibinfo{booktitle}{\emph{2015 IEEE/ACM 37th IEEE International Conference on
  Software Engineering}}, Vol.~\bibinfo{volume}{1}. IEEE,
  \bibinfo{pages}{880--890}.
\newblock


\bibitem[\protect\citeauthoryear{Mundlak}{Mundlak}{1978}]%
        {mundlak1978pooling}
\bibfield{author}{\bibinfo{person}{Yair Mundlak}.}
  \bibinfo{year}{1978}\natexlab{}.
\newblock \showarticletitle{On the pooling of time series and cross section
  data}.
\newblock \bibinfo{journal}{\emph{Econometrica: journal of the Econometric
  Society}} (\bibinfo{year}{1978}), \bibinfo{pages}{69--85}.
\newblock


\bibitem[\protect\citeauthoryear{Murphy, Kery, Alliyu, Macvean, and
  Myers}{Murphy et~al\mbox{.}}{2018}]%
        {murphy2018api}
\bibfield{author}{\bibinfo{person}{Lauren Murphy}, \bibinfo{person}{Mary~Beth
  Kery}, \bibinfo{person}{Oluwatosin Alliyu}, \bibinfo{person}{Andrew Macvean},
  {and} \bibinfo{person}{Brad~A Myers}.} \bibinfo{year}{2018}\natexlab{}.
\newblock \showarticletitle{API designers in the field: Design practices and
  challenges for creating usable APIs}. In \bibinfo{booktitle}{\emph{IEEE
  Symposium on Visual Languages and Human-Centric Computing (VL/HCC)}}. IEEE,
  \bibinfo{pages}{249--258}.
\newblock


\bibitem[\protect\citeauthoryear{Myers, Pane, and Ko}{Myers
  et~al\mbox{.}}{2004}]%
        {Myers2004NaturalPL}
\bibfield{author}{\bibinfo{person}{B. Myers}, \bibinfo{person}{J. Pane}, {and}
  \bibinfo{person}{A. Ko}.} \bibinfo{year}{2004}\natexlab{}.
\newblock \showarticletitle{Natural programming languages and environments}.
\newblock \bibinfo{journal}{\emph{Commun. ACM}}  \bibinfo{volume}{47}
  (\bibinfo{year}{2004}), \bibinfo{pages}{47--52}.
\newblock


\bibitem[\protect\citeauthoryear{Myers and Stylos}{Myers and Stylos}{2016a}]%
        {Myers2016ImprovingAU}
\bibfield{author}{\bibinfo{person}{B. Myers} {and} \bibinfo{person}{Jeffrey
  Stylos}.} \bibinfo{year}{2016}\natexlab{a}.
\newblock \showarticletitle{Improving API usability}.
\newblock \bibinfo{journal}{\emph{Commun. ACM}}  \bibinfo{volume}{59}
  (\bibinfo{year}{2016}), \bibinfo{pages}{62 -- 69}.
\newblock


\bibitem[\protect\citeauthoryear{Myers, Ko, LaToza, and Yoon}{Myers
  et~al\mbox{.}}{2016}]%
        {myers2016programmers}
\bibfield{author}{\bibinfo{person}{Brad~A Myers}, \bibinfo{person}{Amy Ko},
  \bibinfo{person}{Thomas~D LaToza}, {and} \bibinfo{person}{YoungSeok Yoon}.}
  \bibinfo{year}{2016}\natexlab{}.
\newblock \showarticletitle{Programmers are users too: Human-centered methods
  for improving programming tools}.
\newblock \bibinfo{journal}{\emph{Computer}} \bibinfo{volume}{49},
  \bibinfo{number}{7} (\bibinfo{year}{2016}), \bibinfo{pages}{44--52}.
\newblock


\bibitem[\protect\citeauthoryear{Myers and Stylos}{Myers and Stylos}{2016b}]%
        {myers2016improving}
\bibfield{author}{\bibinfo{person}{Brad~A Myers} {and} \bibinfo{person}{Jeffrey
  Stylos}.} \bibinfo{year}{2016}\natexlab{b}.
\newblock \showarticletitle{Improving API usability}.
\newblock \bibinfo{journal}{\emph{Commun. ACM}} \bibinfo{volume}{59},
  \bibinfo{number}{6} (\bibinfo{year}{2016}), \bibinfo{pages}{62--69}.
\newblock


\bibitem[\protect\citeauthoryear{Nakagawa and Schielzeth}{Nakagawa and
  Schielzeth}{2013}]%
        {nakagawa2013general}
\bibfield{author}{\bibinfo{person}{Shinichi Nakagawa} {and}
  \bibinfo{person}{Holger Schielzeth}.} \bibinfo{year}{2013}\natexlab{}.
\newblock \showarticletitle{A general and simple method for obtaining R2 from
  generalized linear mixed-effects models}.
\newblock \bibinfo{journal}{\emph{Methods in Ecology and Evolution}}
  \bibinfo{volume}{4}, \bibinfo{number}{2} (\bibinfo{year}{2013}),
  \bibinfo{pages}{133--142}.
\newblock


\bibitem[\protect\citeauthoryear{Nam, Horvath, Macvean, Myers, and
  Vasilescu}{Nam et~al\mbox{.}}{2019}]%
        {nam2019marble}
\bibfield{author}{\bibinfo{person}{Daye Nam}, \bibinfo{person}{Amber Horvath},
  \bibinfo{person}{Andrew Macvean}, \bibinfo{person}{Brad Myers}, {and}
  \bibinfo{person}{Bogdan Vasilescu}.} \bibinfo{year}{2019}\natexlab{}.
\newblock \showarticletitle{Marble: Mining for boilerplate code to identify API
  usability problems}. In \bibinfo{booktitle}{\emph{International Conference on
  Automated Software Engineering (ASE)}}. IEEE, \bibinfo{pages}{615--627}.
\newblock


\bibitem[\protect\citeauthoryear{Nguyen and Csallner}{Nguyen and
  Csallner}{2015}]%
        {Nguyen2015ReverseEM}
\bibfield{author}{\bibinfo{person}{T. Nguyen} {and} \bibinfo{person}{C.
  Csallner}.} \bibinfo{year}{2015}\natexlab{}.
\newblock \showarticletitle{Reverse Engineering Mobile Application User
  Interfaces with REMAUI (T)}.
\newblock \bibinfo{journal}{\emph{2015 30th IEEE/ACM International Conference
  on Automated Software Engineering (ASE)}} (\bibinfo{year}{2015}),
  \bibinfo{pages}{248--259}.
\newblock


\bibitem[\protect\citeauthoryear{Nowell, Norris, White, and Moules}{Nowell
  et~al\mbox{.}}{2017}]%
        {nowell2017thematic}
\bibfield{author}{\bibinfo{person}{Lorelli~S Nowell}, \bibinfo{person}{Jill~M
  Norris}, \bibinfo{person}{Deborah~E White}, {and} \bibinfo{person}{Nancy~J
  Moules}.} \bibinfo{year}{2017}\natexlab{}.
\newblock \showarticletitle{Thematic analysis: Striving to meet the
  trustworthiness criteria}.
\newblock \bibinfo{journal}{\emph{International Journal of Qualitative
  Methods}} \bibinfo{volume}{16}, \bibinfo{number}{1} (\bibinfo{year}{2017}),
  \bibinfo{pages}{1609406917733847}.
\newblock


\bibitem[\protect\citeauthoryear{Papineni, Roukos, Ward, and Zhu}{Papineni
  et~al\mbox{.}}{2002}]%
        {papineni-etal-2002-bleu}
\bibfield{author}{\bibinfo{person}{Kishore Papineni}, \bibinfo{person}{Salim
  Roukos}, \bibinfo{person}{Todd Ward}, {and} \bibinfo{person}{Wei-Jing Zhu}.}
  \bibinfo{year}{2002}\natexlab{}.
\newblock \showarticletitle{{B}leu: a Method for Automatic Evaluation of
  Machine Translation}. In \bibinfo{booktitle}{\emph{Proceedings of the 40th
  Annual Meeting of the Association for Computational Linguistics (ACL)}}.
  \bibinfo{publisher}{Association for Computational Linguistics},
  \bibinfo{address}{Philadelphia, Pennsylvania, USA},
  \bibinfo{pages}{311--318}.
\newblock
\urldef\tempurl%
\url{https://doi.org/10.3115/1073083.1073135}
\showDOI{\tempurl}


\bibitem[\protect\citeauthoryear{Parisotto, rahman Mohamed, Singh, Li, Zhou,
  and Kohli}{Parisotto et~al\mbox{.}}{2017}]%
        {Parisotto2017NeuroSymbolicPS}
\bibfield{author}{\bibinfo{person}{Emilio Parisotto}, \bibinfo{person}{Abdel
  rahman Mohamed}, \bibinfo{person}{R. Singh}, \bibinfo{person}{L. Li},
  \bibinfo{person}{Dengyong Zhou}, {and} \bibinfo{person}{Pushmeet Kohli}.}
  \bibinfo{year}{2017}\natexlab{}.
\newblock \showarticletitle{Neuro-Symbolic Program Synthesis}.
\newblock \bibinfo{journal}{\emph{5th International Conference on Learning
  Representations (ICLR)}} (\bibinfo{year}{2017}).
\newblock


\bibitem[\protect\citeauthoryear{Ponzanelli, Bacchelli, and Lanza}{Ponzanelli
  et~al\mbox{.}}{2013}]%
        {ponzanelli2013seahawk}
\bibfield{author}{\bibinfo{person}{Luca Ponzanelli}, \bibinfo{person}{Alberto
  Bacchelli}, {and} \bibinfo{person}{Michele Lanza}.}
  \bibinfo{year}{2013}\natexlab{}.
\newblock \showarticletitle{Seahawk: Stack Overflow in the IDE}. In
  \bibinfo{booktitle}{\emph{International Conference on Software Engineering
  (ICSE)}}. IEEE, \bibinfo{pages}{1295--1298}.
\newblock


\bibitem[\protect\citeauthoryear{Ponzanelli, Bavota, Penta, Oliveto, and
  Lanza}{Ponzanelli et~al\mbox{.}}{2014}]%
        {Ponzanelli2014MiningST}
\bibfield{author}{\bibinfo{person}{Luca Ponzanelli}, \bibinfo{person}{G.
  Bavota}, \bibinfo{person}{M.~D. Penta}, \bibinfo{person}{R. Oliveto}, {and}
  \bibinfo{person}{M. Lanza}.} \bibinfo{year}{2014}\natexlab{}.
\newblock \showarticletitle{Mining Stack Overflow to turn the IDE into a
  self-confident programming prompter}. In
  \bibinfo{booktitle}{\emph{International Conference on Mining Software
  Repositories (MSR)}}.
\newblock


\bibitem[\protect\citeauthoryear{Price, Rilofff, Zachary, and Harvey}{Price
  et~al\mbox{.}}{2000}]%
        {price2000naturaljava}
\bibfield{author}{\bibinfo{person}{David Price}, \bibinfo{person}{Ellen
  Rilofff}, \bibinfo{person}{Joseph Zachary}, {and} \bibinfo{person}{Brandon
  Harvey}.} \bibinfo{year}{2000}\natexlab{}.
\newblock \showarticletitle{NaturalJava: a natural language interface for
  programming in Java}. In \bibinfo{booktitle}{\emph{International Conference
  on Intelligent User Interfaces (IUI)}}. \bibinfo{pages}{207--211}.
\newblock


\bibitem[\protect\citeauthoryear{Proksch, Amann, and Nadi}{Proksch
  et~al\mbox{.}}{2018}]%
        {proksch2018enriched}
\bibfield{author}{\bibinfo{person}{Sebastian Proksch}, \bibinfo{person}{Sven
  Amann}, {and} \bibinfo{person}{Sarah Nadi}.} \bibinfo{year}{2018}\natexlab{}.
\newblock \showarticletitle{Enriched event streams: a general dataset for
  empirical studies on in-IDE activities of software developers}. In
  \bibinfo{booktitle}{\emph{Proceedings of the 15th International Conference on
  Mining Software Repositories (MSR)}}. \bibinfo{pages}{62--65}.
\newblock


\bibitem[\protect\citeauthoryear{Radhakrishnan, Srikantan, and
  Lin}{Radhakrishnan et~al\mbox{.}}{2020}]%
        {radhakrishnan2020colloql}
\bibfield{author}{\bibinfo{person}{Karthik Radhakrishnan},
  \bibinfo{person}{Arvind Srikantan}, {and} \bibinfo{person}{Xi~Victoria Lin}.}
  \bibinfo{year}{2020}\natexlab{}.
\newblock \showarticletitle{ColloQL: Robust Text-to-SQL Over Search Queries}.
  In \bibinfo{booktitle}{\emph{Proceedings of the First Workshop on Interactive
  and Executable Semantic Parsing}}. \bibinfo{pages}{34--45}.
\newblock


\bibitem[\protect\citeauthoryear{Raghothaman, Wei, and Hamadi}{Raghothaman
  et~al\mbox{.}}{2016}]%
        {Raghothaman2016SWIMSW}
\bibfield{author}{\bibinfo{person}{Mukund Raghothaman}, \bibinfo{person}{Y.
  Wei}, {and} \bibinfo{person}{Y. Hamadi}.} \bibinfo{year}{2016}\natexlab{}.
\newblock \showarticletitle{SWIM: Synthesizing What I Mean - Code Search and
  Idiomatic Snippet Synthesis}.
\newblock \bibinfo{journal}{\emph{2016 IEEE/ACM 38th International Conference
  on Software Engineering (ICSE)}} (\bibinfo{year}{2016}),
  \bibinfo{pages}{357--367}.
\newblock


\bibitem[\protect\citeauthoryear{Rahman and Roy}{Rahman and Roy}{2014}]%
        {Rahman2014SurfClipseCM}
\bibfield{author}{\bibinfo{person}{M.~M. Rahman} {and} \bibinfo{person}{C.
  Roy}.} \bibinfo{year}{2014}\natexlab{}.
\newblock \showarticletitle{SurfClipse: Context-Aware Meta-search in the IDE}.
\newblock \bibinfo{journal}{\emph{2014 IEEE International Conference on
  Software Maintenance and Evolution}} (\bibinfo{year}{2014}),
  \bibinfo{pages}{617--620}.
\newblock


\bibitem[\protect\citeauthoryear{Rahman, Yeasmin, and Roy}{Rahman
  et~al\mbox{.}}{2014}]%
        {rahman2014towards}
\bibfield{author}{\bibinfo{person}{Mohammad~Masudur Rahman},
  \bibinfo{person}{Shamima Yeasmin}, {and} \bibinfo{person}{Chanchal~K Roy}.}
  \bibinfo{year}{2014}\natexlab{}.
\newblock \showarticletitle{Towards a context-aware IDE-based meta search
  engine for recommendation about programming errors and exceptions}. In
  \bibinfo{booktitle}{\emph{International Conference on Software Analysis,
  Evolution, and Reengineering (SANER)}}. IEEE, \bibinfo{pages}{194--203}.
\newblock


\bibitem[\protect\citeauthoryear{Raychev, Vechev, and Yahav}{Raychev
  et~al\mbox{.}}{2014}]%
        {raychev2014code}
\bibfield{author}{\bibinfo{person}{Veselin Raychev}, \bibinfo{person}{Martin
  Vechev}, {and} \bibinfo{person}{Eran Yahav}.}
  \bibinfo{year}{2014}\natexlab{}.
\newblock \showarticletitle{Code completion with statistical language models}.
  In \bibinfo{booktitle}{\emph{ACM Conference on Programming Language Design
  and Implementation (PLDI)}}. ACM, \bibinfo{pages}{419--428}.
\newblock


\bibitem[\protect\citeauthoryear{Raza, Gulwani, and Milic-Frayling}{Raza
  et~al\mbox{.}}{2015}]%
        {Raza2015CompositionalPS}
\bibfield{author}{\bibinfo{person}{Mohammad Raza}, \bibinfo{person}{Sumit
  Gulwani}, {and} \bibinfo{person}{Natasa Milic-Frayling}.}
  \bibinfo{year}{2015}\natexlab{}.
\newblock \showarticletitle{Compositional Program Synthesis from Natural
  Language and Examples}. In \bibinfo{booktitle}{\emph{Proceedings of the
  Twenty-Fourth International Joint Conference on Artificial Intelligence
  (IJCAI)}}.
\newblock


\bibitem[\protect\citeauthoryear{Rice}{Rice}{1953}]%
        {rice1953classes}
\bibfield{author}{\bibinfo{person}{Henry~Gordon Rice}.}
  \bibinfo{year}{1953}\natexlab{}.
\newblock \showarticletitle{Classes of recursively enumerable sets and their
  decision problems}.
\newblock \bibinfo{journal}{\emph{Trans. Amer. Math. Soc.}}
  \bibinfo{volume}{74}, \bibinfo{number}{2} (\bibinfo{year}{1953}),
  \bibinfo{pages}{358--366}.
\newblock


\bibitem[\protect\citeauthoryear{Richardson, Dominowska, and Ragno}{Richardson
  et~al\mbox{.}}{2007}]%
        {richardson2007predicting}
\bibfield{author}{\bibinfo{person}{Matthew Richardson}, \bibinfo{person}{Ewa
  Dominowska}, {and} \bibinfo{person}{Robert Ragno}.}
  \bibinfo{year}{2007}\natexlab{}.
\newblock \showarticletitle{Predicting clicks: estimating the click-through
  rate for new ads}. In \bibinfo{booktitle}{\emph{Proceedings of the 16th
  International Conference on World Wide Web}}. \bibinfo{pages}{521--530}.
\newblock


\bibitem[\protect\citeauthoryear{Roy, Zhang, Ma, Arnaoudova, Panichella,
  Panichella, Gonzalez, and Mirakhorli}{Roy et~al\mbox{.}}{2020}]%
        {roy2020deeptc}
\bibfield{author}{\bibinfo{person}{Devjeet Roy}, \bibinfo{person}{Ziyi Zhang},
  \bibinfo{person}{Maggie Ma}, \bibinfo{person}{Venera Arnaoudova},
  \bibinfo{person}{Annibale Panichella}, \bibinfo{person}{Sebastiano
  Panichella}, \bibinfo{person}{Danielle Gonzalez}, {and}
  \bibinfo{person}{Mehdi Mirakhorli}.} \bibinfo{year}{2020}\natexlab{}.
\newblock \showarticletitle{DeepTC-Enhancer: Improving the Readability of
  Automatically Generated Tests}. In \bibinfo{booktitle}{\emph{2020 35th
  IEEE/ACM International Conference on Automated Software Engineering (ASE)}}.
  IEEE, \bibinfo{pages}{287--298}.
\newblock


\bibitem[\protect\citeauthoryear{Sadowski, Stolee, and Elbaum}{Sadowski
  et~al\mbox{.}}{2015}]%
        {sadowski2015developers}
\bibfield{author}{\bibinfo{person}{Caitlin Sadowski},
  \bibinfo{person}{Kathryn~T Stolee}, {and} \bibinfo{person}{Sebastian
  Elbaum}.} \bibinfo{year}{2015}\natexlab{}.
\newblock \showarticletitle{How developers search for code: a case study}. In
  \bibinfo{booktitle}{\emph{Proceedings of the 2015 10th Joint Meeting on
  Foundations of Software Engineering (ESEC/FSE)}}. \bibinfo{pages}{191--201}.
\newblock


\bibitem[\protect\citeauthoryear{Sahay, Indamutsa, Ruscio, and
  Pierantonio}{Sahay et~al\mbox{.}}{2020}]%
        {Sahay2020SupportingTU}
\bibfield{author}{\bibinfo{person}{Apurvanand Sahay}, \bibinfo{person}{Arsene
  Indamutsa}, \bibinfo{person}{D.~D. Ruscio}, {and} \bibinfo{person}{A.
  Pierantonio}.} \bibinfo{year}{2020}\natexlab{}.
\newblock \showarticletitle{Supporting the understanding and comparison of
  low-code development platforms}.
\newblock \bibinfo{journal}{\emph{2020 46th Euromicro Conference on Software
  Engineering and Advanced Applications (SEAA)}} (\bibinfo{year}{2020}),
  \bibinfo{pages}{171--178}.
\newblock


\bibitem[\protect\citeauthoryear{Shin, Allamanis, Brockschmidt, and
  Polozov}{Shin et~al\mbox{.}}{2019}]%
        {Shin2019ProgramSA}
\bibfield{author}{\bibinfo{person}{Richard Shin}, \bibinfo{person}{Miltiadis
  Allamanis}, \bibinfo{person}{Marc Brockschmidt}, {and}
  \bibinfo{person}{Oleksandr Polozov}.} \bibinfo{year}{2019}\natexlab{}.
\newblock \showarticletitle{Program Synthesis and Semantic Parsing with Learned
  Code Idioms}.
\newblock \bibinfo{journal}{\emph{33rd Conference on Neural Information
  Processing Systems (NeurIPS)}} (\bibinfo{year}{2019}).
\newblock


\bibitem[\protect\citeauthoryear{Shull, Singer, and Sj{\o}berg}{Shull
  et~al\mbox{.}}{2007}]%
        {shull2007guide}
\bibfield{author}{\bibinfo{person}{Forrest Shull}, \bibinfo{person}{Janice
  Singer}, {and} \bibinfo{person}{Dag~IK Sj{\o}berg}.}
  \bibinfo{year}{2007}\natexlab{}.
\newblock \bibinfo{booktitle}{\emph{Guide to advanced empirical software
  engineering}}.
\newblock \bibinfo{publisher}{Springer}.
\newblock


\bibitem[\protect\citeauthoryear{Solar-Lezama}{Solar-Lezama}{2008}]%
        {Bodk2008ProgramSB}
\bibfield{author}{\bibinfo{person}{Armando Solar-Lezama}.}
  \bibinfo{year}{2008}\natexlab{}.
\newblock \showarticletitle{Program Synthesis by Sketching}.
\newblock


\bibitem[\protect\citeauthoryear{Subramanian, Inozemtseva, and
  Holmes}{Subramanian et~al\mbox{.}}{2014}]%
        {Subramanian2014LiveAD}
\bibfield{author}{\bibinfo{person}{Siddharth Subramanian},
  \bibinfo{person}{Laura Inozemtseva}, {and} \bibinfo{person}{Reid Holmes}.}
  \bibinfo{year}{2014}\natexlab{}.
\newblock \showarticletitle{Live API documentation}.
\newblock \bibinfo{journal}{\emph{International Conference on Software
  Engineering (ICSE)}} (\bibinfo{year}{2014}).
\newblock


\bibitem[\protect\citeauthoryear{Tran, Tran, Nguyen, Nguyen, and Nguyen}{Tran
  et~al\mbox{.}}{2019}]%
        {tran2019does}
\bibfield{author}{\bibinfo{person}{Ngoc Tran}, \bibinfo{person}{Hieu Tran},
  \bibinfo{person}{Son Nguyen}, \bibinfo{person}{Hoan Nguyen}, {and}
  \bibinfo{person}{Tien Nguyen}.} \bibinfo{year}{2019}\natexlab{}.
\newblock \showarticletitle{Does BLEU score work for code migration?}. In
  \bibinfo{booktitle}{\emph{2019 IEEE/ACM 27th International Conference on
  Program Comprehension (ICPC)}}. IEEE, \bibinfo{pages}{165--176}.
\newblock


\bibitem[\protect\citeauthoryear{Tu, Su, and Devanbu}{Tu et~al\mbox{.}}{2014}]%
        {tu2014localness}
\bibfield{author}{\bibinfo{person}{Zhaopeng Tu}, \bibinfo{person}{Zhendong Su},
  {and} \bibinfo{person}{Premkumar Devanbu}.} \bibinfo{year}{2014}\natexlab{}.
\newblock \showarticletitle{On the localness of software}. In
  \bibinfo{booktitle}{\emph{International Symposium on Foundations of Software
  Engineering (ESEC/FSE)}}. \bibinfo{publisher}{ACM},
  \bibinfo{pages}{269--280}.
\newblock


\bibitem[\protect\citeauthoryear{Vadas and Curran}{Vadas and Curran}{2005}]%
        {vadas2005programming}
\bibfield{author}{\bibinfo{person}{David Vadas} {and} \bibinfo{person}{James~R
  Curran}.} \bibinfo{year}{2005}\natexlab{}.
\newblock \showarticletitle{Programming with unrestricted natural language}. In
  \bibinfo{booktitle}{\emph{Proceedings of the Australasian Language Technology
  Workshop 2005}}. \bibinfo{pages}{191--199}.
\newblock


\bibitem[\protect\citeauthoryear{Vinayakarao, Sarma, Purandare, Jain, and
  Jain}{Vinayakarao et~al\mbox{.}}{2017}]%
        {Vinayakarao2017ANNEIS}
\bibfield{author}{\bibinfo{person}{Venkatesh Vinayakarao}, \bibinfo{person}{A.
  Sarma}, \bibinfo{person}{R. Purandare}, \bibinfo{person}{Shuktika Jain},
  {and} \bibinfo{person}{Saumya Jain}.} \bibinfo{year}{2017}\natexlab{}.
\newblock \showarticletitle{ANNE: Improving Source Code Search using Entity
  Retrieval Approach}. In \bibinfo{booktitle}{\emph{WSDM '17}}.
\newblock


\bibitem[\protect\citeauthoryear{Wang, Berant, and Liang}{Wang
  et~al\mbox{.}}{2015}]%
        {wang2015building}
\bibfield{author}{\bibinfo{person}{Yushi Wang}, \bibinfo{person}{Jonathan
  Berant}, {and} \bibinfo{person}{Percy Liang}.}
  \bibinfo{year}{2015}\natexlab{}.
\newblock \showarticletitle{Building a semantic parser overnight}. In
  \bibinfo{booktitle}{\emph{Proceedings of the 53rd Annual Meeting of the
  Association for Computational Linguistics and the 7th International Joint
  Conference on Natural Language Processing (ACL-IJCNLP)}}.
  \bibinfo{pages}{1332--1342}.
\newblock


\bibitem[\protect\citeauthoryear{Wei, Chandrasekaran, Gulwani, and Hamadi}{Wei
  et~al\mbox{.}}{2015}]%
        {wei2015building}
\bibfield{author}{\bibinfo{person}{Yi Wei}, \bibinfo{person}{Nirupama
  Chandrasekaran}, \bibinfo{person}{Sumit Gulwani}, {and}
  \bibinfo{person}{Youssef Hamadi}.} \bibinfo{year}{2015}\natexlab{}.
\newblock \bibinfo{booktitle}{\emph{Building {Bing Developer Assistant}}}.
\newblock \bibinfo{type}{{T}echnical {R}eport}.
  \bibinfo{institution}{MSR-TR-2015-36, Microsoft Research}.
\newblock


\bibitem[\protect\citeauthoryear{Wohlin, Runeson, H{\"o}st, Ohlsson, Regnell,
  and Wessl{\'e}n}{Wohlin et~al\mbox{.}}{2012}]%
        {wohlin2012experimentation}
\bibfield{author}{\bibinfo{person}{Claes Wohlin}, \bibinfo{person}{Per
  Runeson}, \bibinfo{person}{Martin H{\"o}st}, \bibinfo{person}{Magnus~C
  Ohlsson}, \bibinfo{person}{Bj{\"o}rn Regnell}, {and} \bibinfo{person}{Anders
  Wessl{\'e}n}.} \bibinfo{year}{2012}\natexlab{}.
\newblock \bibinfo{booktitle}{\emph{Experimentation in software engineering}}.
\newblock \bibinfo{publisher}{Springer Science \& Business Media}.
\newblock


\bibitem[\protect\citeauthoryear{Xu, Jiang, Yin, Vasilescu, and Neubig}{Xu
  et~al\mbox{.}}{2020}]%
        {xu2020incorporating}
\bibfield{author}{\bibinfo{person}{Frank~F Xu}, \bibinfo{person}{Zhengbao
  Jiang}, \bibinfo{person}{Pengcheng Yin}, \bibinfo{person}{Bogdan Vasilescu},
  {and} \bibinfo{person}{Graham Neubig}.} \bibinfo{year}{2020}\natexlab{}.
\newblock \showarticletitle{Incorporating External Knowledge through
  Pre-training for Natural Language to Code Generation}. In
  \bibinfo{booktitle}{\emph{Annual Meeting of the Association for Computational
  Linguistics (ACL)}}. \bibinfo{publisher}{Association for Computational
  Linguistics}, \bibinfo{pages}{6045--6052}.
\newblock


\bibitem[\protect\citeauthoryear{Yao and Van~Durme}{Yao and Van~Durme}{2014}]%
        {yao2014information}
\bibfield{author}{\bibinfo{person}{Xuchen Yao} {and} \bibinfo{person}{Benjamin
  Van~Durme}.} \bibinfo{year}{2014}\natexlab{}.
\newblock \showarticletitle{Information extraction over structured data:
  Question answering with freebase}. In \bibinfo{booktitle}{\emph{Proceedings
  of the 52nd Annual Meeting of the Association for Computational Linguistics
  (ACL)}}. \bibinfo{pages}{956--966}.
\newblock


\bibitem[\protect\citeauthoryear{Yao, Li, Gao, Sadler, and Sun}{Yao
  et~al\mbox{.}}{2019a}]%
        {yao2019interactive}
\bibfield{author}{\bibinfo{person}{Ziyu Yao}, \bibinfo{person}{Xiujun Li},
  \bibinfo{person}{Jianfeng Gao}, \bibinfo{person}{Brian Sadler}, {and}
  \bibinfo{person}{Huan Sun}.} \bibinfo{year}{2019}\natexlab{a}.
\newblock \showarticletitle{Interactive semantic parsing for if-then recipes
  via hierarchical reinforcement learning}. In
  \bibinfo{booktitle}{\emph{Proceedings of the AAAI Conference on Artificial
  Intelligence (AAAI)}}, Vol.~\bibinfo{volume}{33}.
  \bibinfo{pages}{2547--2554}.
\newblock


\bibitem[\protect\citeauthoryear{Yao, Peddamail, and Sun}{Yao
  et~al\mbox{.}}{2019b}]%
        {Yao2019CoaCorCA}
\bibfield{author}{\bibinfo{person}{Ziyu Yao},
  \bibinfo{person}{Jayavardhan~Reddy Peddamail}, {and} \bibinfo{person}{Huan
  Sun}.} \bibinfo{year}{2019}\natexlab{b}.
\newblock \showarticletitle{CoaCor: Code Annotation for Code Retrieval with
  Reinforcement Learning}.
\newblock \bibinfo{journal}{\emph{The World Wide Web Conference (WWW)}}
  (\bibinfo{year}{2019}).
\newblock


\bibitem[\protect\citeauthoryear{Yao, Weld, Chen, and Sun}{Yao
  et~al\mbox{.}}{2018}]%
        {Yao2018StaQCAS}
\bibfield{author}{\bibinfo{person}{Ziyu Yao}, \bibinfo{person}{Daniel~S. Weld},
  \bibinfo{person}{W. Chen}, {and} \bibinfo{person}{Huan Sun}.}
  \bibinfo{year}{2018}\natexlab{}.
\newblock \showarticletitle{StaQC: A Systematically Mined Question-Code Dataset
  from Stack Overflow}.
\newblock \bibinfo{journal}{\emph{Proceedings of the 2018 World Wide Web
  Conference (WWW)}} (\bibinfo{year}{2018}).
\newblock


\bibitem[\protect\citeauthoryear{Yin, Deng, Chen, Vasilescu, and Neubig}{Yin
  et~al\mbox{.}}{2018}]%
        {yin2018mining}
\bibfield{author}{\bibinfo{person}{Pengcheng Yin}, \bibinfo{person}{Bowen
  Deng}, \bibinfo{person}{Edgar Chen}, \bibinfo{person}{Bogdan Vasilescu},
  {and} \bibinfo{person}{Graham Neubig}.} \bibinfo{year}{2018}\natexlab{}.
\newblock \showarticletitle{Learning to Mine Aligned Code and Natural Language
  Pairs from Stack Overflow}. In \bibinfo{booktitle}{\emph{International
  Conference on Mining Software Repositories}} \emph{(\bibinfo{series}{MSR})}.
  \bibinfo{publisher}{ACM}, \bibinfo{pages}{476--486}.
\newblock
\urldef\tempurl%
\url{https://doi.org/10.1145/3196398.3196408}
\showDOI{\tempurl}


\bibitem[\protect\citeauthoryear{Yin and Neubig}{Yin and Neubig}{2017}]%
        {yin2017syntactic}
\bibfield{author}{\bibinfo{person}{Pengcheng Yin} {and} \bibinfo{person}{Graham
  Neubig}.} \bibinfo{year}{2017}\natexlab{}.
\newblock \showarticletitle{A syntactic neural model for general-purpose code
  generation}.
\newblock \bibinfo{journal}{\emph{Annual Meeting of the Association for
  Computational Linguistics (ACL)}} (\bibinfo{year}{2017}).
\newblock


\bibitem[\protect\citeauthoryear{Yin and Neubig}{Yin and Neubig}{2018}]%
        {yin2018tranx}
\bibfield{author}{\bibinfo{person}{Pengcheng Yin} {and} \bibinfo{person}{Graham
  Neubig}.} \bibinfo{year}{2018}\natexlab{}.
\newblock \showarticletitle{Tranx: A transition-based neural abstract syntax
  parser for semantic parsing and code generation}.
\newblock \bibinfo{journal}{\emph{Conference on Empirical Methods in Natural
  Language Processing (EMNLP), Demo Track}} (\bibinfo{year}{2018}).
\newblock


\bibitem[\protect\citeauthoryear{Yin and Neubig}{Yin and Neubig}{2019}]%
        {yin-neubig-2019-reranking}
\bibfield{author}{\bibinfo{person}{Pengcheng Yin} {and} \bibinfo{person}{Graham
  Neubig}.} \bibinfo{year}{2019}\natexlab{}.
\newblock \showarticletitle{Reranking for Neural Semantic Parsing}. In
  \bibinfo{booktitle}{\emph{Proceedings of the 57th Annual Meeting of the
  Association for Computational Linguistics (ACL)}}.
  \bibinfo{publisher}{Association for Computational Linguistics},
  \bibinfo{address}{Florence, Italy}, \bibinfo{pages}{4553--4559}.
\newblock
\urldef\tempurl%
\url{https://doi.org/10.18653/v1/P19-1447}
\showDOI{\tempurl}


\bibitem[\protect\citeauthoryear{Zavershynskyi, Skidanov, and
  Polosukhin}{Zavershynskyi et~al\mbox{.}}{2018}]%
        {zavershynskyi2018naps}
\bibfield{author}{\bibinfo{person}{Maksym Zavershynskyi}, \bibinfo{person}{Alex
  Skidanov}, {and} \bibinfo{person}{Illia Polosukhin}.}
  \bibinfo{year}{2018}\natexlab{}.
\newblock \showarticletitle{NAPS: Natural program synthesis dataset}.
\newblock \bibinfo{journal}{\emph{2nd Workshop on Neural Abstract Machines \&
  Program Induction (NAMPI), ICML}} (\bibinfo{year}{2018}).
\newblock


\bibitem[\protect\citeauthoryear{Zelle and Mooney}{Zelle and Mooney}{1996}]%
        {zelle1996learning}
\bibfield{author}{\bibinfo{person}{John~M Zelle} {and}
  \bibinfo{person}{Raymond~J Mooney}.} \bibinfo{year}{1996}\natexlab{}.
\newblock \showarticletitle{Learning to parse database queries using inductive
  logic programming}. In \bibinfo{booktitle}{\emph{Proceedings of the national
  conference on artificial intelligence}}. \bibinfo{pages}{1050--1055}.
\newblock


\bibitem[\protect\citeauthoryear{Zettlemoyer and Collins}{Zettlemoyer and
  Collins}{2007}]%
        {zettlemoyer2007online}
\bibfield{author}{\bibinfo{person}{Luke Zettlemoyer} {and}
  \bibinfo{person}{Michael Collins}.} \bibinfo{year}{2007}\natexlab{}.
\newblock \showarticletitle{Online learning of relaxed CCG grammars for parsing
  to logical form}. In \bibinfo{booktitle}{\emph{Proceedings of the 2007 Joint
  Conference on Empirical Methods in Natural Language Processing and
  Computational Natural Language Learning (EMNLP-CoNLL)}}.
  \bibinfo{pages}{678--687}.
\newblock


\bibitem[\protect\citeauthoryear{Zhong, Stern, and Klein}{Zhong
  et~al\mbox{.}}{2020}]%
        {Zhong2020SemanticSF}
\bibfield{author}{\bibinfo{person}{Ruiqi Zhong}, \bibinfo{person}{Mitchell
  Stern}, {and} \bibinfo{person}{D. Klein}.} \bibinfo{year}{2020}\natexlab{}.
\newblock \showarticletitle{Semantic Scaffolds for Pseudocode-to-Code
  Generation}. In \bibinfo{booktitle}{\emph{ACL}}.
\newblock


\bibitem[\protect\citeauthoryear{Zhong, Xiong, and Socher}{Zhong
  et~al\mbox{.}}{2017}]%
        {zhong2017seq2sql}
\bibfield{author}{\bibinfo{person}{Victor Zhong}, \bibinfo{person}{Caiming
  Xiong}, {and} \bibinfo{person}{Richard Socher}.}
  \bibinfo{year}{2017}\natexlab{}.
\newblock \showarticletitle{Seq2sql: Generating structured queries from natural
  language using reinforcement learning}.
\newblock \bibinfo{journal}{\emph{arXiv preprint arXiv:1709.00103}}
  (\bibinfo{year}{2017}).
\newblock


\end{thebibliography}


\newpage
\appendix
\section{User Study Environment Design}
\label{app:environment_design}
To control the user study's development environment for different users as much as possible, and to enable data collection and activity recording outside the IDE (e.g. web browsing activity during the development), we design a complete virtual machine-based environment for users to access remotely and perform the user study on.
We build the virtual machine based on a lot of open source software, including Ubuntu 18.04 operating system\footnote{\url{https://releases.ubuntu.com/18.04/}} with XFCE 4.1 desktop environment.\footnote{\url{https://www.xfce.org/}}
The virtual machine software is VirtualBox 6.1.10,\footnote{\url{https://www.virtualbox.org/wiki/Downloads}} and we use Vagrant software\footnote{\url{https://www.vagrantup.com/}} for automatic virtual machine provisioning.

Inside the Linux virtual machine, we install and configure a set of programs for data collection and workflow control during the user study:
\begin{enumerate}
    \item \textbf{Python environment.} Python 3.6\footnote{\url{https://www.python.org/}} is installed inside the VM, alongside with pip package manager and several commonly used Python packages for the user study tasks.
    The user is free to install any additional packages they need during the development.
    \item \textbf{IDE with plugin.} PyCharm Community Edition 2020.1, with the plugin described in Section~\ref{sec:plugin} is installed.
    This provides consistent Python development environment for the user study and  the testing of the code generation and retrieval.
    The plugin also handles various data collection processes inside the IDE.
    \item \textbf{Man-in-the-middle proxy.} We install \texttt{mitmproxy}\footnote{\url{https://mitmproxy.org/}} in the VM, along with our customized script sending logs back to our server.
    This infrastructure enables interception and data collection of both HTTP and secured HTTPS requests.
    With this we can collect users' complete web browsing activities during the user study.
    \item \textbf{Web browser.} We install Firefox browser,\footnote{\url{https://www.mozilla.org/en-US/firefox/}} configured to use the proxy mentioned above so that all users' browsing activities could be logged for analysis.
    \item \textbf{Keylogger.} We develop a program that runs in the background during the user study, and logs all the user's keystrokes along with the timestamps to our server.
    With the keylogger we can collect data outside the IDE about the users' activities.
    This data is useful for mining and analyzing developer activity patterns in terms of keyboard operations, for example copy and pasting shortcuts.
    \item \textbf{User study control scripts.} We provide users a handful of scripts for easy and fully automatic retrieval, start and submission of the tasks.
    The scripts allow user to check their completion status of the whole study, as well as to pause and resume during a task for a break.
    All the user's task start, pause, resume, and submission events are logged so that the completion time of each task for the user could be calculated.
\end{enumerate}

\section{Pre-test Survey Details}
\label{app:pre_survey_details}

For each of the prospective participants, we asked them about two parts of the information in a pre-study survey, apart from personal information for contact purposes.
The first is regarding programming experience, used to determine if the participants have enough expertise in Python as well as the categories of tasks that we designed.
The questions are:
\begin{enumerate}
    \item Which of the following best describes your current career status: Student (computer science), Student (other field), Software Engineer, Data Scientist, Researcher, Other.
    \item How do you estimate your programming experience? (1: very inexperienced to 5: very experienced)
    \item How experienced are you with Python? (1: very inexperienced to 5: very experienced)
    \item How experienced are you with each of the following tasks in Python? (1: very inexperienced to 5: very experienced) Basic Python, File, OS, Web Scraping, Web Server \& Client, Data Analysis \& Machine Learning, Data Visualization.
\end{enumerate}
The second part of the information is about their development preferences, used to ask for their preferences with IDE and assistive tools.
The questions are:
\begin{enumerate}
    \item What editor/IDE do you use for Python projects? Vim, Emacs, VSCode, PyCharm, Jupyter Notebook, Sublime Text, other.
    \item Do you use any assistive tools or plugins to improve your coding efficiency? Some examples are code linting, type checking, snippet search tools, etc. If yes, what are they?
\end{enumerate}

\section{Participants Programming Experience}
\label{app:particpants_experience}
\begin{figure}[t]
     \centering
     \begin{subfigure}[b]{0.3\textwidth}
         \centering
         \includegraphics[width=\textwidth]{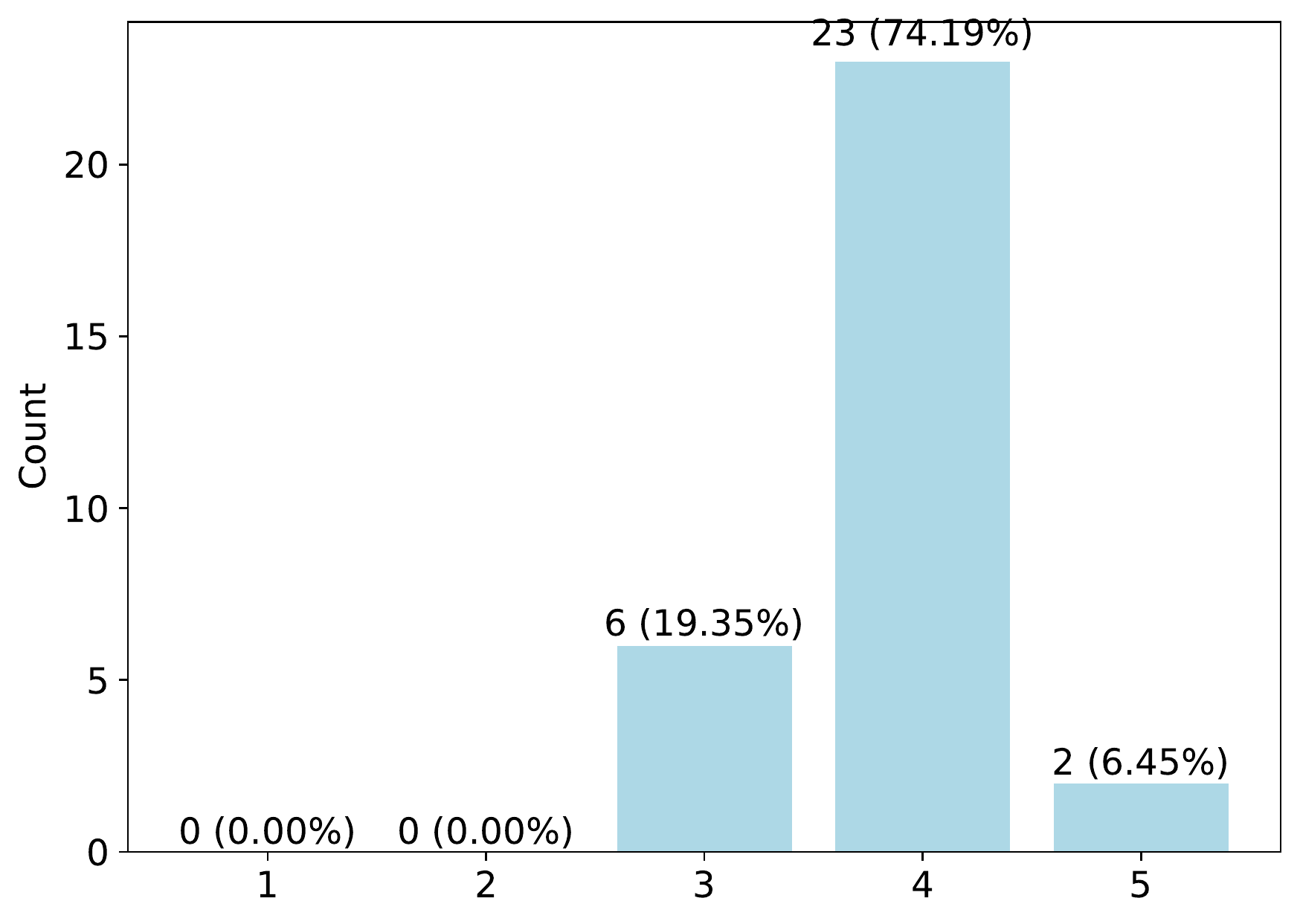}
         \caption{Overall Python Experience}
         \label{fig:exp_python_all}
     \end{subfigure}
     \hfill
     \begin{subfigure}[b]{0.3\textwidth}
         \centering
         \includegraphics[width=\textwidth]{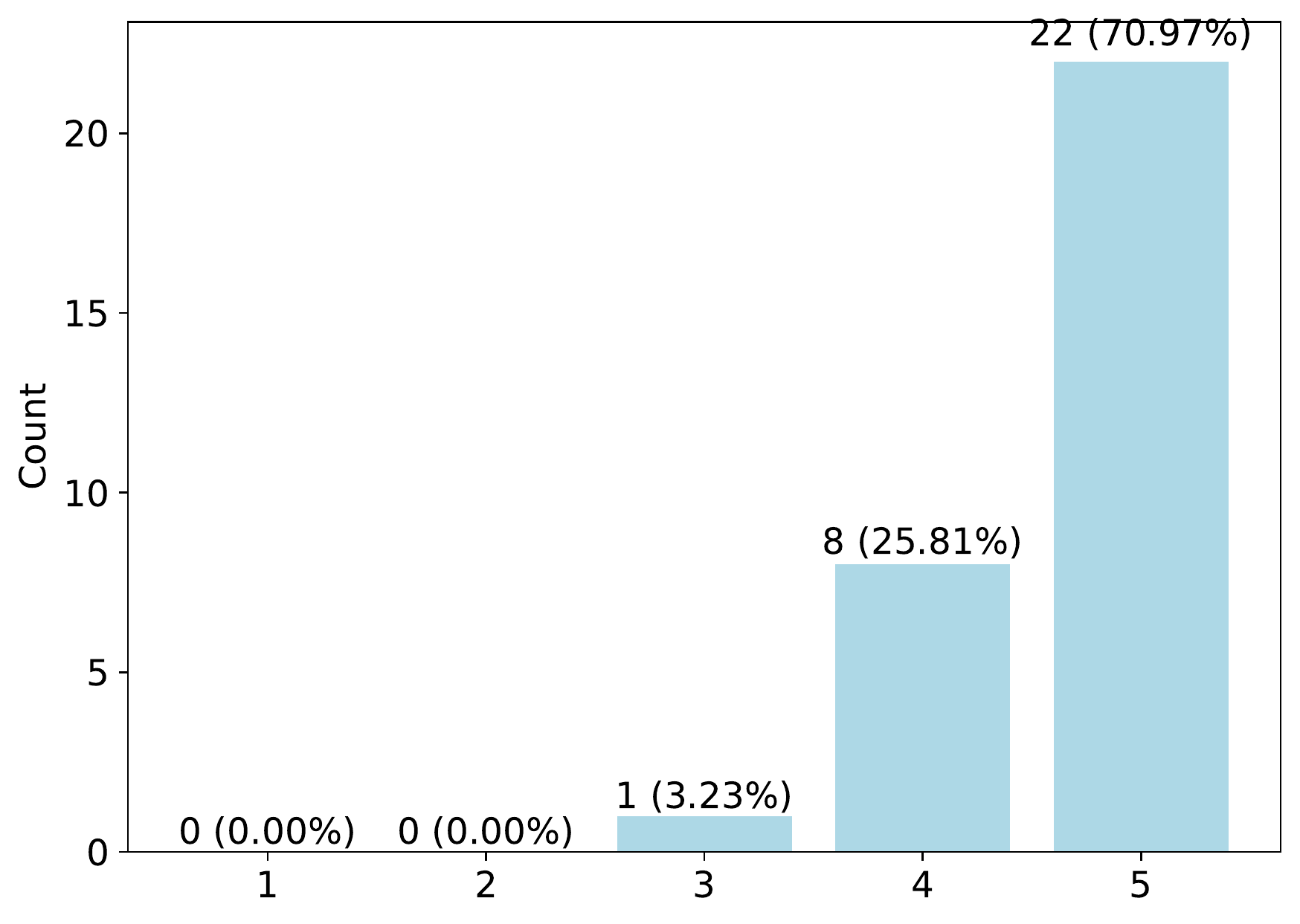}
         \caption{Basic Python}
         \label{fig:exp_1}
     \end{subfigure}
     \hfill
     \begin{subfigure}[b]{0.3\textwidth}
         \centering
         \includegraphics[width=\textwidth]{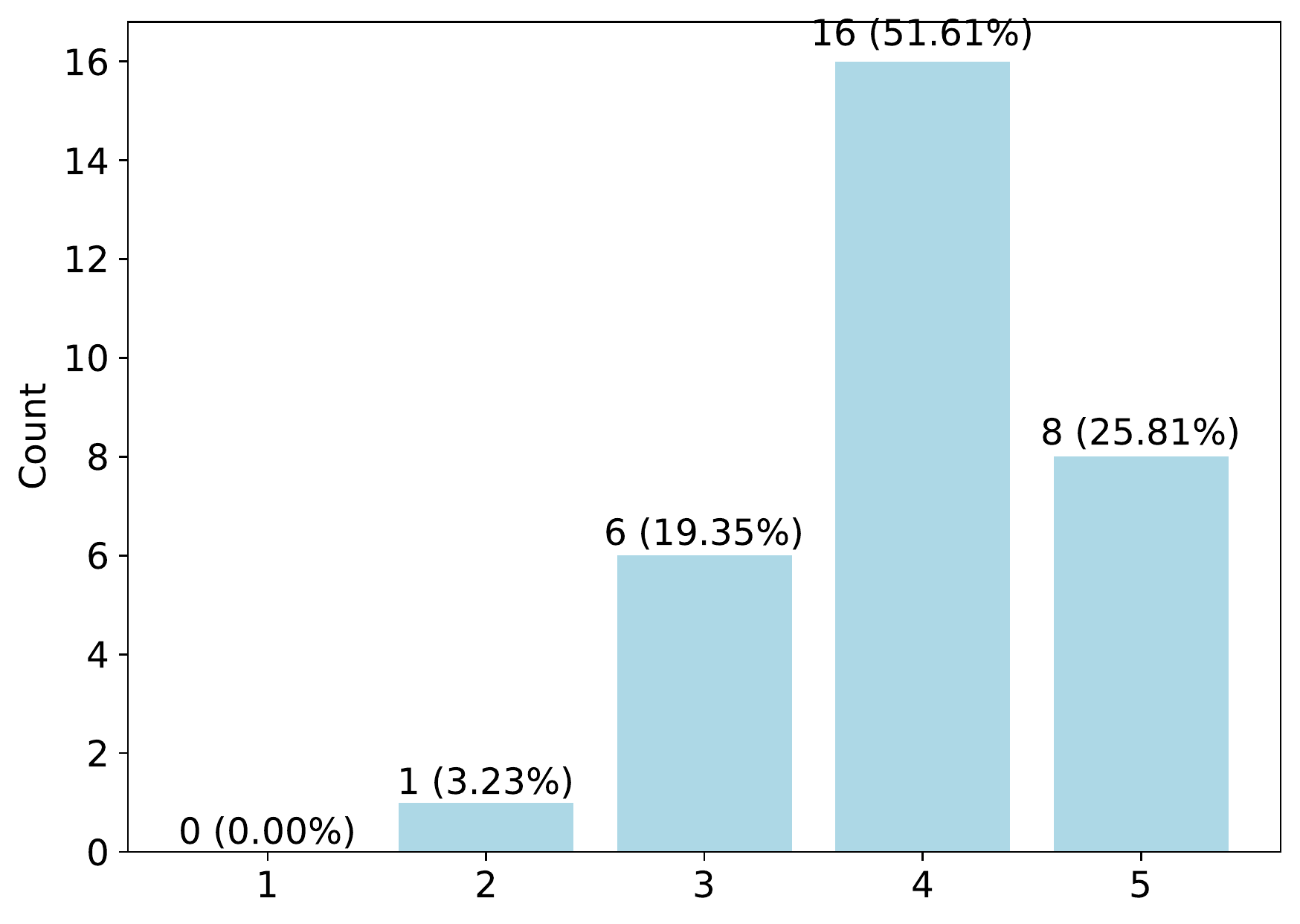}
         \caption{File}
         \label{fig:exp_2}
     \end{subfigure}
     \\
     \begin{subfigure}[b]{0.3\textwidth}
         \centering
         \includegraphics[width=\textwidth]{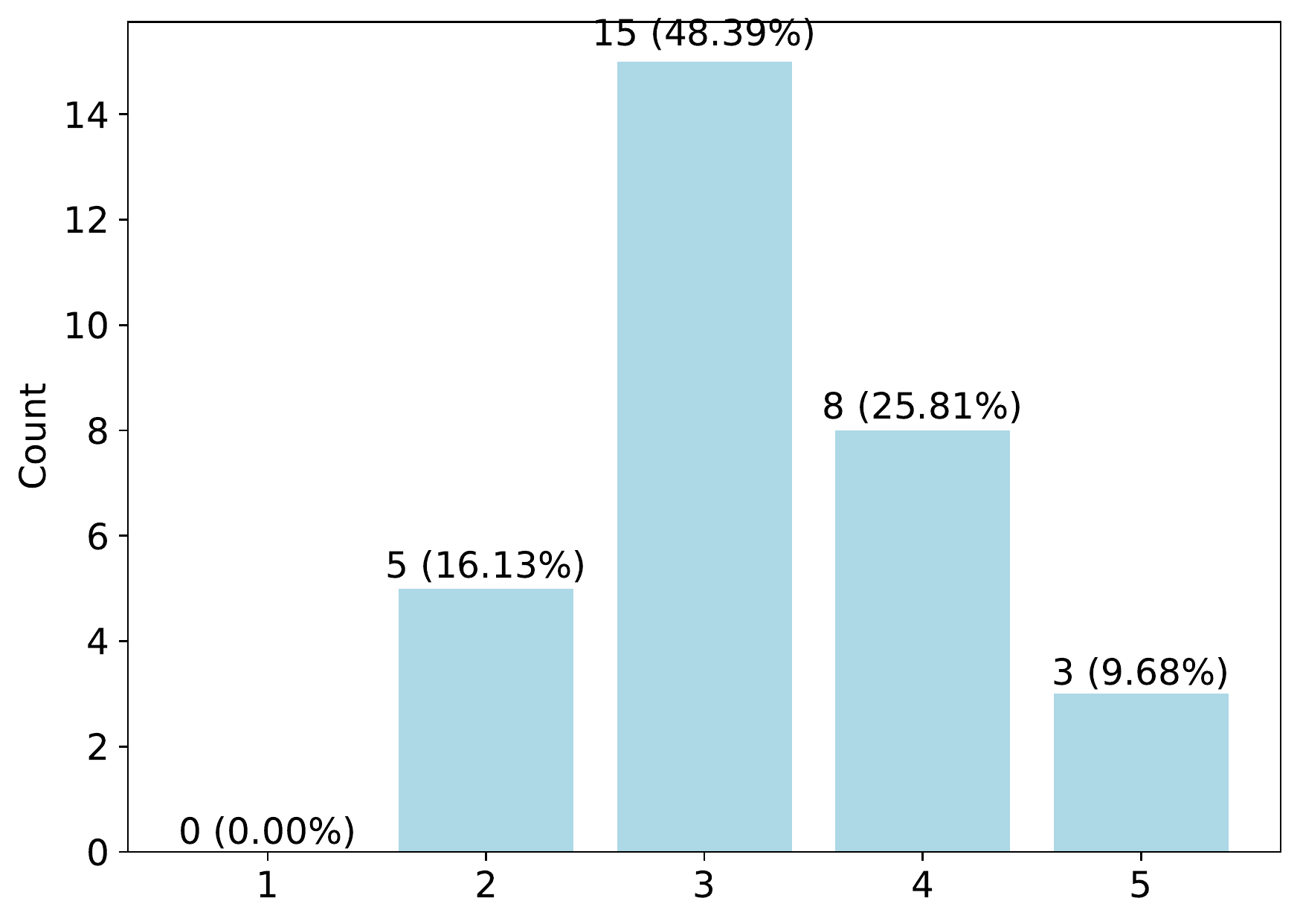}
         \caption{OS}
         \label{fig:exp_3}
     \end{subfigure}
     \hfill
      \begin{subfigure}[b]{0.3\textwidth}
         \centering
         \includegraphics[width=\textwidth]{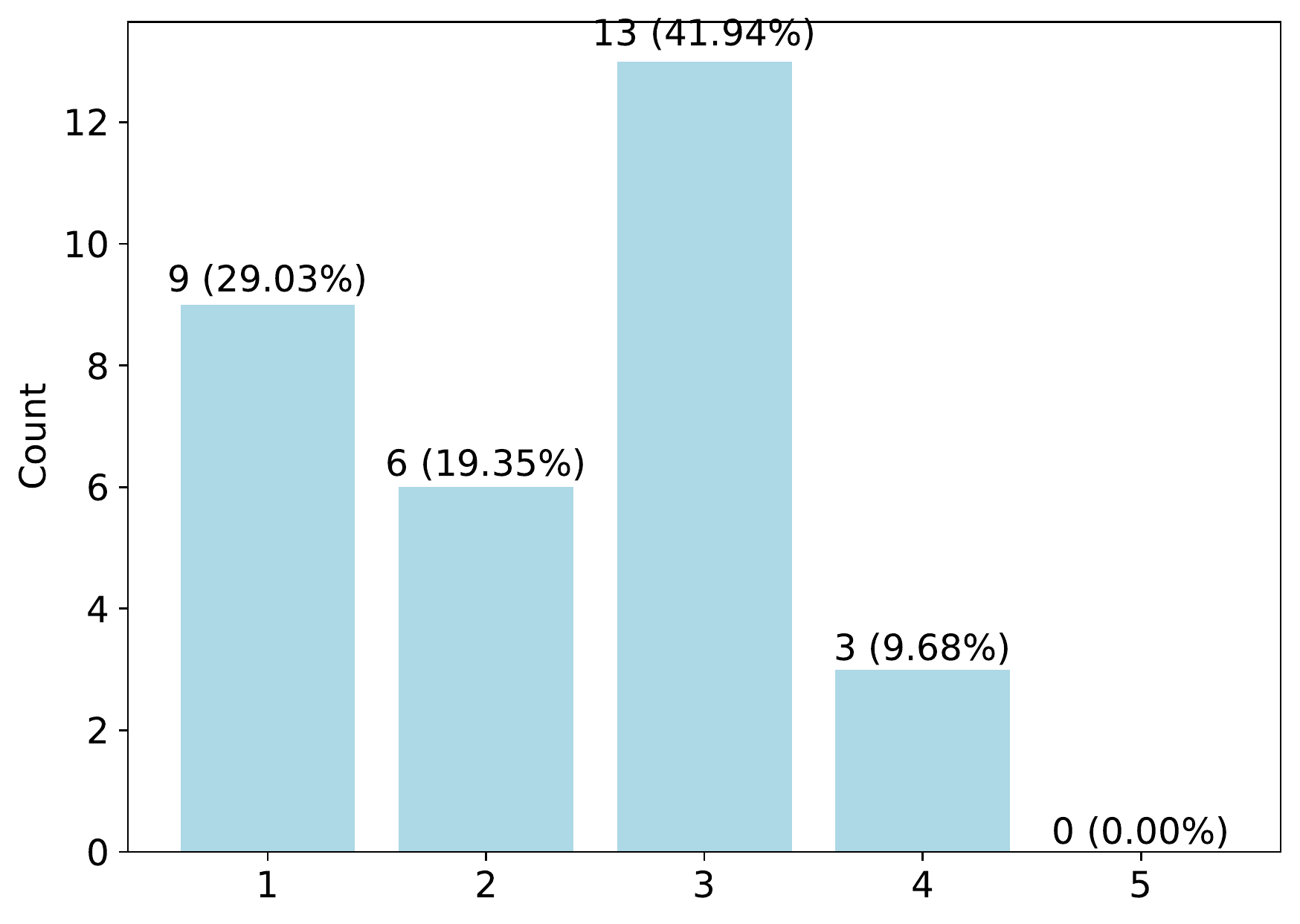}
         \caption{Web Scraping}
         \label{fig:exp_4}
     \end{subfigure}
     \hfill
     \begin{subfigure}[b]{0.3\textwidth}
         \centering
         \includegraphics[width=\textwidth]{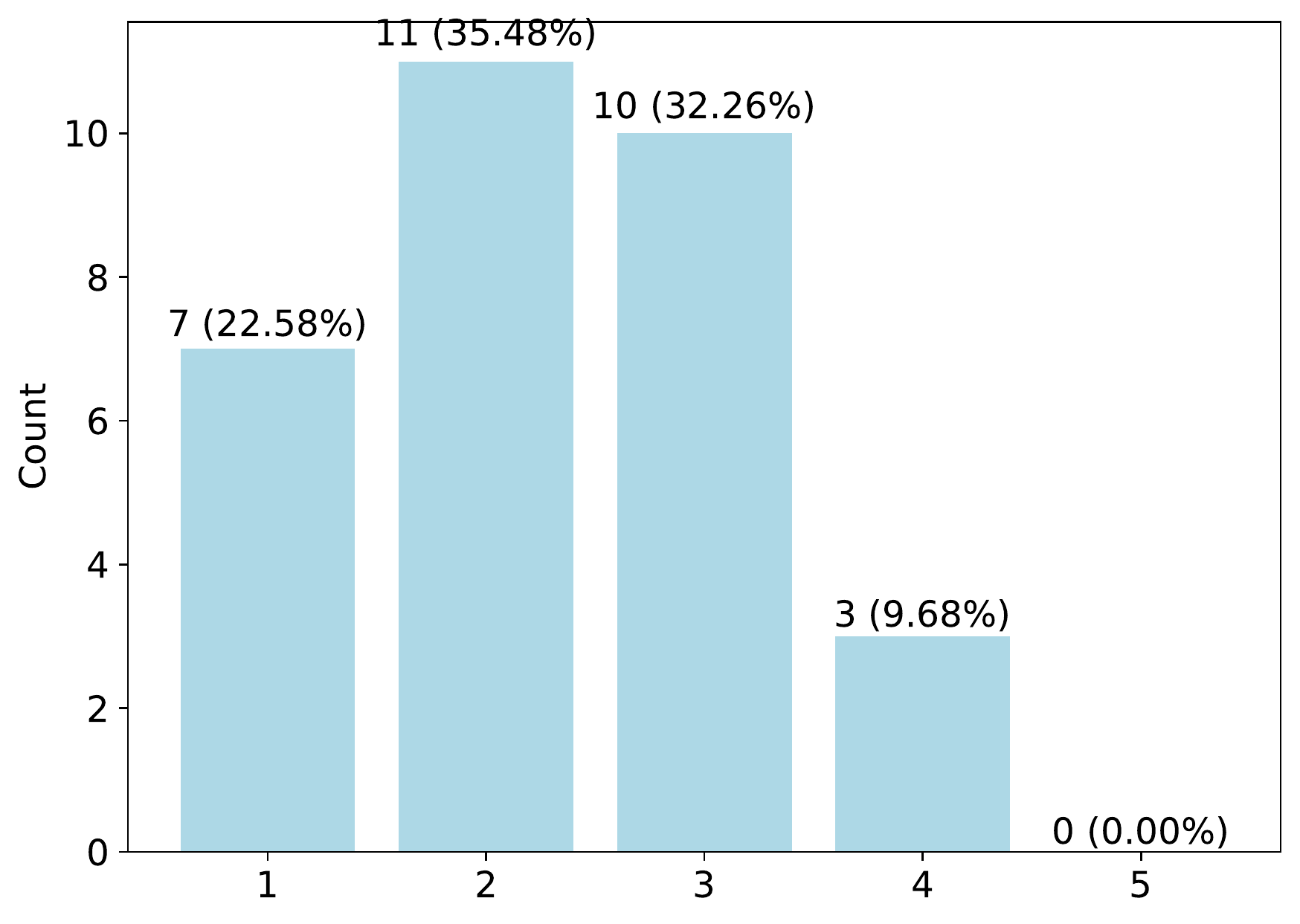}
         \caption{Web Server \& Client}
         \label{fig:exp_5}
     \end{subfigure}
     \\
     \begin{subfigure}[b]{0.3\textwidth}
         \centering
         \includegraphics[width=\textwidth]{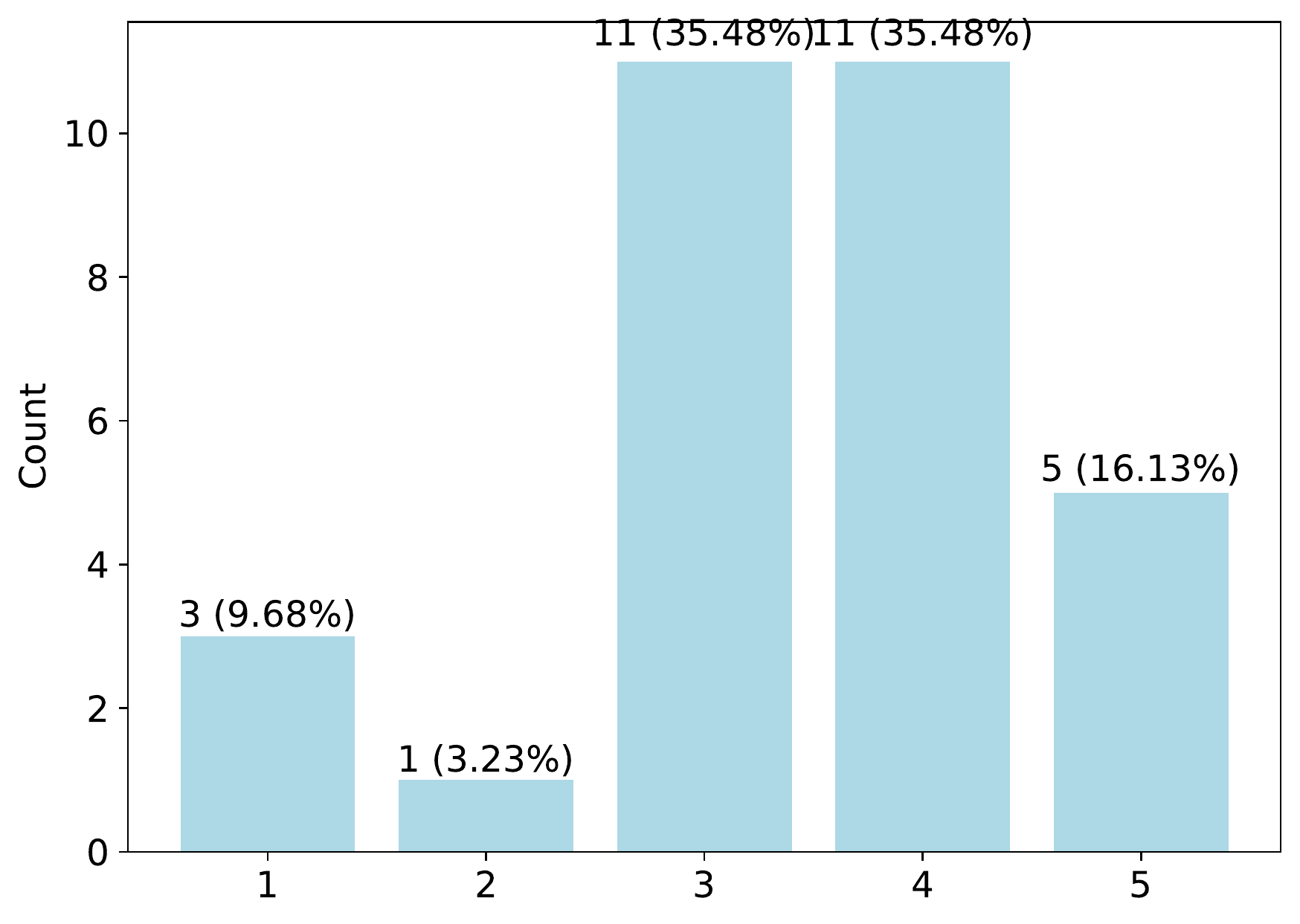}
         \caption{Data Analysis \& Machine Learning}
         \label{fig:exp_6}
     \end{subfigure}
     \quad
     \begin{subfigure}[b]{0.3\textwidth}
         \centering
         \includegraphics[width=\textwidth]{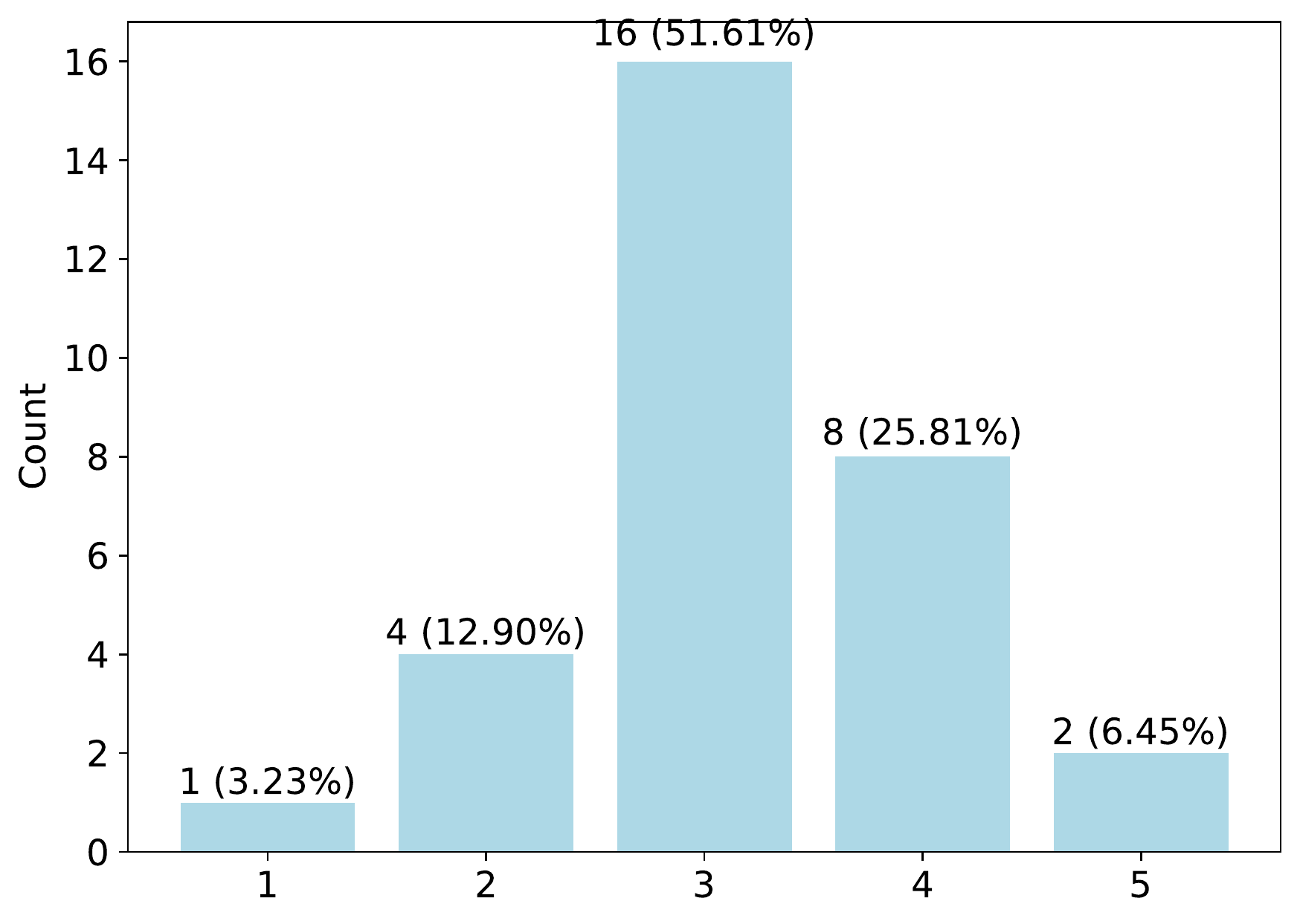}
         \caption{Data Visualization}
         \label{fig:exp_7}
     \end{subfigure}
        \caption{The experience and expertise for overall Python programming and 7 specific areas that we design different tasks for, from all the participants that completed the survey. 1 represents very inexperienced and 5 represents very experienced.}
        \label{fig:participants_experience}
\end{figure}

The detailed participants' programming experience responded in the survey is shown in Figure~\ref{fig:participants_experience}.

\section{Post-study Survey Details}
\label{app:post_survey_details}
After each task, we ask the following questions to all users (disregarding using the plugin or not) about the task design, self-assessment, as well as the help needed during the process:
\begin{enumerate}
    \item How difficult did you feel about the task? (1: very easy to 5: very hard)
    \item How would you evaluate your performance on the task?  (1: very bad to 5: very good)
    \item How often did you need to look for help during the task, including web search, looking up API references, etc.?  (1: not at all to 5: very often)
\end{enumerate}

For users that completed the current task with plugin enabled, the following additional questions about the plugin user experience are asked:
\begin{enumerate}
    \item How do you think the plugin impacted your efficiency timewise, if at all? (1: hindered significantly, to 3: neither hindered nor helped, to 5: helped significantly)
    \item How do you think the plugin impacted your quality of life, with respect to ease of coding, concentration, etc., if at all? (1: hindered significantly, to 3: neither hindered nor helped, to 5: helped significantly)
\end{enumerate}

After all assigned tasks are completed for the user, we ask them to complete a form about the overall experience with the user study and the evaluation of the plugin, as well as soliciting comments and suggestions.
\begin{enumerate}
    \item What did you think of the tasks assigned to you in general?
\item Overall, how was your experience using this plugin? (1: very bad to 5: very good)
\item What do you think worked well, compared with your previous ways to solve problems during programming?
\item What do you think should be improved, compared with your previous ways to solve problems during programming?
\item Do you have any other suggestions/comments for the plugin?
\end{enumerate}

\section{Plugin Effect on Code Complexity Metrics}
\label{app:code_complexity}
We also analyze the plugin's effect on code complexity metrics, following the same methods used in Section~\ref{sec:results:task-performance}.
We measure two standard proxies for code complexity of the Python 
programs produced by our study participants in each of their assigned 
tasks, \ie the number of source lines of code (SLOC) and McCabe's 
cyclomatic complexity (CC), a measure of the number of linearly independent 
paths through a program's source code~\cite{McCabe1976ACM};
in real programs, CC depends a lot on the ``if''-statements, as well as 
conditional loops, and whether these are nested.
The two measures tend to be correlated, but not strongly enough to 
conclude that CC is redundant with SLOC~\cite{landman2016empirical}.
We use the open-source library Radon\footnote{\url{https://github.com/rubik/radon}} 
to calculate CC.

One could expect that code produced by our NL2Code plugin may be more 
idiomatic (possibly shorter and less complex) than code written by the 
participants themselves.

\begin{figure}[t]
     \centering
    \includegraphics[width=\textwidth]{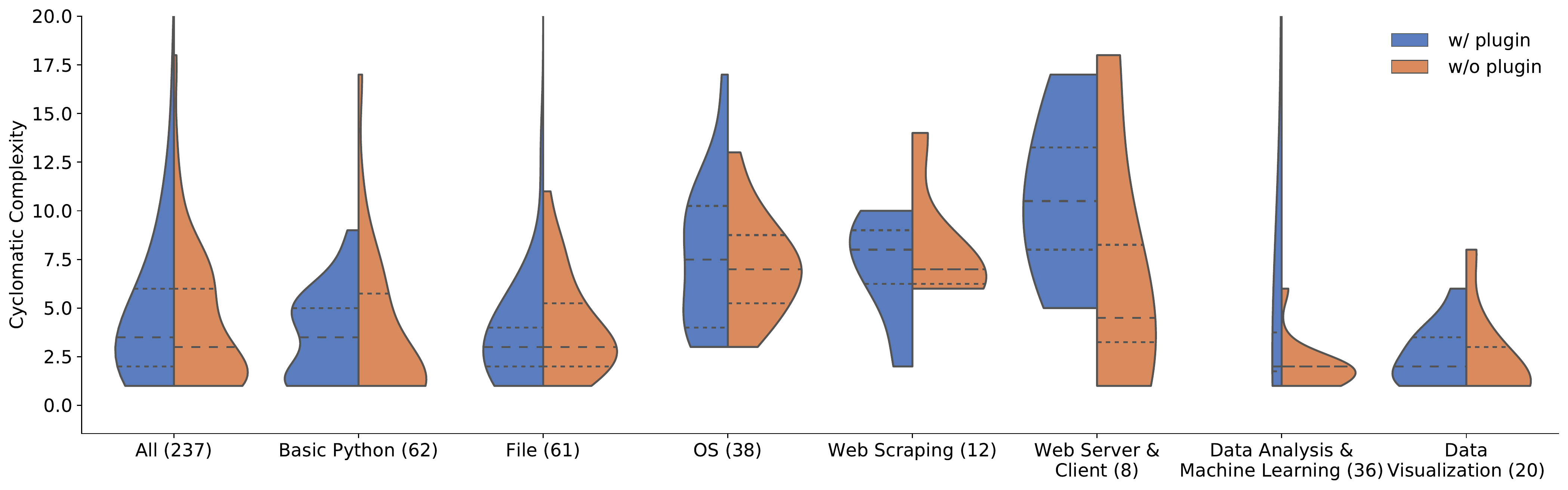}
        \caption{Distributions of cyclomatic complexity values across tasks
        and conditions. The horizontal dotted lines represent 25\% and 75\% quartiles, and the dashed lines represent medians.} 
        \label{fig:cc_category}
\end{figure}

\begin{figure}[t]
     \centering
    \includegraphics[width=\textwidth]{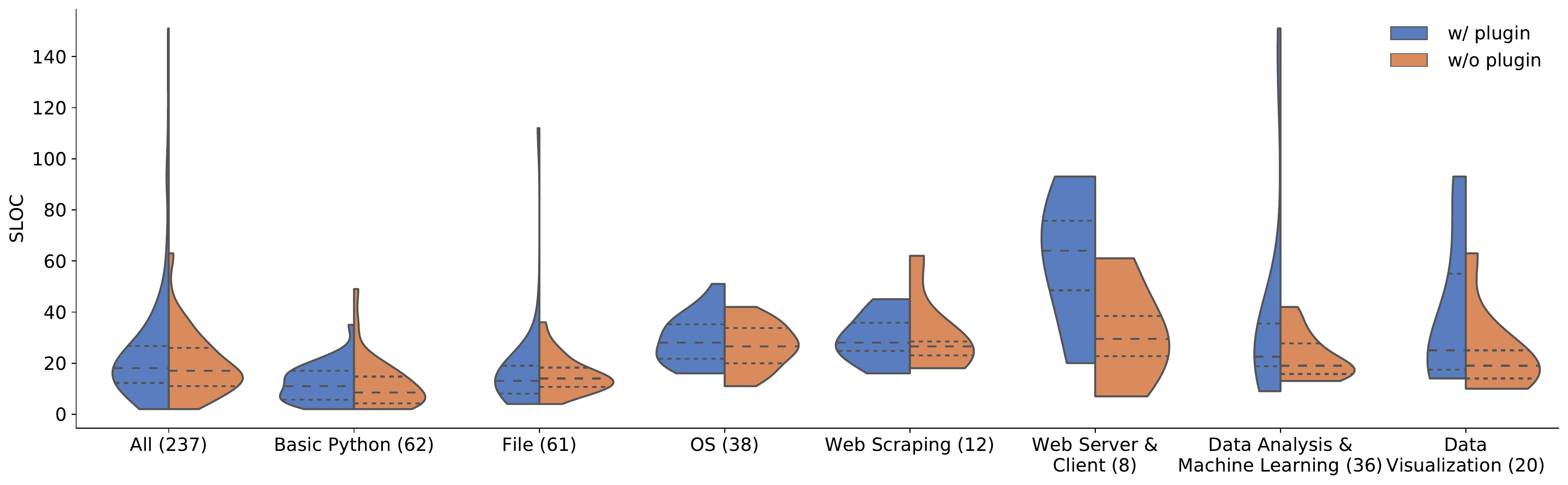}
        \caption{Distributions of SLOC values across tasks
        and conditions. The horizontal dotted lines represent 25\% and 75\% quartiles, and the dashed lines represent medians.}
        \label{fig:sloc_category}
\end{figure}

Figure~\ref{fig:cc_category} shows the distributions of CC values 
across tasks and conditions.
Figure~\ref{fig:sloc_category} shows the distributions of SLOC values 
across tasks and conditions.

Table~\ref{tbl:task-performance-code-complexity} summarizes our default specification mixed-effects regressions with CC and SLOC variables included;
the models with our second specification (de-meaned task experience)
are shown in Appendix~\ref{app:demeaned-models}.
The models fit the data reasonably well ($R^2_c = 50\%$ for SLOC, $R^2_c = 27\%$ for CC).

Analyzing the models we make the following observations.
There is no statistically 
significant difference between the two conditions in cyclomatic complexity values (model~(4)).
That is, the code written by users in the plugin condition appears statistically 
indistinguishably as correct and as complex from the code written by users 
in the control group.

We note a small effect of using the plugin on code length (model~(3)).
On average, the code written by users in the plugin condition is \textasciitilde 4
source lines of code longer than the code written by users without using
the plugin.
However, this effect is quite small, smaller than the standard deviation
of the random {\small \texttt{user}} intercept (\textasciitilde 6 source lines of code).

\section{NL2Code Plugin Query Syntax}
\label{app:plugin-syntax}

For the best results from the code generation model, we also instruct the users to write queries as expected by the model with the following rules:
\begin{itemize}
    \item Quote variable names in the query with grave accent marks: ... \`{}variable\_name\`{} ...
    \item Quote string literals with regular quotation marks: ... ``Hello World!'' ...
    \item Example query 1: open a file ``yourfile.txt'' in write mode.
    \item Example query 2: lowercase a string \`{}text\`{} and remove non-alphanumeric characters aside from space.
\end{itemize}

\section{Task Performance Models (De-meaned Specification)}
\label{app:demeaned-models}
Table~\ref{tbl:task-performance-demean} summarizes our alternative specification (de-meaned task experience)
mixed-effects regressions for two response variables in the main article, plug two response variables (CC and SLOC) introduced in Appendix~\ref{app:code_complexity}.

{\footnotesize 
\begin{table}[t] \centering 
  \caption{LMER task performance models (default specification, w/ code complexity metrics).} 
  \label{tbl:task-performance-code-complexity} 
\begin{tabular}{@{\extracolsep{5pt}}lD{.}{.}{-2} D{.}{.}{-2} D{.}{.}{-2} D{.}{.}{-2} } 
\\[-1.8ex]\toprule
\\[-1.8ex] 
 & \multicolumn{4}{c}{\textit{Dependent variable:}} \\ 
\cline{2-5} 
\\[-1.8ex] & \multicolumn{1}{c}{Completion time} & \multicolumn{1}{c}{Correctness score} & \multicolumn{1}{c}{SLOC} & \multicolumn{1}{c}{CC} \\ 
\\[-1.8ex] & \multicolumn{1}{c}{(1)} & \multicolumn{1}{c}{(2)} & \multicolumn{1}{c}{(3)} & \multicolumn{1}{c}{(4)}\\ 
\midrule \\[-1.8ex] 
 Experience & -195.62 & 0.07 & -0.62 & -0.21 \\ 
  & (183.11) & (0.24) & (1.61) & (0.46) \\ 
  Uses plugin & 15.76 & 0.44 & 4.16^{**} & 0.73 \\ 
  & (196.11) & (0.30) & (1.91) & (0.58) \\ 
  Constant & 3,984.51^{***} & 5.88^{***} & 27.15^{***} & 5.64^{***} \\ 
  & (838.07) & (1.03) & (7.40) & (1.95) \\ 
 \midrule \\[-1.8ex] 
Observations & 224 & 237 & 237 & 237 \\ 
Num users & 31 & 31 & 31 & 31 \\ 
Num tasks & 14 & 14 & 14 & 14 \\ 
sd(user) & 1489.25 & 0.82 & 6.16 & 1.18 \\ 
sd(task) & 1104.7 & 1.14 & 12.65 & 2.33 \\ 
R2m & 0.004 & 0.008 & 0.011 & 0.006 \\ 
R2c & 0.642 & 0.289 & 0.502 & 0.27 \\ 
Akaike Inf. Crit. & \multicolumn{1}{c}{3,987.14} & \multicolumn{1}{c}{1,106.66} & \multicolumn{1}{c}{2,002.42} & \multicolumn{1}{c}{1,417.27} \\ 
Bayesian Inf. Crit. & \multicolumn{1}{c}{4,007.61} & \multicolumn{1}{c}{1,127.46} & \multicolumn{1}{c}{2,023.23} & \multicolumn{1}{c}{1,438.08} \\ 
\bottomrule \\[-1.8ex] 
\textit{Note:}  & \multicolumn{4}{r}{$^{*}$p$<$0.1; $^{**}$p$<$0.05; $^{***}$p$<$0.01} \\ 
\end{tabular} 
\end{table}  }

{\footnotesize 
\begin{table}[h] \centering 
  \caption{LMER task performance models (de-meaned experience, w/ code complexity metrics).} 
  \label{tbl:task-performance-demean} 
\begin{tabular}{@{\extracolsep{5pt}}lD{.}{.}{-2} D{.}{.}{-2} D{.}{.}{-2} D{.}{.}{-2} } 
\\[-1.8ex]\toprule 
\\[-1.8ex] 
 & \multicolumn{4}{c}{\textit{Dependent variable:}} \\ 
\cline{2-5} 
\\[-1.8ex] & \multicolumn{1}{c}{Completion time} & \multicolumn{1}{c}{Correctness score} & \multicolumn{1}{c}{SLOC} & \multicolumn{1}{c}{CC} \\ 
\\[-1.8ex] & \multicolumn{1}{c}{(1)} & \multicolumn{1}{c}{(2)} & \multicolumn{1}{c}{(3)} & \multicolumn{1}{c}{(4)}\\ 
\midrule \\[-1.8ex] 
 Experience BTW & -478.55 & -0.04 & -1.47 & 0.04 \\ 
  & (566.62) & (0.43) & (2.98) & (0.74) \\ 
  Experience WI & -166.14 & 0.12 & -0.30 & -0.35 \\ 
  & (191.33) & (0.29) & (1.87) & (0.56) \\ 
  Uses plugin & 14.47 & 0.44 & 4.15^{**} & 0.74 \\ 
  & (196.07) & (0.30) & (1.90) & (0.58) \\ 
  Constant & 5,142.42^{**} & 6.32^{***} & 30.59^{**} & 4.62 \\ 
  & (2,348.61) & (1.77) & (12.60) & (3.07) \\ 
 \midrule \\[-1.8ex] 
Observations & 224 & 237 & 237 & 237 \\ 
Num users & 31 & 31 & 31 & 31 \\ 
Num tasks & 14 & 14 & 14 & 14 \\ 
sd(user) & 1482.32 & 0.81 & 6.15 & 1.17 \\ 
sd(task) & 1107.9 & 1.13 & 12.69 & 2.32 \\ 
R2m & 0.012 & 0.008 & 0.012 & 0.007 \\ 
R2c & 0.643 & 0.287 & 0.504 & 0.269 \\ 
Akaike Inf. Crit. & \multicolumn{1}{c}{3,988.86} & \multicolumn{1}{c}{1,108.56} & \multicolumn{1}{c}{2,004.30} & \multicolumn{1}{c}{1,419.09} \\ 
Bayesian Inf. Crit. & \multicolumn{1}{c}{4,012.74} & \multicolumn{1}{c}{1,132.84} & \multicolumn{1}{c}{2,028.58} & \multicolumn{1}{c}{1,443.36} \\ 
\bottomrule \\[-1.8ex] 
\textit{Note:}  & \multicolumn{4}{r}{$^{*}$p$<$0.1; $^{**}$p$<$0.05; $^{***}$p$<$0.01} \\ 
\end{tabular} 
\end{table}  }

\section{User Queries}
\label{app:plugin-queries}

{\scriptsize \renewcommand*{\arraystretch}{0.8}
\begin{longtable}{lLL}
\caption{Unique successful user queries to the NL2Code plugin, per task, 
    for the \numparticipants study participants. Queries for
    which the participant chose a snippet produced by the code 
    generation model are shown in boldface, and in the remainder a retrieved snippet was used.} \label{tbl:queries} \\
\toprule
Task & Queries & \\
\midrule
\endfirsthead
\caption[]{(continued)}\\
\toprule
Task & Queries & \\
\midrule
\endhead
\bottomrule
\endfoot
T1-1 & \textbf{call \texttt{\`}pick$\_$with$\_$replacement\texttt{\`}} & \textbf{how to generate random letter} \\
& \textbf{create a dictionary with keys \texttt{\`}random$\_$letters\texttt{\`} and values \texttt{\`}random$\_$numbers\texttt{\`}} & import library random \\
& \textbf{create dictionary} & \textbf{list to dict} \\
& create empty dictionary & loop on numbers from 0 to 100 \\
& \textbf{create list "a$\_$list"} & \textbf{loop over a range of \texttt{\`}count\texttt{\`}} \\
& defaultdict & merge 2 dictionaries \\
& \textbf{dictionary of characters and int} & \textbf{pair characters in \texttt{\`}characters\texttt{\`} and numbers in \texttt{\`}numbers\texttt{\`}} \\
& for loop on range 100 & \textbf{print \texttt{\`}dic\texttt{\`} keys on each line} \\
& generat integers 1-20 & \textbf{print \texttt{\`}dic\texttt{\`} keys sorted} \\
& generate 100 integers (1-20 inclusive). & \textbf{print \texttt{\`}dic\texttt{\`} sorted by keys} \\
& \textbf{generate 100 random lower-cased leters} & print a to z \\
& generate 100 random lowercase letters & \textbf{print list} \\
& \textbf{generate 100 random numbers} & print list as string \\
& \textbf{generate 100 random numbers from 1 to 20} & print list elements \\
& generate a rondom lower case character & print without newline \\
& generate char lower case & random \\
& generate dict & random character between a and z \\
& generate list of random charachters & random characters \\
& \textbf{generate lowercase char} & \textbf{random integer between 1 and 20} \\
& generate random & \textbf{random number} \\
& \textbf{generate random between 0 and 20} & \textbf{random sample with replacement} \\
& generate random charachter & randomly generate 100 letters \\
& \textbf{generate random int} & \textbf{randomly pick an item from \texttt{\`}seq\texttt{\`}} \\
& generate random letters & \textbf{rearrange dictionary keys into alphabetic order} \\
& generate random lower case letters & \textbf{sort a list} \\
& generate random nu,ber & \textbf{sort a list into ascending order} \\
& \textbf{generate random number} & \textbf{sort a list x into ascending order} \\
& \textbf{generate random numbers} & \textbf{sort dict by key} \\
& \textbf{generate random numbers between 1-20 inclusive} & \textbf{sort key of dict} \\
& \textbf{get a random letter} & sort list \\
& \textbf{given list \texttt{\`}letters\texttt{\`} and \texttt{\`}integers\texttt{\`}, create a dicitonary such that the values in \texttt{\`}letters\texttt{\`} are keys and values in \texttt{\`}integers\texttt{\`} are values} & \textbf{sort list 'values' into ascending order} \\
& how to append value in dict & squence of integers from 1 to 20 inclusive \\
& how to check if a  key is in a dictionay & zip 2 lists \\
& \textbf{how to generate random int in range between 1 and 20} & \textbf{zip \texttt{\`}hundred$\_$characters\texttt{\`} with \texttt{\`}hundred$\_$numbers\texttt{\`}} \\
T1-2 & add a week to a datetime & get gmt timezone \\
& \textbf{add days to time} & get now one week from now \\
& \textbf{assign current date and time to \texttt{\`}now\texttt{\`}} & \textbf{get the current date in utc} \\
& \textbf{change date format} & \textbf{get the current time in utc} \\
& \textbf{change datetime format of \texttt{\`}week$\_$date\texttt{\`} to mm-dd-yyyy hh:mm} & \textbf{get the date and time a week from now in gmt} \\
& \textbf{convert \texttt{\`}week$\_$date\texttt{\`} to GMT timezone and assign to \texttt{\`}GMT$\_$week$\_$date\texttt{\`}} & get time and date \\
& convert date timezone & \textbf{get time and date in gmt in \texttt{\`}date\texttt{\`}} \\
& date from 7 days & \textbf{get time and date one week from now} \\
& date gmt & \textbf{get time now} \\
& date now & gmt \\
& \textbf{datetime} & gmt time 24 \\
& \textbf{display \texttt{\`}week$\_$date\texttt{\`} in format mm-dd-yyyy hh:mm} & import datetime \\
& format datetime & \textbf{import time} \\
& format datetime 24 hour & mm-dd-yyyy \\
& \textbf{format time} & \textbf{print current date time} \\
& get current datetime & \textbf{print date and time in GMT in 24hr format} \\
& get date 7 days from today & \textbf{print datetime in mm-dd-yyyy hh:mm  format} \\
& \textbf{get date and time in gmt} & \textbf{time add} \\
& \textbf{get date and time one week from now} & time and date \\
& \textbf{get date time one week from now} & \textbf{time and date in certain} \\
& get datetime & \textbf{timedelta} \\
T2-1 & \textbf{copy column from "data.csv" file to another "output.csv"} & \textbf{new line} \\
& copy column from "data.csv" to "output.csv" & \textbf{number of columns of csv} \\
& create 'output.csv' csv file & open "data.csv" file \\
& csv write & \textbf{open a csv file \texttt{\`}data.csv\texttt{\`} and read the data} \\
& csv writer & \textbf{open csv} \\
& cvs & \textbf{open csv file \texttt{\`}data.csv\texttt{\`}} \\
& cvs files & \textbf{open csv file with read and write} \\
& \textbf{delete a column in csv} & \textbf{open file} \\
& delete column from csv & \textbf{pandas read csv} \\
& delete column from csv file & \textbf{pandas read csv named "data.csv"} \\
& delete first and last column in csv file & print csv without row numbers \\
& delete first and last column of \texttt{\`}df\texttt{\`} & python make dir \\
& \textbf{delete first and last row from the dataframe \texttt{\`}df\texttt{\`}} & \textbf{read "data.csv" file} \\
& \textbf{delete first row from dataframe \texttt{\`}df\texttt{\`}} & \textbf{read csv file "data.csv"} \\
& \textbf{delete row in csv} & \textbf{read csv file using pandas} \\
& delete the first column in csv file \texttt{\`}df\texttt{\`} & read csv pure python \\
& file to csv & read cvs \\
& \textbf{get current path} & remove columns from csv file and save it to another csv file \\
& get specific columns by index in pandas data frame & \textbf{remove first column from csv file} \\
& headers in a dataframe & save \texttt{\`}df\texttt{\`} to a file \texttt{\`}output.csv\texttt{\`} in a new directory \texttt{\`}example$\_$output\texttt{\`} \\
& how to delete a column in a dataframe python & \textbf{save dataframe to csv} \\
& how to delete columns in dataframe & save pandas dataframe to a file \\
& how to save a dataframe in csv file & \textbf{save this dataframe to a csv} \\
& if dir exist & \textbf{write \texttt{\`}output\texttt{\`} to csv file} \\
& if directory "output" exists & \textbf{write csv \texttt{\`}output$\_$f\texttt{\`} to file "output/output.csv"} \\
& make directory & \textbf{write output to csv file "output.csv"} \\
& make directory "output" if it doesn't exist & \textbf{write to csv file} \\
T2-2 & change directory & \textbf{list files in folder} \\
& \textbf{change directory to "data"} & \textbf{list of filenames from a folder} \\
& check file encoding & \textbf{move file to other directory} \\
& \textbf{check if directory exists} & \textbf{normalize newlines to \textbackslash n} \\
& convert binary decoded string to ascii & \textbf{open file} \\
& convert file encoding & \textbf{open text file} \\
& convert file to utf & read a file and iterate over its contents \\
& convert latin-1 to utf-8 & \textbf{read all files under a folder} \\
& \textbf{convert str to utf-8} & \textbf{read file} \\
& \textbf{convert text file encoding} & read ISO-8859-15 \\
& convert text files from encoding ISO-8859-15 to encoding UTF-8. & readline encoding \\
& copy a file & redirect \\
& \textbf{copy file} & \textbf{remove header} \\
& \textbf{copy file \texttt{\`}ddd.png\texttt{\`}} & remove heading white space \\
& copy file to other folder & text normalize newlines to \textbackslash n \\
& covert file to utf & traverse a directory \\
& \textbf{find character} & \textbf{travverse list of files} \\
& \textbf{get all files in directory} & \textbf{trim heading whitespace} \\
& get the file extension & \textbf{trim the heading and trailing whitespaces and blank lines for all text files} \\
& \textbf{iterating files in a folder} & unkown encoding \\
& \textbf{list all text files in the data directory} & write to file \\
& list files in directory \\
T3-1 & \textbf{check if \texttt{\`}file\texttt{\`} is a directory} & match regex year month day \\
& \textbf{check if string has specific pattern} & \textbf{move file} \\
& \textbf{copy a file to dist} & move files from directory to directory \\
& copy all files and directories from one folder to another & recursive copy files and folders \\
& \textbf{copy directory to another directory} & \textbf{recursively iterate over all files in a directory} \\
& copy directory to directory & \textbf{regex dd-mm-yy} \\
& copy directory tree from source to destination & regex digit python \\
& \textbf{copy file from \texttt{\`}src$\_$path\texttt{\`} to \texttt{\`}dest$\_$path\texttt{\`}} & regex for date \\
& copy files & \textbf{regex replace capture group} \\
& copy files and directories under \texttt{\`}data\texttt{\`} directory & regexp date \\
& copy files creating directory & rename file \\
& copy files from folder & \textbf{rename file with regex} \\
& create file & \textbf{rename files} \\
& create folder & replace pattern in string \\
& datetime to string & \textbf{search all matches in a string} \\
& extract year month day from string regex & \textbf{search for pattern "\%d\%d-\%d\%d" in \texttt{\`}file\texttt{\`}} \\
& get all files and folders & walk all files in a directory \\
& get the files that inside the folders & walk all nested files in the directory "data" \\
& \textbf{list all filepaths in a directory} & \textbf{walke all files in a directory} \\
& make a folder recersively & write to file \\
T3-2 & \textbf{add entry to json file} & \textbf{load json file} \\
& \textbf{check if file \texttt{\`}output$\_$file\texttt{\`} exists} & load json from a file \\
& check if file ends with .json & \textbf{read a json file named \texttt{\`}f\texttt{\`}} \\
& convert dict to string & sorting a dictionary by key \\
& \textbf{convert list to dictionary} & write into txt file \\
& import json parsing library & write json in \texttt{\`}ret\texttt{\`} to file \texttt{\`}outfile\texttt{\`} \\
T4-1 & find all bold text from html \texttt{\`}soup\texttt{\`} & parse all hyperlinks from \texttt{\`}r\texttt{\`} using bs4 \\
& find all hrefs from \texttt{\`}soup\texttt{\`} & visit \texttt{\`}url\texttt{\`} and extract hrefs using bs4 \\
& find all red colored text from html \texttt{\`}soup\texttt{\`} & visit the given url \texttt{\`}url\texttt{\`} and extract all hrefs from there \\
& go to a url & \textbf{visit the url \texttt{\`}url\texttt{\`}} \\
& how to get page urls beautifulsoup \\
T4-2 & create directory & \textbf{regex []} \\
& download an image request & save dict to csv \\
& extract imafe from html & \textbf{save table beautifulsoup} \\
& \textbf{http reques get html} \\
T5-1 & \textbf{add json file to a list} & check email correctness \\
T5-2 & argparse subprogram & \textbf{print format} \\
& \textbf{exit program} & request with params \\
& \textbf{gET request to "https://jsonplaceholder.typicode.com/posts" with argument userId} \\
T6-1 & a list of dictionary to pandas dataframe & pandas change dataframe column name \\
& add a new column to a dataframe row & pandas create buckets by column value \\
& \textbf{average by group pandas} & pandas dropnan \\
& \textbf{cast a float to two decimals} & pandas get average of column \\
& \textbf{cast a list to a dataframe} & pandas group by \\
& \textbf{column to integer pandas} & \textbf{pandas join dataframes} \\
& \textbf{create a dataframe from a list} & pandas join series into dataframes \\
& csv & pandas output csv \\
& csv write & pandas print with two decimals \\
& delete coloumn pd & \textbf{pandas read from csv} \\
& df set column to 7 decimals & \textbf{pandas round value} \\
& filter df with two conditions & pandas save csv two decimal \\
& filter values in pandas df & \textbf{pandas to csv} \\
& find unique data from csv & pandas to csv decimal \\
& \textbf{findall} & pandas write df to csv \\
& floating data in csv group in digit & pandas write to csv file \\
& \textbf{format output to 2 decimal} & pandas write to file decimal \\
& \textbf{get average of row values in pandas dataframe} & \textbf{read csv} \\
& get average value from group of data in csv & \textbf{read csv file} \\
& get the head of dataframe \texttt{\`}df\texttt{\`} & remove repeated column in csv file \\
& group by range pandas & \textbf{rename column pandas} \\
& group of data  from csv & rename pandas df columns \\
& \textbf{how to combine 2 lists into a dictionary} & \textbf{round a variable to 2dp} \\
& how to remove an item from a list using the index & save \texttt{\`}compan$\_$df\texttt{\`} dataframe to a file \\
& import pandas & save \texttt{\`}compand$\_$df\texttt{\`} dataframe to a file \\
& list to an entry in pandas dataframe & \textbf{sort dataframe \texttt{\`}jdf\texttt{\`} by \texttt{\`}scores\texttt{\`}} \\
& \textbf{load csv file with pandas} & \textbf{sort dataframe \texttt{\`}jdf\texttt{\`} by the values of column 'scores'} \\
& loop files recursive & \textbf{sort pandas dataframe} \\
& \textbf{newline space} & standard deviation from group of data in csv \\
& pandas add new column based on row values & two deciaml place \\
& \textbf{pandas calculate mean} & \textbf{write \texttt{\`}final$\_$data\texttt{\`} to csv file "price.csv"} \\
T6-2 & cross validation in scikit learn & multinomial logistic regression model \\
& cross validation mean accuracy & \textbf{numpy load from csv} \\
& disable warnings & run 5-fold accuracy \\
& how to determine cross validation mean in scikit learn & set numpy random seed to 0 \\
& how to split dataset  in scikit learn & sklearn 5 fold cross validation \\
& how to split dataset in scikit learn & sklearn 5-fold cross validation \\
& linear regressor 5 folder cross validation & sklearn cross validation x, y for 5 folds \\
& load wine dataset & sklearn ignore warnings \\
T7-1 & how to choose plot size in inches & plt set x axis tick range \\
& how to choose plot title in matplotlib & plt set xtick font size \\
& how to create ascatter plot using matplotlib & \textbf{reformat date} \\
& \textbf{how to draw scatter plot for data in csv file} & \textbf{save plot as image} \\
& \textbf{plt create figure with size} & \textbf{save plt figure} \\
& plt date as x axis & \textbf{scatter} \\
& plt set x axis label & scatter plot purple \\
T7-2 & bar graph side by side & \textbf{plot bar} \\
& bar plot with multiple bars per label & plot size \\
& get height of bars in subplot bar gaphs & plot title \\
& get labels above bars in subplots & \textbf{plt ax legend} \\
& \textbf{group pandas df by two columns} & plt ax xlabel \\
& \textbf{horizontal subplot} & plt create 3 subplots \\
& import matplotlib & plt set title for subplot figure \\
& matplotlib grouped bar chart & plt set x tick labels \\
& matplotlib multiple histograms & plt show values on bar plot \\
& matplotlib theme & \textbf{pyplot subplots} \\
& pandas dataframe from csv & select row pandas \\
& \textbf{pandas dataframe groupby column} \\

\bottomrule
\end{longtable}
 }

\section{Randomly Sampled User Queries for the Oracle Analysis}
\label{app:oracle-queries}

{\scriptsize \renewcommand*{\arraystretch}{0.8}
\begin{longtable}{lLL}
\caption{Sampled user queries for the oracle analysis. Queries for
    which the user chose a snippet from the code 
    generation model are shown in boldface. $\bullet$ denotes 
    queries ``good enough'' on their own; $\circ$ denotes queries
    good enough given the rest of the source file as context; the
    former is a strict subset of the latter.} \label{tbl:queries-sampled} \\
\toprule
Task & Queries & \\
\midrule
\endfirsthead
\caption[]{(continued)}\\
\toprule
Task & Queries & \\
\midrule
\endhead
\bottomrule
\endfoot
T1-1 & \textbf{call \texttt{\`}pick$\_$with$\_$replacement\texttt{\`} $\circ$} & defaultdict \\
& \textbf{generate lowercase char $\bullet\circ$} & for loop on range 100 $\bullet\circ$ \\
& \textbf{generate random between 0 and 20 $\bullet\circ$} & generate char lower case \\
& \textbf{random sample with replacement $\bullet\circ$} & generate random letters $\bullet\circ$ \\
& \textbf{sort key of dict $\bullet\circ$} & random characters \\
T1-2 & \textbf{change datetime format of \texttt{\`}week$\_$date\texttt{\`} to mm-dd-yyyy hh:mm $\bullet\circ$} & format datetime \\
& \textbf{convert \texttt{\`}week$\_$date\texttt{\`} to GMT timezone and assign to \texttt{\`}GMT$\_$week$\_$date\texttt{\`} $\bullet\circ$} & get gmt timezone $\circ$ \\
& \textbf{print datetime in mm-dd-yyyy hh:mm  format $\bullet\circ$} & get now one week from now $\bullet\circ$ \\
& date now $\bullet\circ$ & get time and date $\bullet\circ$ \\
T2-1 & \textbf{remove first column from csv file $\bullet\circ$} & how to delete columns in dataframe $\circ$ \\
& csv writer & open "data.csv" file $\bullet\circ$ \\
& how to delete a column in a dataframe python $\circ$ \\
T2-2 & traverse a directory $\circ$ \\
T3-1 & \textbf{copy a file to dist $\circ$} & recursive copy files and folders $\circ$ \\
& match regex year month day & regexp date \\
T4-2 & download an image request & save dict to csv \\
T5-2 & \textbf{exit program $\bullet\circ$} & argparse subprogram \\
T6-1 & \textbf{load csv file with pandas $\bullet\circ$} & how to remove an item from a list using the index $\bullet\circ$ \\
& \textbf{pandas round value $\circ$} & pandas create buckets by column value \\
& \textbf{pandas to csv} & pandas group by \\
& \textbf{read csv file $\bullet\circ$} & pandas output csv $\circ$ \\
& \textbf{rename column pandas $\circ$} & pandas to csv decimal $\circ$ \\
& filter df with two conditions & pandas write df to csv \\
T6-2 & load wine dataset \\
T7-1 & \textbf{plt create figure with size $\bullet\circ$} & \textbf{scatter $\circ$} \\
T7-2 & \textbf{plt ax legend $\circ$} & plt create 3 subplots $\bullet\circ$ \\
& bar plot with multiple bars per label $\circ$ \\

\bottomrule
\end{longtable}
 }

\end{document}